\documentclass[10pt,journal,compsoc]{IEEEtran}
\usepackage[utf8]{inputenc}
\usepackage[T1]{fontenc}
\usepackage{amsmath,amssymb}
\usepackage{amsthm}
\usepackage{booktabs}
\usepackage{multirow}
\usepackage{array}
\usepackage{graphicx}
\usepackage{xcolor}
\usepackage{tikz}
\usetikzlibrary{arrows.meta,positioning,shapes.misc,calc,fit,backgrounds,decorations.pathreplacing}
\usepackage{url}
\usepackage{cite}
\usepackage{microtype}
\makeatletter
\let\supp@maketitle\maketitle
\let\supp@@maketitle\@maketitle
\let\supp@thanks\thanks
\edef\supp@textfloatsep{\the\textfloatsep}
\edef\supp@floatsep{\the\floatsep}
\edef\supp@dbltextfloatsep{\the\dbltextfloatsep}
\edef\supp@dblfloatsep{\the\dblfloatsep}
\newcommand{\BeginPaperSupplement}{%
  \clearpage
  \setcounter{page}{1}%
  \setcounter{section}{0}%
  \setcounter{subsection}{0}%
  \setcounter{subsubsection}{0}%
  \setcounter{figure}{0}%
  \setcounter{table}{0}%
  \setcounter{equation}{0}%
  \setcounter{theorem}{0}%
  \setcounter{proposition}{0}%
  \setcounter{definition}{0}%
  \setcounter{remark}{0}%
  \setcounter{footnote}{0}%
  \setlength{\textfloatsep}{\supp@textfloatsep}%
  \setlength{\floatsep}{\supp@floatsep}%
  \setlength{\dbltextfloatsep}{\supp@dbltextfloatsep}%
  \setlength{\dblfloatsep}{\supp@dblfloatsep}%
  \let\maketitle\supp@maketitle
  \let\@maketitle\supp@@maketitle
  \let\thanks\supp@thanks
  \def\@IEEEtitleabstractindextext{}%
}
\makeatother
\graphicspath{{figures/}}

\theoremstyle{plain}
\newtheorem{theorem}{Theorem}
\newtheorem{proposition}{Proposition}
\theoremstyle{definition}
\newtheorem{definition}{Definition}
\newtheorem{remark}{Remark}

\newcommand{\sortM}{\mathsf{M}}
\newcommand{\sortA}{\mathsf{A}}
\newcommand{\inv}{\mathsf{I}}
\newcommand{\op}[1]{\mathrm{#1}}
\newcommand{\calU}{\mathcal{U}}
\newcommand{\calF}{\mathcal{F}}
\newcommand{\R}{\mathbb{R}}
\newcommand{\Q}{\mathbb{Q}}

\begin{document}
\bstctlcite{TKDEbstcontrol}
\title{Adjustment-Anchor Dependence in Formulaic Alpha Languages:\\ A Certified Grammar and a Re-Anchoring Audit}

\author{Zhengkai~Tu,
        Xinyue~Xiang,
        and~Jimmy~Huang
\IEEEcompsocitemizethanks{
\IEEEcompsocthanksitem Z.~Tu is with The Chinese University of Hong Kong, Hong Kong SAR, China.
\IEEEcompsocthanksitem X.~Xiang is with the University of Michigan, Ann Arbor, MI, USA.
\IEEEcompsocthanksitem J.~Huang is with York University, Toronto, ON, Canada. Corresponding author: J.~Huang (jhuang@yorku.ca).}
}

\markboth{IEEE Transactions on Knowledge and Data Engineering}{Tu \MakeLowercase{\textit{et al.}}: Adjustment-Anchor Dependence in Formulaic Alpha Languages}

\IEEEtitleabstractindextext{%
\begin{abstract}
Several public scripts for formulaic alpha mining store prices adjusted to the snapshot's last day, so a past day's stored price changes with every later split or dividend and formulas that do not cancel the adjustment factor depend on future corporate actions. We formalise a change of adjustment anchor as a per-stock group action and give a two-sorted type system that soundly certifies anchor invariance from a formula's syntax, under a declared field-adjustment contract and idealised operator semantics; its second sort admits log-returns, and a linear-time checker plugs into any miner. In a re-anchoring audit on Chinese A-shares, pools that five published methods mined in our benchmark pipeline score a mean information coefficient (IC) 0.005 higher on the stored representation than on a cutoff-anchored one, about 14\% of their IC. The mean signed change for certified pools is about $-0.001$, and pools mined on past-anchored data stay sensitive. On the Chinese development folds, certification shows no detectable predictive cost or gain; the estimated minimum detectable difference is about 0.017 IC across cells on pool ICs of 0.03--0.04. We also trace an erroneous index-membership entry to public Qlib data snapshots and show that re-mining exposes the effect of such data defects, which re-evaluating fixed pools understates.
\end{abstract}
}

\maketitle
\IEEEdisplaynontitleabstractindextext
\IEEEpeerreviewmaketitle

\section{Introduction}\label{sec:intro}

\begin{figure}[!t]
\centering
\includegraphics[width=\columnwidth]{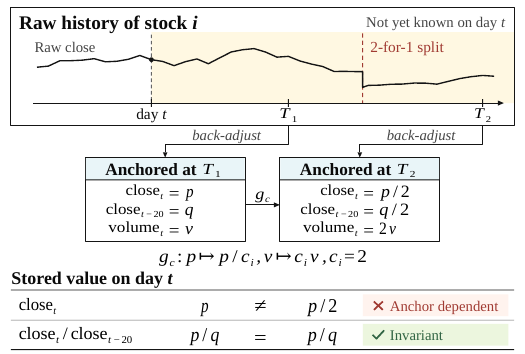}
\caption{The anchor problem under end-anchored adjustment. The same raw history of stock $i$ is stored twice, anchored at $T_1$ before a 2-for-1 split and at $T_2$ after it; the anchor group action $g_c$ with $c_i = 2$ maps one snapshot onto the other. The stored close on day $t$ is $p$ in one and $p/2$ in the other, so a bare price depends on a corporate action that had not yet happened on day $t$. The ratio $\op{close}_t/\op{close}_{t-20}$ is identical in both.}
\label{fig:problem}
\end{figure}

\IEEEPARstart{A}{djusted} prices are raw prices multiplied by a cumulative split-and-dividend factor. Several public collection scripts for formulaic alpha mining normalise that factor at the last day of the snapshot. The BaoStock script released with AlphaGen~\cite{yu2023alphagen}, and redistributed with AlphaForge and AlphaSAGE~\cite{shi2025alphaforge,chen2026alphasage}, does exactly this (Section~S5; items numbered with S are in the Supplementary Material). The close it stores for a day in 2015 therefore changes every time the snapshot is refreshed after a later split or dividend (Fig.~\ref{fig:problem}). A ratio of two prices of the same stock is unaffected. A bare price, a power of a price or a cross-sectional rank of price levels is affected, and a factor built from them depends on corporate actions that had not happened on the day it is evaluated. Formulaic alpha miners search over exactly such expressions, score them against forward returns and keep a pool of weakly correlated factors for a combiner. The question is which formulas computed on their output depend on the anchor.

\begin{figure*}[!t]
\centering
\includegraphics[width=\textwidth]{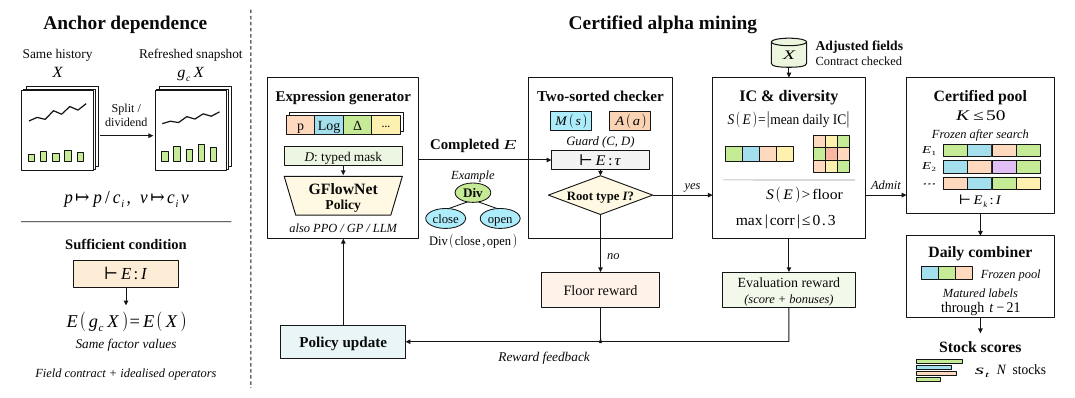}
\caption{Overview. Left: A later split or dividend changes the stored history: the refreshed snapshot is the group action $g_c$ on the stored fields, and a certified formula, $\vdash E : \inv$, returns the same values on both snapshots under the field contract and the idealised operator semantics (Proposition~\ref{prop:protection}, Theorem~\ref{thm:soundness}). Right: The generator builds an expression, the checker certifies it before any evaluation, a rejected expression receives the floor reward, and a certified one is scored on the discovery data ($S(E)$: absolute mean daily IC), admitted to the pool past the IC floor and the correlation limit, and combined daily on matured labels.}
\label{fig:framework}
\end{figure*}

Three things have to hold before a factor value can be trusted. The stored fields must be adjusted the way the provider declares, which we call the \emph{field contract} and check on the values. The formula must not depend on the anchor, which a \emph{symbolic certificate} establishes under idealised operator semantics. And the executed code must behave like the formula, which we record as a \emph{numerical validation status} after re-anchoring. Our own pipeline once broke the contract (Section~\ref{sec:data}), and fixing the data does not settle the second point: pools mined on data anchored in the past still change when the anchor moves (Section~\ref{sec:exp-anchor}).

A miner adds a further exposure, because it selects its output by scoring the discovery data. A mislisted index ticker in public Qlib data snapshots, touching at most 0.1\% of the discovery stock-days, changed the pools our miner returned, and re-evaluating the fixed pools understated it (Section~\ref{sec:exp-sh000905}). Data problems compound downstream in other pipelines in the same way~\cite{sambasivan2021cascades,northcutt2021labelerrors}.

We ask three questions. Can a composable certificate of invariance to the anchor be built from a formula's syntax (RQ1)? What does enforcing it cost a miner (RQ2)? How far does a change in the discovery data travel into the pool a miner returns, compared with its effect on evaluating a pool already found (RQ3)? Fig.~\ref{fig:framework} links anchor dependence to certified mining.

We reduce immunity to later corporate actions to invariance under a per-stock group action that divides every price of a stock by a constant and multiplies its volume by it. A two-sorted type system certifies that invariance from the syntax, soundly but not completely (Section~\ref{sec:grammar}). Its additive sort records the shift into which a logarithm turns the scale, so it admits log-returns, which exponent bookkeeping alone rejects. We attach the checker to a GFlowNet miner and to three other generators and measure its cost on Chinese A-shares under frozen stage-level protocols (Sections~\ref{sec:protocol} and~\ref{sec:exp}); the S\&P~500 serves mainly as a second provider for the audit.

\textbf{Contributions.}
(1) \textit{A certified grammar and checker.} We reduce immunity to later corporate actions to invariance under a per-stock group action (Proposition~\ref{prop:protection}) and give a two-sorted type system that certifies it from the syntax, with soundness and equivariance theorems (Theorems~\ref{thm:soundness} and~\ref{thm:equivariance}) and a precision result for its root rule (Proposition~\ref{prop:rootexact}). The checker runs in linear time and attaches unchanged to GFlowNet, reinforcement-learning, genetic-programming and LLM generators. An event-level check of the field contract and a float64 reference evaluator tie the certificate to the stored data and to the executed code (Sections~\ref{sec:data} and~\ref{sec:exp-anchor}).
(2) \textit{A re-anchoring audit.} Pools that five published methods (AlphaGen, AlphaSAGE, AlphaForge, AlphaQCM and genetic programming) mined in our earlier benchmark pipeline score a mean IC 0.005 higher on the stored end-anchored representation than on a cutoff-anchored one, about 14\% of their stored IC, while the mean change for certified pools is $-0.001$. Pools mined on past-anchored data stay sensitive to the anchor (Section~\ref{sec:exp-anchor}).
(3) \textit{Re-mining as a data-quality audit.} A mislisted index in public Qlib snapshots, and injected defects that mimic it, change re-mined pools far more than re-evaluating fixed pools reveals, and about two to three times more than a matched $10^{-6}$ perturbation (Section~\ref{sec:exp-sh000905}).
(4) \textit{Empirical assessment of predictive cost.} On 80 clean units, certification has no detectable effect on predictive accuracy; the estimated minimum detectable difference is about 0.017 IC over cells (Section~\ref{sec:exp-main}); results on later windows, the S\&P~500 and other generators are mixed (Section~\ref{sec:exp-transfer}).

\section{Related Work}
\label{sec:related}

\textbf{Look-ahead bias by construction.}
Practitioners re-anchor the series at its first date~\cite{portfoliooptimizer2022adjusted} or adjust past bars at run time~\cite{realtest2026split}. The closest formal work is Fonseca's type-and-effect system~\cite{fonseca2026lookahead}, which verifies look-ahead freedom as temporal non-interference: a decision at epoch $t$ may consume only values whose declared availability time is at most $t$. The stamps are declared at ingestion and trusted. The close stored under a day in 2015 therefore passes the check although it is a function of later splits and dividends, and stamping it correctly requires a history of vintages that the snapshot does not keep. Our certificate needs no availability history. Fonseca's bias taxonomy~\cite{fonseca2026pointintime} names price-data forward leakage as a bias class without saying which formulas inherit it; leakage benchmarks and constrained agent pipelines~\cite{zhang2026whenalpha,huang2026hypotheses} address other parts of backtesting correctness. Backtest overfitting~\cite{bailey2014pseudo,bailey2017probability} and the rewriting of database history~\cite{ljungqvist2009rewriting} motivate point-in-time data and frozen protocols.

\textbf{Data fixes and the certificate.} Anchoring at the listing date (backward adjustment, \emph{hfq}, AlphaGen's current default) or keeping a point-in-time database of vintages removes the future dependence of a stored snapshot. The certificate complements both: it names the formulas on an existing end-anchored snapshot that the anchor cannot affect, it adds the formula check that an undocumented convention needs beside the value check (Section~\ref{sec:data}), and a pool mined on past-anchored data stays tied to where the data begin (Section~\ref{sec:exp-anchor}).

\textbf{Data validation.} Deequ~\cite{schelter2018deequ} and TFX Data Validation~\cite{breck2019datavalidation} check declarative constraints on the values of a dataset, as our field contract does (Section~\ref{sec:data}). The certificate answers a complementary question, whether a formula on data that pass these checks depends on the anchor.

\textbf{Typed expression search.}
In symbolic regression, typing enforces physical units. Dimensionally consistent genetic programming goes back to Keijzer and Babovic~\cite{keijzer1999dimensionally}, $\Phi$-SO~\cite{tenachi2023phiso} masks unit-inconsistent tokens during reinforcement learning, and unit-aware genetic programming costs little accuracy~\cite{reuter2024unitaware}. Kennedy's dimension types~\cite{kennedy1997parametricity} prove by relational parametricity that well-typed programs are invariant under changes of units. Atkey, Johann and Kennedy~\cite{atkey2013abstraction} generalise this to types indexed by algebraic structures, including groups of scalings and translations, and derive invariance theorems from parametricity. Our two sorts specialise that framework to financial data, one scaling per stock with the translation it induces under a logarithm. On top of the specialisation we contribute the model that links the adjustment anchor to look-ahead (Proposition~\ref{prop:protection}) and the root rule that turns the bookkeeping into a certificate checkable in linear time. We also add the data contract and checks that make it hold on a real provider, and its use as a search constraint inside miners. Dimension types accept any unit-consistent program, whereas our root rule demands the identity type. SymLang~\cite{baig2026symlang} types a grammar with dimension vectors, parity and physical-symmetry classes, AlphaCFG~\cite{yang2026alphacfg} restricts alpha search with a context-free grammar without typing, and Alpha$^2$~\cite{xu2024alpha2}, the closest typed prior work, keeps two categories (currency for prices, unit for volume) and forbids addition across them. It has no exponent bookkeeping or root constraint and states no invariance property; our reconstruction of its rules is a control arm (Section~\ref{sec:coarse}). Table~\ref{tab:novelty} summarises what each line of work provides and what this paper adds. Market microstructure invariance~\cite{kyle2016invariance} concerns scaling laws of markets, Villar \textit{et al.}~\cite{villar2023dimensionless} canonicalise to dimensionless features, and generalisation results for compact groups~\cite{elesedy2021provably} do not apply, as $(\R_+)^{|\calU|}$ is not compact.

\begin{table*}[!t]
\caption{Closest prior work and what this paper adds on dependence on the adjustment anchor.}
\label{tab:novelty}
\centering
\scriptsize
\setlength{\tabcolsep}{3pt}
\renewcommand{\arraystretch}{1.0}
\begin{tabular}{@{}>{\raggedright\arraybackslash}p{0.17\textwidth}>{\raggedright\arraybackslash}p{0.22\textwidth}>{\raggedright\arraybackslash}p{0.25\textwidth}>{\raggedright\arraybackslash}p{0.32\textwidth}@{}}
\toprule
Line of work & What it provides & What it leaves open here & What this paper adds \\
\midrule
Re-anchoring remedies~\cite{portfoliooptimizer2022adjusted,realtest2026split} & Re-anchor at the first date, or adjust bars at run time & Change the data; silent on formulas over an existing end-anchored snapshot & Certifies from the syntax which formulas the anchor cannot affect \\
Look-ahead type system~\cite{fonseca2026lookahead} & Look-ahead freedom from declared availability stamps & An end-anchored value is stamped wrongly without the vintage history & No availability history; immunity reduces to invariance (Proposition~\ref{prop:protection}) \\
Unit-typed regression, dimension and algebraically indexed types~\cite{keijzer1999dimensionally,tenachi2023phiso,reuter2024unitaware,kennedy1997parametricity,atkey2013abstraction} & Unit exponents; invariance from parametricity & No model of the anchor; any root type; no data contract or search & Additive sort for log-returns; root type $\inv$ required; soundness, equivariance and root-rule precision (Theorems~\ref{thm:soundness}, \ref{thm:equivariance}, Proposition~\ref{prop:rootexact}) \\
Typed formula search~\cite{baig2026symlang,yang2026alphacfg,xu2024alpha2} & Two categories (Alpha$^2$); a grammar (AlphaCFG); physical symmetries (SymLang) & No model of the adjustment-anchor action or of its non-interference guarantee & Anchor-specific exponent and log-shift typing with a root invariance rule; Alpha$^2$-style control arm (Section~\ref{sec:coarse}); four miners \\
\bottomrule
\end{tabular}
\end{table*}

\textbf{Platform and evaluation of mined factors.}
AlphaGen~\cite{yu2023alphagen} (proximal policy optimisation (PPO) over reverse-Polish tokens), AlphaForge~\cite{shi2025alphaforge} (a daily re-fitted linear combiner), AlphaQCM~\cite{zhu2025alphaqcm} (distributional reinforcement learning) and AlphaSAGE~\cite{chen2026alphasage} (a GFlowNet~\cite{bengio2021gflownet} over a relational-graph encoder~\cite{schlichtkrull2018rgcn}) form our platform. Across five equity markets, FactorBench~\cite[Section~4.3]{wang2026factorbench} reports negative aggregate residual IC for AlphaGen, AlphaForge and AlphaSAGE despite positive raw IC, while AlphaQCM retains most of its raw IC after neutralisation. A third-party re-run puts AlphaSAGE at a CSI300 IC of 0.033 with a seed standard deviation of 0.016~\cite{dhan2026alpharjm}, several times what the original papers report, which our protocol (Section~\ref{sec:protocol}) is designed around.

\section{Preliminaries}
\label{sec:prelim}

\subsection{Formulaic Alphas and the Search}
\label{sec:alphas}

For readers outside finance, a \emph{factor} is a formula that scores every stock every day. Its \emph{IC} is the cross-sectional Pearson correlation between those scores and the next 20-day return, averaged over days. A \emph{pool} is a set of factors a miner has kept, and a \emph{combiner} turns a pool into one score per stock and day. Let $\calU$ be a stock universe and $X_{\le t}$ the stored fields $\op{open}$, $\op{high}$, $\op{low}$, $\op{close}$, $\op{vwap}$ and $\op{volume}$ through trading day $t$. A formulaic alpha is an expression tree $E$ over the fields, numeric constants $\{\pm 30, \pm 10, \pm 5, \pm 2, \pm 1, \pm 0.5, \pm 0.01\}$, window lengths $\{10, 20, 30, 40, 50\}$ and the 36 operators listed in Table~\ref{tab:rules}. Expressions are written in reverse Polish notation with at most 20 tokens. A pool holds at most $K = 50$ expressions in order.

The miner (Section~\ref{sec:search}) samples token sequences from a GFlowNet policy and scores each completed expression by the absolute mean daily Pearson IC on the discovery window. An expression is admitted to the pool if its absolute IC exceeds a floor and its maximal absolute correlation with the pool members is at most 0.3. Prediction uses the frozen pool through a combiner that is re-fitted daily on realised labels (Section~\ref{sec:combiners}).

\subsection{Adjusted Prices and the Anchor Convention}
\label{sec:anchor}

Vendors deliver raw prices $p^{\mathrm{raw}}_{t,i}$ and a cumulative adjustment factor. In the Chinese provider used here, the stored adjusted price is $p_{t,i} = p^{\mathrm{raw}}_{t,i}\, F_{t,i}$ with $F_{t,i} = A_{t,i}/A_{T_{\mathrm{end}},i}$, where $A$ is the cumulative split-and-dividend factor and $T_{\mathrm{end}}$ is the last day of the snapshot. Volume is stored as $v_{t,i} = v^{\mathrm{raw}}_{t,i}/F_{t,i}$ and VWAP as $\op{vwap}^{\mathrm{raw}}_{t,i}\, F_{t,i}$. We call this declared treatment of each field the \emph{field contract}. Prices are \emph{end-anchored}: $F_{t,i}$ depends on corporate actions in $(t, T_{\mathrm{end}}]$.

BaoStock's forward adjustment (qfq), rescaled as in the scripts of Section~\ref{sec:intro}, is end-anchored; backward adjustment (hfq), AlphaGen's current default, is anchored at the listing date, and Qlib's U.S. bundle at each stock's first valid day (Sections~S5 and~\ref{sec:exp-anchor}).

Changing the anchor, or refreshing the snapshot, multiplies all prices of stock $i$ by a constant $1/c_i$ and all volumes by $c_i$, where $c_i$ depends on the corporate actions between the two anchors (Fig.~\ref{fig:problem}).

\begin{definition}[anchor group action]
\label{def:group}
For $c = (c_i)_{i \in \calU} \in (\R_+)^{|\calU|}$, let $g_c$ act on the field space by $p \mapsto p / c_i$ for every price field ($\op{open}$, $\op{high}$, $\op{low}$, $\op{close}$, $\op{vwap}$) and by $v \mapsto v \cdot c_i$ for $\op{volume}$, on every day. The action is constant in time and varies across stocks. An expression $E$ is \emph{anchor-invariant} if $E(g_c X) = E(X)$ pointwise for all $c$ on the domain of $E$.
\end{definition}

The convention here scales volume for cash dividends exactly as for splits. A provider that adjusts volume for splits only has a different field contract, under which volume is not of type $\sortM(+1)$, so a user should check the convention before relying on that type. Under the split-only convention, running the checker for the split and the dividend scaling and requiring both to accept keeps 96--97\% of the factors in the certified pools certified; the rest multiply a price by volume (Section~S16.1). Price times volume is invariant under the convention here; bare prices, bare volume and powers of a price are not.

\begin{proposition}[sufficient condition for immunity to the anchor]
\label{prop:protection}
Let the provider be end-anchored and honour the field contract. Consider two histories $\omega$ and $\omega'$ whose raw data and corporate actions are identical up to and including day $t$ and differ only after $t$. Then the stored fields through $t$ satisfy $X_{\le t}(\omega') = g_c X_{\le t}(\omega)$ with $c_i = A_{T_{\mathrm{end}},i}(\omega')/A_{T_{\mathrm{end}},i}(\omega)$.
(i)~If $E(g_c X)_{t,i} \ne E(X)_{t,i}$ for this induced $c$, then the stored value $E_{t,i}$ differs between two histories that are identical through $t$, so $E_{t,i}$ is not determined by the information $\calF_t$ available at $t$.
(ii)~If $E$ is anchor-invariant, then $E_{t,i}$ is the same on both histories for every such pair.
\end{proposition}
We call the composed guarantee, field contract plus root type $\inv$, \emph{conditional anchor non-interference}: conditional on the contract, and restricted to the channel through the anchor.

\begin{proof}
Because $A_{t,i}$ is common to both histories, $p_{t,i}(\omega') = p^{\mathrm{raw}}_{t,i} A_{t,i}/A_{T_{\mathrm{end}},i}(\omega') = p_{t,i}(\omega)/c_i$, and likewise $v_{t,i}(\omega') = c_i\, v_{t,i}(\omega)$, for every day up to $t$. Both statements follow from Definition~\ref{def:group}.
\end{proof}

Invariance is sufficient, not necessary, and it addresses the channel through the anchor; a field that breaks the contract, a mis-dated input and a label that looks ahead are separate routes.

\section[A Certified Adjustment-Invariant Grammar]{A Certified Adjustment-Invariant\\ Grammar}\label{sec:grammar}

\subsection{Two Sorts}\label{sec:sorts}

\begin{figure}[!t]
\centering
\includegraphics[width=\columnwidth]{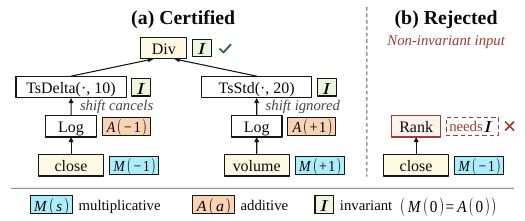}
\caption{Two typing derivations drawn as trees, with each node's type beside it. (a) The worked example of this section. The logarithm turns the scale of a price or a volume into a shift; $\op{TsDelta}$ cancels the shift and $\op{TsStd}$ ignores it, so both operands are invariant and their quotient is certified. (b) $\op{Rank}(\op{close})$ is rejected because $\op{Rank}$ accepts only an invariant input.}
\label{fig:typing}
\end{figure}

We assign each sub-expression a \emph{type} that describes how its value transforms under the anchor group action $g_c$ of Definition~\ref{def:group}. There are two sorts. Under the multiplicative sort $\sortM(s)$, with scale exponent $s \in \Q$, the value transforms as $x \mapsto c_i^{\,s} x$. Under the additive sort $\sortA(a)$, with log-shift coefficient $a \in \Q$, it transforms as $x \mapsto x + a \log c_i$. The types $\sortM(0)$ and $\sortA(0)$ coincide and denote invariance, written $\inv$. Price fields have type $\sortM(-1)$, volume has type $\sortM(+1)$, and constants and window lengths have type $\inv$. Table~\ref{tab:rules} gives the typing rules. The judgement $\vdash E : \tau$ either succeeds or fails with a reason, and $E$ is \emph{certified} if and only if $\vdash E : \inv$ (two derivations drawn as trees: Fig.~\ref{fig:typing}). We use three terms with distinct meanings. \emph{Certified} is this syntactic judgement, \emph{invariant} is the semantic property of Definition~\ref{def:group}, and \emph{numerically invariant} is the outcome of a finite re-anchoring test on executed float32 code (Section~\ref{sec:exp-anchor}).

\begin{table}[!t]
\caption{Typing rules of the two-sorted checker (summary; complete rule set: Table~S1). ``$\sortM(s)$ only'' means that an $\sortA(a \ne 0)$ input is rejected.}
\label{tab:rules}
\centering
\scriptsize
\setlength{\tabcolsep}{3pt}
\renewcommand{\arraystretch}{1.1}
\begin{tabular}{@{}>{\raggedright\arraybackslash}p{0.39\columnwidth}>{\raggedright\arraybackslash}p{0.24\columnwidth}>{\raggedright\arraybackslash}p{0.32\columnwidth}@{}}
\toprule
Operator & Input type(s) & Output \\
\midrule
price field / volume / constant & -- & $\sortM(-1)$ / $\sortM(+1)$ / $\inv$ \\
$\op{Abs}$, $\op{Inv}$, $\op{Sign}$ & $\sortM(s)$ only & $\sortM(s)$, $\sortM(-s)$, $\inv$ \\
$\op{Log}$ & $\sortM(s)$ only & $\sortA(s)$ \\
$\op{Rank}$, $\op{SLog1p}$ & $\inv$ only & $\inv$ \\
$\op{Add}$, $\op{Sub}$, $\op{Greater}$, $\op{Less}$ & $\inv$, $\inv$ & $\inv$ \\
$\op{Add}$, $\op{Sub}$, $\op{Greater}$, $\op{Less}$ & $\sortM(s)$, $\sortM(r)$ & $\sortM(s)$ iff $s = r$, else rejected \\
$\op{Add}$, $\op{Sub}$ & $\sortA(a)$, $\sortA(b)$ & $\sortA(a \pm b)$ \\
$\op{Greater}$, $\op{Less}$ & $\sortA(a)$, $\sortA(b)$ & $\sortA(a)$ iff $a = b$, else rejected \\
$\op{Add}$, $\op{Sub}$, $\op{Greater}$, $\op{Less}$ & $\sortM(s \ne 0)$, $\sortA(a \ne 0)$ & rejected \\
$\op{Mul}$ & $\sortM(s)$, $\sortM(r)$ & $\sortM(s + r)$ \\
$\op{Div}$ & $\sortM(s)$, $\sortM(r)$ & $\sortM(s - r)$ \\
$\op{Mul}$ & $\sortA(a)$, constant $q$ & $\sortA(aq)$ \\
$\op{Div}$ & $\sortA(a)$, constant $q \ne 0$ & $\sortA(a/q)$ \\
$\op{Pow}$ & $\sortM(s)$, constant $q$ & $\sortM(sq)$; non-constant exponent only if both $\inv$ \\
$\op{Ref}$, $\op{TsMean}$, $\op{TsMax}$, $\op{TsMin}$, $\op{TsMed}$, $\op{TsWMA}$, $\op{TsEMA}$ & $\tau$ & $\tau$ \\
$\op{TsSum}$ (window $w$) & $\sortM(s)$ / $\sortA(a)$ & $\sortM(s)$ / $\sortA(aw)$ \\
$\op{TsStd}$, $\op{TsMad}$, $\op{TsMaxDiff}$, $\op{TsMinDiff}$, $\op{TsMinMaxDiff}$, $\op{TsDelta}$ & $\sortM(s)$ / $\sortA(a)$ & $\sortM(s)$ / $\inv$ \\
$\op{TsVar}$ & $\sortM(s)$ / $\sortA(a)$ & $\sortM(2s)$ / $\inv$ \\
$\op{TsSkew}$, $\op{TsKurt}$, $\op{TsRank}$ & any & $\inv$ \\
$\op{TsIr}$, $\op{TsDiv}$, $\op{TsPctChange}$ & $\sortM(s)$ only & $\inv$ \\
$\op{TsCov}$ & $\sortM(s)$ or $\sortA$, $\sortM(r)$ or $\sortA$ & $\sortM(s' + r')$, $s' = s$ for $\sortM(s)$ and $0$ for $\sortA$ \\
$\op{TsCorr}$ & any, any & $\inv$ \\
\bottomrule
\end{tabular}
\end{table}

An invariant operand of a binary operator is read as $\sortM(0)$ or $\sortA(0)$ to match the other operand, so a constant added to a price is rejected and a constant added to a log-price is a shifted log-price (Table~S1). The additive sort separates this system from exponent bookkeeping. $\log \op{close}$ has type $\sortA(-1)$ and is rejected at the root, whereas $\op{TsDelta}(\log \op{close}, 10)$, $\log \op{close} - \log \op{open}$ and $\op{TsStd}(\log \op{close}, 20)$ are certified. A derivation reads bottom up. In $\op{Div}(\op{TsDelta}(\log \op{close}, 10), \op{TsStd}(\log \op{volume}, 20))$, $\op{close} : \sortM(-1)$ gives $\log \op{close} : \sortA(-1)$, and $\op{TsDelta}$ removes the shift, giving $\inv$. Likewise $\op{volume} : \sortM(+1)$ gives $\log \op{volume} : \sortA(+1)$, which $\op{TsStd}$ maps to $\inv$, and the quotient of two invariant operands is $\inv$, so the expression is certified. $\op{Rank}(\op{close})$ fails at its root, because $\op{Rank}$ accepts only $\inv$. In the certified pools of the original-membership confirmatory stage (Section~S25), 4.0\% of the factors (44 unique expressions, each a shift-free statistic of a log-price or log-volume) are certified only by the two-sorted system, and 9.4\% under the typed action mask.

\subsection{Theorems}\label{sec:theorems}

The following results concern the algebraic semantics of the operators without implementation-specific absolute numerical guards, which we call the \emph{idealised operator semantics}. Concretely, $\op{Greater}$ and $\op{Less}$ are pointwise maximum and minimum, $\op{Log}$ is defined on positive arguments, a division on non-zero divisors, $\op{Pow}$ where the power is real, and a statistic that divides by a window's standard deviation on windows with non-zero variance. Elsewhere the value is non-finite, and non-finite values propagate (Section~S1). We also assume fixed, value-independent rolling weights ($\op{TsWMA}$ and $\op{TsEMA}$ normalise theirs), identical handling of missing values on both sides of the group action, and full windows, so that incomplete windows lie outside the domain of $E$. One excluded guard is the fixed variance floor of the implemented $\op{TsCorr}$. It replaces the normalisation by 1 when a window's sum of squared deviations falls below $10^{-6}$, so scaling the prices can turn a correlation into an unnormalised covariance (Section~\ref{sec:exp-anchor}; counterexample: Section~S16.4).

\begin{theorem}[soundness]
\label{thm:soundness}
If $\vdash E : \inv$, then for every $c \in (\R_+)^{|\calU|}$ and every $(t,i)$ in the domain of $E$, $E(g_c X)_{t,i} = E(X)_{t,i}$, and the sets of non-finite outputs coincide.
\end{theorem}

\begin{theorem}[equivariance]
\label{thm:equivariance}
If $\vdash E : \sortM(s)$, then $E(g_c X)_{t,i} = c_i^{\,s}\, E(X)_{t,i}$. If $\vdash E : \sortA(a)$, then $E(g_c X)_{t,i} = E(X)_{t,i} + a \log c_i$.
\end{theorem}

Theorem~\ref{thm:equivariance} describes how every typable expression transforms, and Theorem~\ref{thm:soundness} is its case $s = 0$ or $a = 0$, stated first because it is the certificate. Both are proved by structural induction on $E$, rule by rule (full case-by-case proof: Section~S1). A typical case is $\op{TsStd}$. On a window ending at $t$, every value of an operand of type $\sortM(s)$ is multiplied by $c_i^{\,s}$ and every value of type $\sortA(a)$ is shifted by $a \log c_i$. The standard deviation scales with the first map and ignores the second, so $\op{TsStd}$ maps $\sortM(s)$ to $\sortM(s)$ and $\sortA(a)$ to $\inv$, and a window has zero variance before the action if and only if it has zero variance after it. The rules compose through every operator, including those where the exponent changes ($\op{Pow}$, $\op{TsVar}$, $\op{TsSum}$ on the additive sort) and those where the obvious extension of unit typing fails. A system that treats the logarithm of a scaled quantity as dimensionless would certify $\op{Rank}(\log \op{close})$ and $\op{TsIr}(\log \op{close}, 20)$, yet a per-stock shift reorders the stocks in the first and changes the second, whose numerator shifts while its denominator does not. The rules use only per-entity positive scalings of the fields, so $k$ independent per-entity factors give $\mathbb{Q}^k$-graded sorts with the same proofs.

\subsection{Incompleteness}\label{sec:incomplete}

Among typable expressions, the root rule loses only expressions with trivial values. This is a statement about the root rule; completeness of the whole language is a separate matter, discussed below.
\begin{proposition}[precision of the root rule]\label{prop:rootexact}
Under the idealised operator semantics, let $\vdash E : \tau$ and fix an input $X$. If $\tau = \sortA(a)$ with $a \ne 0$, then $E(g_c X)_{t,i} = E(X)_{t,i}$ for every $c$ and every $(t,i)$ at which $E(X)$ is finite if and only if $E(X)$ is finite at no $(t,i)$. If $\tau = \sortM(s)$ with $s \ne 0$, the same holds if and only if $E(X)_{t,i} = 0$ at every $(t,i)$ at which $E(X)$ is finite.
\end{proposition}

The proof is in Section~S1.

\begin{remark}[expressiveness]\label{prop:separation}
The expressions certified by the single-sort system (the $\sortM$ rules alone, with $\op{Log}$ admitted only on $\inv$) form a strict subset of those certified by the two-sorted system: a single-sort derivation is a two-sorted derivation without $\sortA$, and $\op{TsDelta}(\log \op{close}, 10)$ is certified only by the latter.
\end{remark}
By Proposition~\ref{prop:rootexact}, an invariant expression that the checker rejects is either untypable at an internal node or typable with a trivial value, such as $\op{Sub}(\op{close}, \op{close})$ of type $\sortM(-1)$.

The system is a sufficient condition, so a rejected expression may still be invariant. A non-affine cancellation such as the expanded polynomial of $(\log \op{close} - \log \op{open})^2$ is untypable although the whole cancels the shift. A degenerate expression whose sub-expression is constant or saturated on the data range is numerically invariant under the tested re-anchorings on that panel only (81 of the 2,000 arm-A factors of the original-membership confirmatory pools, Section~S16.1).

Of 1,709 unique expressions in 40 earlier unrestricted pools, a single-sort system certifies 22.1\%, the two-sorted system 23.1\% and the Alpha$^2$-style coarse typing 88.3\% (Section~S1).

\subsection{A Coarse Control}\label{sec:coarse}

To separate invariance from grammar pruning, we reconstruct the Alpha$^2$ rule set~\cite{xu2024alpha2} from the examples in its paper, as its full rules are not published (Section~S16.1). It accepts 88.3\% of the unique expressions above, provides no guarantee of anchor invariance, and is the coarse-typed control (arm B).

\section{Certified Search}\label{sec:search}

\begin{figure*}[!t]
\centering
\includegraphics[width=\textwidth]{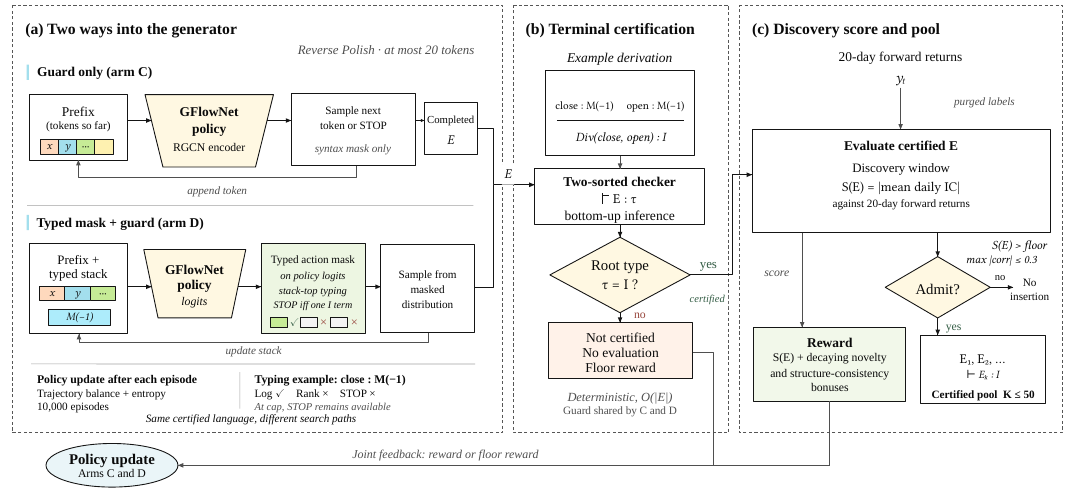}
\caption{Certified search. (a) Arm C samples the next token under the syntax mask and checks the completed expression; arm D also masks the policy's logits for tokens that fail typing on the stack top, including STOP while the stack is not one term of type $\inv$ (at the length limit STOP stays available). Both arms search the same certified language. (b) The terminal guard types the completed expression bottom up and tests the root; a rejected expression is not evaluated and receives the floor reward. (c) A certified expression is scored on the discovery window ($S(E)$: absolute mean daily IC against the purged 20-day forward return), admitted to the pool past the IC floor and the correlation limit, and rewarded with $S(E)$ plus decaying novelty and structure-consistency bonuses; the reward trains the policy by trajectory balance.}
\label{fig:search}
\end{figure*}

The type checker of Section~\ref{sec:grammar} can enter the miner after an expression is complete, as a rejection guard, or during construction, as a typed action mask (Fig.~\ref{fig:search}).

\subsection{Generator}\label{sec:generator}

We use the GFlowNet generator of AlphaSAGE~\cite{chen2026alphasage} unchanged. Its forward policy builds an expression of at most 20 tokens by choosing the next token or termination, and it is trained for 10,000 episodes with trajectory balance~\cite{whitammer2022trajectory} and an entropy regulariser. The reward is the pool-admission score (absolute mean daily Pearson IC on the discovery window) plus linearly decaying novelty and structure-consistency terms. Invalid terminal states get a floor reward (state encoding and optimiser: Section~S16.2).

\subsection{Rejection Guard (Arm C)}\label{sec:guard}

The guard places the checker between completion and reward. A completed expression that is not certified is rejected before evaluation and receives the floor reward, like a syntactically invalid one. Checking is linear in the size of the expression (6~ms for the 2,000 pool factors of arm A on one laptop core); the guard's real cost is the trajectories it rejects (Section~S10).

\subsection{Typed Action Mask (Arm D)}\label{sec:mask}

The mask moves the checker into the environment, where the generator already masks syntactically invalid tokens. Each applicable operator token is typed on the top of the operand stack and masked if typing fails. Termination is masked unless the stack holds one expression of type $\inv$; at the length limit it stays available, and the guard remains as a safety net. Because the rules are compositional, mask plus guard accepts exactly the language of the guard alone, so the two differ only in how the policy reaches it (Sections~S10 and~S16.2).

\section{Evaluation Protocol}\label{sec:protocol}

\subsection{Data and the Corrected Provider}\label{sec:data}

We use daily open, high, low, close, volume and VWAP of the CSI300 and CSI500 constituents in Qlib format~\cite{yang2020qlib}, from 2010-01-04 to 2026-08-21 (1,730 stocks in the union universe). The CSI500 membership file is the one in the v2 and v3 data snapshots distributed through Qlib's download utility, with the end dates of open spells extended by the AlphaGen collection script. It lists the CSI500 index itself, SH000905, as a constituent from 2007-01-15 to 2012-06-25 (Sections~\ref{sec:exp-sh000905} and~S5).

\textbf{The inconsistent build.} An earlier build of ours stored adjusted open, high, low and close but unadjusted VWAP and volume, which breaks the field contract of Section~\ref{sec:anchor}. VWAP/close, a certified ratio, then carried the inverse of the cumulative adjustment factor and predicted 20-day CSI300 returns with an IC of 0.042 (0.003 after repair). The earlier pipeline's pools fell from a CSI300 IC of 0.0749 to 0.0477 once the data were repaired (provenance: Section~S5, Fig.~S4). We multiplied VWAP by the price factor, divided volume by it and froze the corrected provider with a hash manifest.

\textbf{Checking the field contract.} A range test requires an adjusted VWAP to lie inside the adjusted daily $[\mathrm{low}, \mathrm{high}]$ interval. On all constituent stock-days in 2010--2024, index tickers excluded, the inconsistent build fails it on 86\% and the corrected provider on one stock-day. The mislisted index ticker of Section~\ref{sec:exp-sh000905} fails it on every one of its days. An event-level check compares the stored data on the 19,867 ex-dates of 2010--2026 with exchange-reported quantities that bypass our adjustment code (Section~S5). The stored adjusted return matches the exchange's percentage change, computed from the ex-rights reference price, within $10^{-3}$ on all but five ex-dates and on all but 15 of the 6.08 million other stock-days. On ex-dates with a factor change below 10\%, a proxy for cash dividends, the share count implied by volume, factor and turnover rate moves by less than half the factor change in 99.8\% of 17,161 cases. An unadjusted volume would move it by the whole change.

Unless stated otherwise, China-market pools were discovered and evaluated on the corrected provider. Section~\ref{sec:exp-sh000905} also uses variants of it: one without the mislisted index ticker and, on that variant, VWAP defects of divisor 400 for two drawn stocks and a divisor-4 defect and a $10^{-6}$ perturbation for the first (Sections~S5 and~S17). The retrospective comparison (Section~S4) re-evaluates existing external pools discovered on the reference snapshot, and the S\&P~500 experiment uses the public Qlib US bundle.

\subsection{Folds, Labels and Purging}\label{sec:folds}

Development uses four rolling folds, f2015\_16, f2017\_18, f2019\_20 and f2021, named after their validation years, with discovery windows that all start on 2010-01-04 (Table~S14). The target is the 20-day close-to-close forward return, a ratio of two closes of one stock and therefore anchor-invariant. Labels are purged 20 trading days before each cutoff. Two later windows (2022--2024 and 2025--2026) were exposed during earlier exploratory work, so they serve as retrospective evidence (Section~S16.4).

\subsection{Combiners}\label{sec:combiners}

A pool is turned into a daily score by one of two combiners, both using only labels realised by the prediction day, that is, matured statistics through $t-21$. The \textit{adaptive combiner} is AlphaForge's~\cite{shi2025alphaforge} with the AlphaSAGE release's thresholds~\cite{chen2026alphasage} and is the primary combiner in every stage protocol. Each day it selects the first 20 factors in pool order with $|\mathrm{RankIC}| > 0.015$ and $|\mathrm{RankICIR}| > 0.15$ on matured statistics, or the single best by $|\mathrm{RankICIR}|$ if none qualifies. It weights them by an expanding-window ordinary least-squares (OLS) fit on daily-normalised factor values. The \textit{equal-weight combiner} weights every factor equally, signed by its matured mean IC. Because the adaptive combiner's thresholds make it precision-sensitive, the main comparison is also reported under equal weighting and in float64 (Section~S24); float32 stays the primary evaluator because every frozen stage protocol specified it.

\subsection{Metrics and Statistical Units}\label{sec:metrics}

\begin{table}[!t]
\caption{Search arms and the data they were mined on. \emph{Orig.}: original CSI500 membership file (confirmatory stage); \emph{clean}: without the mislisted ticker. Units are fold $\times$ market $\times$ seed pools; evaluation is laptop float32 unless Table~\ref{tab:summary} says otherwise.}
\label{tab:arms}
\centering
\scriptsize
\setlength{\tabcolsep}{2pt}
\begin{tabular}{@{}l>{\raggedright\arraybackslash}p{0.37\columnwidth}>{\raggedright\arraybackslash}p{0.30\columnwidth}l@{}}
\toprule
Arm & Search & Data (units) & Section \\
\midrule
A & unrestricted grammar & orig.\ (40); clean (80) & 7.1--7.5 \\
C & certified guard & orig.\ (40); clean (80) & 7.1--7.5 \\
B & coarse, Alpha$^2$-style guard & orig.\ (40); clean (40) & 7.3 \\
S & single-sort guard & orig.\ (40); clean (40) & 7.3 \\
R & random guard, rate 0.896 & orig.\ (40) & 7.3 \\
D, Cm & mask plus guard; re-run of C & orig.\ (40 each) & 7.3 \\
AH & unrestricted grammar & past-anchored clean (40) & 7.1 \\
C, A & as above & clean with a field defect (20 per defect) & 7.5 \\
\bottomrule
\end{tabular}
\end{table}

An \emph{arm} is one search configuration (Table~\ref{tab:arms}), a \emph{cell} is a fold $\times$ market, and a \emph{unit} is one pool, a cell $\times$ seed. For each pool and combiner we report the mean daily IC over the validation window and the exposure-neutralised IC, which residualises the signal each day on 60-day beta, 20-day volatility, 20--240-day momentum, 5-day reversal and log liquidity. The statistical unit is a pool, giving 40 units per arm in the confirmatory stages and 80, ten seeds per cell, in the clean main comparison; factors are never treated as samples. Paired $t$ and Wilcoxon signed-rank $p$-values are descriptive and uncorrected for multiple comparisons, because seeds within a cell share the market and the window. As the 20-day labels overlap, we also report a moving-block bootstrap over trading days and cluster-robust intervals over cells and folds (Section~S13). The evaluator is platform-sensitive: per unit, adaptive IC differs between our cloud and laptop environments, and between float32 and float64, by an SD of 0.005--0.010, because a rounding difference can flip a threshold selection of the adaptive combiner (Section~S24). Each paired comparison therefore uses one environment, one precision and one evaluator, the laptop float32 environment unless stated.

\subsection{Frozen Stage-Level Protocols}\label{sec:frozen}

We worked in stages, each with a protocol whose text, decision rule, thresholds, code hashes and analysis script were recorded in a freeze file before the stage's outcomes were computed. Fixed-seed re-runs reproduce pools byte for byte, as did the control runs of every re-mining stage (Sections~S16.5, S17, S23 and~S25). The two confirmatory rules for the cost of the constraint were as follows.

\textit{Main comparison} (Section~\ref{sec:exp-main}): the criterion is met if the mean adaptive $\Delta\mathrm{IC}$ of certified minus unrestricted search is at least $-0.005$ and the corresponding check on the neutralised IC passes. Attribution requires $(\mathrm{C} - \mathrm{B}) \ge (\mathrm{B} - \mathrm{A}) - 0.002$.

\textit{Matched-pruning control} (Section~\ref{sec:exp-controls}): for the tested random scheme, the pruning explanation is rejected if and only if $\mathrm{C} - \mathrm{R} \ge +0.005$ and $\mathrm{R} - \mathrm{A} \le +0.005$ in adaptive mean IC over the paired units. Arm S was pre-specified as descriptive (Section~S6).

The frozen rules of the other confirmatory stages (Sections~\ref{sec:exp-controls} and~\ref{sec:exp-transfer}, and the size-matched analysis) and every deviation from the frozen protocols, with its timing relative to the first look at the affected outcomes, are in the Supplementary Material (Section~S6).

\section{Experiments}\label{sec:exp}

Section~\ref{sec:exp-anchor} takes the certificate to data (RQ1), Sections~\ref{sec:exp-main}--\ref{sec:exp-transfer} measure what the constraint costs a miner (RQ2), and Section~\ref{sec:exp-sh000905} follows changes in the discovery data through the search (RQ3). Confirmatory stages follow the frozen rules of Section~\ref{sec:frozen} and Section~S6, and Table~\ref{tab:summary} summarises the principal paired comparisons.

\begin{table}[!t]
\caption{Paired comparisons of Section~\ref{sec:exp} (adaptive combiner, validation windows, laptop float32 evaluation except the confirmatory stage, which is cloud; $+0.0096$ on the laptop, Section~S25). Tags: \emph{orig.}, mined with the original CSI500 membership file; \emph{clean}, without the mislisted ticker. $n$: units (c: cells); $\Delta$IC: mean paired difference; CI: 95\% interval over units; $>0$: units positive. Other generators, official miners and later windows: Tables~S11, S12 and S16; Sections~S16.4 and S19--S21. Rule: ``met'' means the stage's pre-specified engineering rule is met, not statistical significance or non-inferiority; ``desc.'': no rule was set.}
\label{tab:summary}
\centering
\scriptsize
\setlength{\tabcolsep}{2pt}
\renewcommand{\arraystretch}{0.88}
\begin{tabular}{@{}lrrrrl@{}}
\toprule
Comparison & $n$ & $\Delta$IC & 95\% CI & $>0$ & Rule \\
\midrule
\multicolumn{6}{@{}l}{\textit{Main comparison (Section~\ref{sec:exp-main})}} \\
\textbf{C $-$ A, clean, 10 seeds} & 80 & $\mathbf{+0.0044}$ & $[-0.0013{,}{+}0.0101]$ & 48/80 & met \\
C $-$ A, orig., cloud & 40 & $+0.0114$ & $[+0.0045{,}{+}0.0184]$ & 29/40 & met \\
\addlinespace
\multicolumn{6}{@{}l}{\textit{Controls and representation baseline (Sections~\ref{sec:exp-anchor}, \ref{sec:exp-controls})}} \\
C $-$ B (coarse), clean & 40 & $+0.0089$ & $[+0.0003{,}{+}0.0176]$ & 26/40 & post hoc \\
C $-$ R (random), orig. & 40 & $+0.0181$ & $[+0.0090{,}{+}0.0273]$ & 31/40 & met \\
R $-$ A, orig. & 40 & $-0.0085$ & $[-0.0161{,}{-}0.0010]$ & 17/40 & met \\
C $-$ AH, past-anchored & 40 & $+0.0070$ & $[+0.0010{,}{+}0.0129]$ & 29/40 & desc. \\
AH $-$ A, past-anchored & 40 & $-0.0015$ & $[-0.0079{,}{+}0.0049]$ & 21/40 & desc. \\
\addlinespace
\multicolumn{6}{@{}l}{\textit{Discovery-data changes, $\Delta$(C $-$ A) (Section~\ref{sec:exp-sh000905})}} \\
Injection /400, 2 tickers & 40 & $+0.0027$ & $[-0.0042{,}{+}0.0095]$ & 23/40 & desc. \\
Injection /4 & 20 & $+0.0060$ & $[-0.0006{,}{+}0.0125]$ & 14/20 & desc. \\
Perturbation $\times(1{+}10^{-6})$ & 20 & $+0.0006$ & $[-0.0030{,}{+}0.0042]$ & 8/20 & desc. \\
\bottomrule
\end{tabular}
\end{table}

\subsection{Certification and Re-Anchoring}\label{sec:exp-anchor}

Certified pools barely move when the anchor changes, and unrestricted pools do. We recomputed every factor of the 40 certified (arm C) and 40 unrestricted (arm A) pools of the confirmatory stage under re-anchored copies of the provider and re-ran the full combiner in float32. Simulated stock-specific anchors, at the fold's 50\% and 90\% training-calendar quantile dates, vary the corporate-action history and hold the observations fixed, as in Proposition~\ref{prop:protection}; a global scaling of every stock by one constant serves as a placebo for each. The \emph{real} anchor is the fold's discovery cutoff $T_c$, with $c_i = A_{T_c,i}/A_{T_{\mathrm{end}},i}$, which gives the vintage a snapshot taken at the cutoff would have stored (Section~S22).

\begin{figure}[!t]
\centering
\includegraphics[width=\columnwidth]{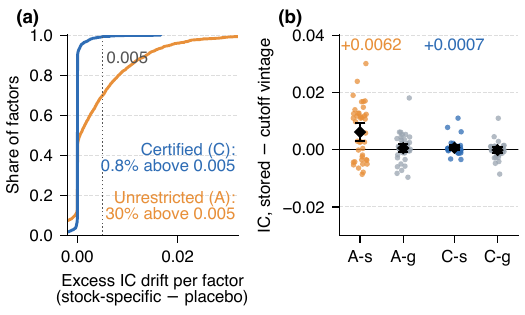}
\caption{Re-anchoring drift of the 40 unrestricted (A) and 40 certified (C) pools of the original-membership confirmatory stage (Table~\ref{tab:summary}; Section~S25). (a)~Distribution of the factor-level excess IC drift under the simulated stock-specific anchors (stock-specific minus global placebo, larger of the two anchors) for the 2,000 factors of each arm; 30\% of the unrestricted factors and 0.8\% of the certified ones exceed 0.005. (b)~Real anchor: signed difference in validation IC between the stored end-anchored representation and the cutoff-anchored one, one point per pool, under the stock-specific cutoff anchor (s) and the matched global placebo (g), with the mean and its 95\% interval.}
\label{fig:drift}
\end{figure}

\begin{table}[!t]
\caption{Drift under the simulated anchors, 40 pools per arm (larger of the two anchors per factor or pool). \emph{Shipped}: the implemented operators in float32 (cloud). \emph{Reference}: float64 with a relative variance floor in $\op{TsCorr}$ (laptop). Excess drift is stock-specific minus global-placebo IC drift; elementwise match: equal finite masks and relative error below $10^{-4}$. The shipped operators on the laptop give the same factor shares and a certified pool drift of 0.0018 (placebo 0.0018; Section~S8). Real anchor: Fig.~\ref{fig:drift}(b), Table~S22.}
\label{tab:drift}
\centering
\scriptsize
\setlength{\tabcolsep}{3pt}
\begin{tabular}{@{}lcccc@{}}
\toprule
& \multicolumn{2}{c}{Shipped, float32} & \multicolumn{2}{c}{Reference, float64} \\
\cmidrule(lr){2-3}\cmidrule(l){4-5}
Statistic & A & C & A & C \\
\midrule
\multicolumn{5}{@{}l}{\textit{Pool IC drift, simulated anchors}} \\
\quad stock-specific, mean & 0.0148 & \textbf{0.0011} & 0.0143 & 0.0027 \\
\quad stock-specific, max & 0.036 & 0.0058 & 0.031 & 0.013 \\
\quad global placebo, mean & 0.0056 & 0.0011 & 0.0051 & 0.0028 \\
\addlinespace
\multicolumn{5}{@{}l}{\textit{Factors}} \\
\quad corr.\ $< 0.9$, stock-specific & 43.1\% & 12.2\% & 38.9\% & \textbf{0.55\%} \\
\quad corr.\ $< 0.9$, global placebo & 18.4\% & 12.5\% & 13.6\% & 0.70\% \\
\quad elementwise match, both anchors & 13.3\% & 34.9\% & 24.9\% & 75.1\% \\
\quad excess IC drift, mean & 0.0038 & \textbf{0.00007} & 0.0039 & $-0.00004$ \\
\quad excess IC drift $> 0.005$ & 30\% & 0.8\% & 31\% & 0.35\% \\
\bottomrule
\end{tabular}
\end{table}

Under simulated anchors the unrestricted pools drift 2.7 times as much as under the placebo, while the certified pools drift no more than under it (Table~\ref{tab:drift}, Fig.~\ref{fig:drift}(a)). What drift remains for certified factors comes from the implementation. In the shipped float32 code, 12.2\% of them correlate below 0.9 with their re-anchored values, mostly through the finite precision of $\op{TsKurt}$ and the fixed variance floor of $\op{TsCorr}$. A reference evaluator in float64, with a floor relative to the window's scale, brings this down to 0.55\%, all ill-conditioned expressions such as $\op{high} \times \op{Inv}(\op{high})$, and leaves the unrestricted factors as anchor-sensitive as before (Table~\ref{tab:drift}, Section~S8). Under that evaluator the certified pools' IC moves as much under the placebo as under re-anchoring, which points to the combiner's threshold selection rather than to the anchor, on the confirmatory and the re-mined CSI500 pools alike (Sections~S8 and~S18).

On the real anchor, unrestricted pools score a mean adaptive IC 0.0062 higher on the stored end-anchored representation than on the cutoff-anchored one. Certified pools score 0.0007 higher, about the same as under the global placebo (Fig.~\ref{fig:drift}(b)). We call this difference a pool's \emph{re-anchoring sensitivity}: how much its measured IC depends on the representation of the same data. Factor by factor, the mean signed difference is close to zero outside the ten bare fields, although 13--24\% of the factors in each family lose more than 0.005 of absolute IC, so the pool-level difference comes from how the combiner selects and weights factors whose values moved (Section~S22).

\textbf{Published methods.}\label{sec:exp-published} The dependence is not specific to our miner. We ran the same audit on pools that five published methods (AlphaGen, AlphaSAGE, AlphaForge, AlphaQCM and AlphaGen's genetic-programming baseline) mined in our earlier benchmark pipeline, five seeds per market, evaluated on the previously exposed 2022--2024 window (Section~S4). Their pools score a mean adaptive IC 0.0050 $[0.0030, 0.0070]$ higher on the stored representation than on the cutoff-anchored one, about 14\% of their IC, with 42 of 50 pools positive; certified pools on the same window have a mean signed difference of $-0.0010$ IC between the two representations. A sizeable part of what these methods measure on end-anchored data therefore depends on the representation. The audit concerns these pools, not the numbers reported in the original papers (Section~S22).

\textbf{Past-anchored data.}\label{sec:exp-hfq} Past anchoring removes the future dependence of a snapshot, yet formulas mined on it stay sensitive to the anchor. For unrestricted pools mined on a past-anchored provider built by the group action (arm AH; Section~S23), the standard deviation over the 40 paired units of the IC change between the two representations is 0.0193, against 0.0042 for certified pools, a ratio of about 4.6; on that provider the certified pools are ahead by $+0.0070$ (Table~\ref{tab:summary}). Their factors depend on where the data begin.

\textbf{A second provider.}\label{sec:exp-usaudit} On the public Qlib US bundle~\cite{yang2020qlib} (S\&P~500, anchored at each stock's first day), 71\% of the unrestricted factors show excess anchor drift above 0.005, against 3.4\% in certified pools (Sections~S16.4 and~S25).

\textbf{Hand-crafted feature sets.}\label{sec:exp-alpha158} With Qlib's $10^{-12}$ division guards read as numerical guards, 153 of the 158 Alpha158 features are certified, most because their designers divided by a price or a sum (Table~S15). Of Kakushadze's 101 Formulaic Alphas~\cite{kakushadze2016alpha101}, the checker certifies 17 and rejects 84, and numerical re-anchoring shows deviations for 82 of the 84, mostly through a cross-sectional rank of a price level (Section~S12).

\subsection{Predictive Cost}\label{sec:exp-main}

On clean data, certification has no detectable effect on predictive accuracy. The comparison is certified (C) against unrestricted (A) search, the AlphaSAGE generator with and without the guard, over $4$ folds $\times$ $2$ markets $\times$ $10$ seeds $= 80$ units (Section~S23).

\begin{figure*}[!t]
\centering
\includegraphics[width=\textwidth]{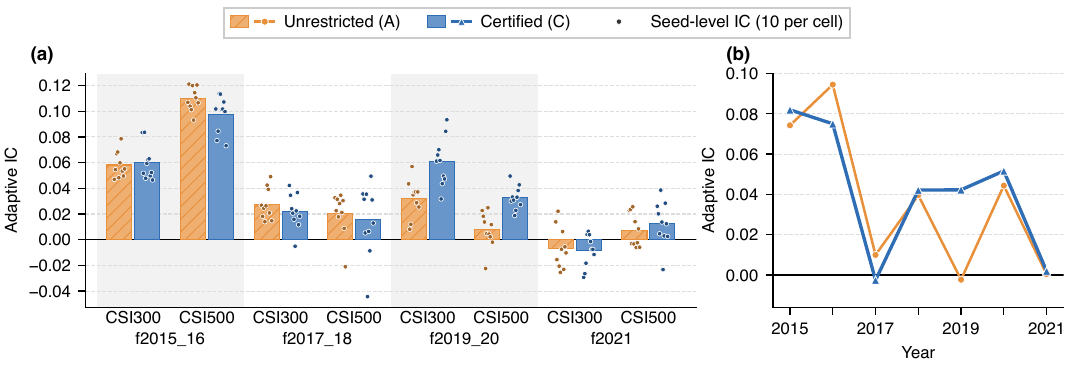}
\caption{Main comparison on clean data under the adaptive combiner, laptop evaluation; ten seeds per fold $\times$ market cell. (a)~Adaptive IC per cell for the unrestricted (A) and certified (C) arms, with seed-level points. (b)~Yearly adaptive IC of the two arms from 2015 to 2021. The difference sits in f2019\_20, the window that contains the 2020 crash; in 2016--2017 the certified arm is below the unrestricted arm. The 80 sorted paired differences: Fig.~S9.}
\label{fig:main}
\end{figure*}

\begin{table}[!t]
\caption{Certified minus unrestricted search (C $-$ A) on clean data, $4$ folds $\times$ $2$ markets $\times$ $10$ seeds $= 80$ units, laptop evaluation, by combiner and market, and under a float64 build of the evaluator (last row); seed blocks: Section~S13. Units and cells count positive paired differences; the cell- and fold-level intervals are paired-$t$ intervals over the 8 cell means and the 4 fold means (tests: Section~S13).}
\label{tab:paired}
\centering
\scriptsize
\setlength{\tabcolsep}{2pt}
\renewcommand{\arraystretch}{0.95}
\begin{tabular}{@{}lrrrrr@{}}
\toprule
Comparison & $n$ & $\Delta$IC & 95\% CI & Units & Cells \\
\midrule
C $-$ A, adaptive & 80 & $\mathbf{+0.0044}$ & $[-0.0013{,}{+}0.0101]$ & 48/80 & 4/8 \\
\quad cell-level interval & 8c & & $[-0.0080{,}{+}0.0168]$ & & \\
\quad fold-level interval & 4f & & $[-0.0199{,}{+}0.0287]$ & & \\
\quad CSI300 & 40 & $+0.0057$ & $[-0.0026{,}{+}0.0140]$ & 23/40 & 2/4 \\
\quad CSI500 & 40 & $+0.0031$ & $[-0.0050{,}{+}0.0112]$ & 25/40 & 2/4 \\
C $-$ A, neutralised & 80 & $-0.0005$ & $[-0.0060{,}{+}0.0049]$ & 44/80 & 3/8 \\
C $-$ A, equal weight & 80 & $+0.0061$ & $[-0.0016{,}{+}0.0137]$ & 44/80 & 4/8 \\
C $-$ A, float64 eval. & 80 & $+0.0048$ & $[-0.0010{,}{+}0.0106]$ & 48/80 & 5/8 \\
\bottomrule
\end{tabular}
\end{table}

The certified search is ahead by $+0.0044$ adaptive IC, positive in 4 of 8 cells and concentrated in f2019\_20. Over the 8 cell means the interval is $[-0.0080, +0.0168]$, and it is wider over folds because the two markets of a fold share a period (Table~\ref{tab:paired}; tests in Section~S13). At the observed variability, the estimated minimum detectable difference over the 8 cell means is about 0.017 IC at 80\% power, so the design has limited power for smaller effects. The mean passes the engineering rule fixed in advance (Section~\ref{sec:frozen}), while the cell interval still allows losses beyond its tolerance. The source of the larger difference in f2019\_20, the window of the 2020 crash, is unresolved; the certified pools are steadier across years (Fig.~\ref{fig:main}(b)) and their single factors are weaker, so the difference arises in combination (Section~S16.4).

\subsection{Grammar Controls and Search Efficiency}\label{sec:exp-controls}

The certified search leads the coarse-typing control; the tested random-pruning control does not reproduce its pools, and the length-matched analyses do not separate invariance from expression composition. On 40 clean units it is ahead of an Alpha$^2$-style coarse guard (arm B) by 0.0089 $[+0.0003, +0.0176]$ under the adaptive combiner, with neutralised and equal-weight intervals that include zero (Section~S25). Because B is below A, the pre-specified attribution criterion holds trivially. The single-sort guard lands between A and C, which places the additive sort's measurable contribution in the language it certifies. A random guard at the same rejection rate (arm R) lowers pool IC on the original file. The certified pools are shorter (6.8 against 10.1 tokens), and their factors' out-of-sample IC changes sign less often (Wilcoxon $p = 0.031$). Matching on length and operator composition leaves that difference in place, with an interval that includes zero (Sections~S6, S14 and~S15).\label{sec:exp-mechanism} The tested diagnostics do not support the hypothesis that non-invariant factors act as stock identifiers through the adjustment factor (Section~S15).\label{sec:exp-mask} A typed action mask (Section~\ref{sec:mask}) fills the pool in 3,050 instead of 5,300 episodes but missed its pre-specified quality criterion on CSI500 (Section~S10).

\subsection{Integration and Transfer}\label{sec:exp-transfer}

The guard attaches unchanged to three other generators, with no consistent difference in prediction. On AlphaGen's PPO generator it rejects 96--99\% of candidates and cuts a run from about 7~h to 1~h (Section~S16.4). With AlphaGen's genetic-programming baseline and its LLM generator (Claude Haiku~4.5 or Opus~5.5, identical prompts), no difference is detectable except one exploratory positive block that comes from weak unguarded pools (Sections~S19--S21). In a TopK-Dropout backtest net of costs, certified minus unrestricted is $+0.014$ a year with an interval that includes zero, and every arm trails the equal-weight benchmark (Section~S20).

\textbf{Official miners and a supervised reference.}\label{sec:exp-matched} Against the official AlphaSAGE~\cite{chen2026alphasage} and AlphaGen~\cite{yu2023alphagen} code, re-run in their published configurations on our data and evaluator (80 runs), the certified search has 5 and 3 of 8 cells non-negative. The first is one cell short of the frozen criterion (Table~S11, Section~S13). Alpha158 + LightGBM~\cite{yang2020qlib,ke2017lightgbm} predicts best by a wide margin (61--92\% higher raw IC) but produces no formulas that can be read, certified or audited one by one.

\textbf{Later windows.}\label{sec:exp-later} Re-mined on the two later windows with everything frozen, the certified search was ahead under the adaptive combiner but short of the frozen rule, and the equal-weight difference changed sign between the windows (Section~\ref{sec:folds}, Table~S12). The S\&P~500 experiment\label{sec:exp-us} is mainly an audit: neither arm carries detectable signal, and under equal weighting the certified pools are below the unrestricted ones ($-0.0111$ $[-0.0176, -0.0045]$; Table~S16), as on 2022--2024 (Table~S12).

\subsection{Discovery Sensitivity to Data Perturbations}\label{sec:exp-sh000905}

\begin{table}[!t]
\caption{The two channels of Section~\ref{sec:exp-sh000905}. No validation day contains a change; in the fixed-pool channel the combiner's matured statistics from 2010 do.}
\label{tab:channels}
\centering
\scriptsize
\setlength{\tabcolsep}{3pt}
\begin{tabular}{@{}l>{\raggedright\arraybackslash}p{0.40\columnwidth}>{\raggedright\arraybackslash}p{0.30\columnwidth}@{}}
\toprule
Channel & Pools compared & Evaluation data \\
\midrule
Re-mined & mined with and without the change, same seeds & clean for both \\
Fixed pools & one set (clean pools; for the ticker, those mined with it) & with and without the change \\
\bottomrule
\end{tabular}
\end{table}

\begin{figure}[!t]
\centering
\includegraphics[width=\columnwidth]{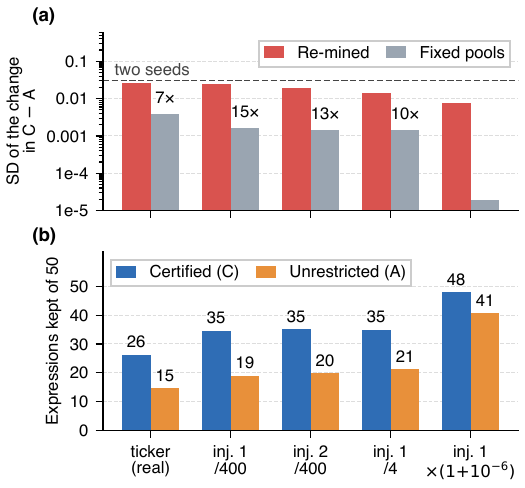}
\caption{A data change through the two channels on CSI500: the mislisted index ticker, the stored VWAP of two randomly drawn constituents divided by 400, the first of them divided by 4, and the first multiplied by $1 + 2^{-20}$. (a)~Standard deviation over the 20 fold $\times$ seed units of the change in C $-$ A when the fixed pools are re-evaluated with and without the change (for the ticker, the original pools mined with it; otherwise the clean pools) and when both arms are re-mined on it (re-mined), with the re-mined/fixed ratio. The ratio is omitted for the perturbation, whose fixed-pool SD is near zero. The dashed line is the SD of the difference in C $-$ A between two seeds of the same cell. (b)~Expressions of the original 50 kept by the re-mined pool, unit mean. Intervals: Table~S17.}
\label{fig:remine}
\end{figure}

Re-mining on defective discovery data changes the selected pools far more than re-scoring the fixed pools reveals. We measure the size against two references: a matched $10^{-6}$ perturbation of the same field, which the defects exceed about two to three times, and a change of seed, which they do not reach. The direction of the mean effect is not resolved.

The CSI500 membership file in the v2 and v3 data snapshots distributed through Qlib's download utility lists the index itself, SH000905, as a constituent from 2007 to mid-2012, with a close about 1,500 times the median member's (Section~S5). It touches 0.05--0.10\% of our discovery stock-days and no validation day. We measure its effect in our pipeline only; which snapshots individual published studies used is outside this check. We re-evaluated the existing pools without the ticker and re-mined the 40 CSI500 pools of arms A and C without it, always evaluating re-mined pools on clean data so that only their discovery data differ (Table~\ref{tab:channels}, Section~S25). To test whether the pattern is specific to the ticker, separate runs divide the stored VWAP of one of two randomly drawn CSI500 members by 400 over the same span, or that of the first by 4 (Section~S17).

Fig.~\ref{fig:remine} shows the sizes. After re-mining, the change in C $-$ A has an SD of 0.014--0.026 under the four defects, against 0.0014--0.0038 when the fixed pools are re-scored and 0.031 between two seeds of one cell (Table~S17, Section~S26). The certified search keeps more of its original expressions in 17--20 of 20 units per defect. Removing the ticker moves the discovery IC of certified factors about a quarter as much as that of the non-certified factors in the same pools (Section~S5); the theorem covers scale, not membership, so this is an empirical observation. The mean change in C $-$ A cannot be distinguished from zero: the /4 injection has a positive point estimate of $+0.0060$ with an interval that includes zero, and detecting 0.011 at these spreads would need about 37 independent units~\cite{agarwal2021precipice}.

\textbf{A small numerical perturbation.} To separate the defects from the search's general sensitivity, we multiplied the first injected ticker's stored VWAP by $1 + 2^{-20}$, a relative change of about $10^{-6}$, over the same span, and re-mined both arms on the 20 units (Section~S17). Even this changes 39 of 40 pools, but much less than the defects do: the change in C $-$ A has an SD of 0.0077, and the pools keep 48 (C) and 41 (A) of their 50 expressions (Fig.~\ref{fig:remine}). Re-scoring the fixed pools on the perturbed data changes almost nothing. The defects therefore act beyond the search's sensitivity to tiny input changes, at least for this stock, span and perturbation size.
\section{Discussion and Limitations}\label{sec:discussion}

\textbf{Scope of the guarantee.} The certificate covers formulas, on data that honour the field contract and under idealised operator semantics; our inconsistent build shows the first condition failing (VWAP/close leaked through an unadjusted field), and the float64 reference evaluator cuts the share of certified factors that correlate below 0.9 with their re-anchored values from 12.2\% to 0.55\% (elementwise agreement 75.1\%), so numerical validation stays distinct from the symbolic guarantee (Section~\ref{sec:exp-anchor}).

\textbf{Recommendations for practice.} Check every formula statically and evaluate in float64 with scale-relative floors. Check the data first: no index code among its own constituents, fields that match the declared adjustment event by event, and a closer look at any series whose price is hundreds of times the cross-sectional median. After a data correction, re-mine with several seeds; re-scoring the old pools understates the change.

\textbf{Predictive cost.} Invariance excludes price levels, share-count turnover and other expressions that depend on the adjustment base, including real signals such as the low-price effect. Within the resolution of our design this exclusion has no measurable effect on the Chinese folds, in either direction. Elsewhere the picture is mixed: equal weighting favours the unrestricted pools on the S\&P~500 and on 2022--2024, and the supervised Alpha158 model predicts better (Section~\ref{sec:exp-transfer}). The case for the certificate therefore rests on the audit.

\textbf{Limitations.} The engineering rule of the main comparison is a threshold on the mean, not a non-inferiority bound, and the cell interval allows losses beyond it. Re-anchoring sensitivity compares two representations of the same data; the cutoff vintage is not a point-in-time series, so it does not measure a leakage gain. The controls bound the explanation of the small positive difference without isolating invariance as its cause. Arm R, the typed mask, the matched comparison and the later windows use pools mined with the original membership file, no length rule of equal strength exists, and whether shorter factors resist overfitting was not tested. The event-level contract check relies on the exchange's reference prices and turnover rates and does not verify dividend amounts. The primary predictive comparison covers one country, one data source and two index universes. The discovery analysis covers one real membership error, one injected defect at two magnitudes and one perturbation, all on CSI500, and a predictive test on a second market needs licensed survivorship-free data such as CRSP. Eight fold $\times$ market cells with dependent seeds limit inference, and the PPO, GP and LLM comparisons cover 6--10 pairs each. The certified grammar was chosen after exploratory analyses on earlier pools, so the stages do not form one study frozen in advance. Survivorship and vendor revisions remain unaddressed.

\textbf{Future work.} LLM-based and agentic miners propose formulas faster than they can be reviewed by hand, and a linear-time checker can guard their output; whether it changes what such agents learn to propose is open. The construction extends to other per-entity transformations of stored data, such as a change of lot size, and a second market with point-in-time, survivorship-free data would show how far the audit generalises.

\section{Conclusion}\label{sec:conclusion}

A formula evaluated on end-anchored prices can depend on corporate actions that had not yet happened. Under a declared field contract and idealised operator semantics, a two-sorted type system certifies from the syntax alone that a formula cannot. The audit shows why this matters: pools that five published methods mined on such data score a mean 0.005 IC higher than on a cutoff-anchored representation, about 14\% of their stored IC, while the mean change for certified pools is $-0.001$. The same audit found an erroneous index entry in public Qlib snapshots. Defects like it reshape the pools a miner selects, and re-evaluating old pools hides most of the effect, so data corrections should be checked by re-mining. On the Chinese development folds, certification had no detectable effect on predictive accuracy at the resolution of our design. Its benefit is a guarantee that can be checked one expression at a time.

\section*{Acknowledgments}
The authors designed the study, the type system and the experiments, analysed and interpreted all results, and designed and drew Figs.~2 and~4. Anthropic's Claude Fable~5.1 and Claude Opus~5.5, used through Claude Code under the authors' direction, wrote, tested and ran the analysis, experiment and figure scripts, and assisted in language editing and stylistic polishing of the manuscript. The scripts and per-unit result files are released in the repository. The authors reviewed all scripts, checked every reported number against the per-unit result files, and are responsible for all content. Claude Haiku~4.5 and Claude Opus~5.5 also serve as generators in the LLM experiment (Section~S19).

\section*{Code and Data Availability}
Code, frozen protocols and freeze files, hash manifests, the final factor pools and the per-unit evaluation outputs behind every table are at \mbox{\url{github.com/xiangxinyue/anchor-invariant-alpha}}. The checker is one file, \texttt{checker/scale\_checker\_v2.py}, whose only dependency is AlphaGen's expression classes; the README's quick start runs its 43 unit cases and certifies an expression in a few lines. The data provider is not redistributed; its SHA-256 manifest lets a rebuilt provider be verified file by file.

\bibliographystyle{IEEEtran}
\bibliography{refs}

\begin{IEEEbiographynophoto}{Zhengkai Tu}
is a Research Assistant in the Department of Decision Sciences, CUHK Business School, The Chinese University of Hong Kong. Zhengkai Tu received the M.Eng. degree from the University of Toronto in 2025 and a bachelor's degree from York University in 2023. Research interests include graph neural networks, information retrieval and financial time series.
\end{IEEEbiographynophoto}

\vspace{-2\baselineskip}
\begin{IEEEbiographynophoto}{Xinyue Xiang}
received the B.Sc. degree from the University of Alberta in 2021, the M.S. degree in Computer Science from the Georgia Institute of Technology in 2025, and the Master of Applied Data Science degree from the University of Michigan in 2026. Research interests include data quality, quantitative finance and computational social science.
\end{IEEEbiographynophoto}

\vspace{-2\baselineskip}
\begin{IEEEbiographynophoto}{Jimmy Huang}
is a Professor in the School of Information Technology, York University, Toronto, and holds a Tier 1 York Research Chair in Big Data Analytics. Jimmy Huang received the Ph.D. degree from City, University of London, was General Chair of ACM SIGIR 2020, and is an IEEE Fellow and a Fellow of the Canadian Academy of Engineering. Research interests include information retrieval, artificial intelligence and big data analytics.
\end{IEEEbiographynophoto}

\BeginPaperSupplement
\makeatletter
\setlength{\@fptop}{0pt}
\setlength{\@fpsep}{12pt plus 2pt}
\setlength{\@fpbot}{0pt plus 1fil}
\setlength{\@dblfptop}{0pt}
\setlength{\@dblfpsep}{12pt plus 2pt}
\setlength{\@dblfpbot}{0pt plus 1fil}
\makeatother

\renewcommand{\thesection}{S\arabic{section}}
\makeatletter\def\thesectiondis{\thesection}\makeatother
\renewcommand{\thetable}{S\arabic{table}}
\renewcommand{\thefigure}{S\arabic{figure}}
\title{Supplementary Material for ``Adjustment-Anchor Dependence in Formulaic Alpha Languages:\\ A Certified Grammar and a Re-Anchoring Audit''}

\author{Zhengkai~Tu,
        Xinyue~Xiang,
        and~Jimmy~Huang
\IEEEcompsocitemizethanks{
\IEEEcompsocthanksitem Z.~Tu is with The Chinese University of Hong Kong, Hong Kong SAR, China.
\IEEEcompsocthanksitem X.~Xiang is with the University of Michigan, Ann Arbor, MI, USA.
\IEEEcompsocthanksitem J.~Huang is with York University, Toronto, ON, Canada. Corresponding author: J.~Huang (jhuang@yorku.ca).}
}

\markboth{IEEE Transactions on Knowledge and Data Engineering (Supplementary Material)}{Tu \MakeLowercase{\textit{et al.}}: Adjustment-Anchor Dependence in Formulaic Alpha Languages}

\maketitle

\noindent This document contains the material referred to from the main paper. Section, table and figure numbers without the prefix S refer to the main paper.

\medskip
\noindent{\footnotesize
\renewcommand{\arraystretch}{1.05}
\begin{tabular}{@{}>{\raggedright\arraybackslash}p{0.26\columnwidth}>{\raggedright\arraybackslash}p{0.68\columnwidth}@{}}
\toprule
Topic & Sections \\
\midrule
Grammar and theory (main Sections 3--5) & S1 typing rules, proofs, unit cases and coverage; S12 typing of the 101 Formulaic Alphas; S14 expression length; S16 additional method and evaluation details, with the reproducibility record \\
Data and protocol (main Section 6) & S5 provenance of the inconsistent build and origin of the SH000905 line; S6 protocol records and deviations; S11 compute by stage \\
Certification and re-anchoring (7.1) & S8 numerical diagnosis; S9 figures; S18 audit of the re-mined CSI500 pools; S22 real anchors; S23 past-anchored baseline \\
Predictive cost and controls (7.2--7.3) & S2 size-matched subsets; S3 combiner ablation; S7 cloud-instance rows; S10 typed action mask; S13 statistical detail; S15 mechanism; S24 float64 evaluator; S25 additional paired comparisons \\
Integration and transfer (7.4) & S4 retrospective comparison, 2022--2024; S19 LLM generator; S20 portfolio backtest; S21 genetic-programming generator \\
Discovery sensitivity (7.5) & S17 injected defects and the numerical perturbation; S26 seed-to-seed reference \\
\bottomrule
\end{tabular}}
\medskip

\section{Typing Rules, Unit Cases and Coverage Details}\label{supp:rules}

\textbf{Typing rules.} Table~\ref{tab:rulesfull} gives the complete typing rules of the two-sorted checker; Table~2 of the main paper is the summary. Exponents and coefficients are exact rationals. Constants are converted with a denominator bound of $10^6$. This rounds a short decimal such as 0.01 to its intended rational 1/100 and is lossless for every constant of the search grammar. A constant whose rational approximation is zero, or differs from its floating-point value by more than a relative $10^{-12}$, would be typed with a different exponent or coefficient from the one the evaluator uses; an exponent of $10^{-7}$, for instance, is rounded to zero. The search grammar has a fixed set of constants, all of which convert within that tolerance, so typing and evaluation agree in every experiment. For other uses, the released checker provides \texttt{strict\_check} (\texttt{checker/constant\_guard.py}), which rejects any expression containing such a constant before typing it.

\begin{table*}[!t]
\caption{Complete typing rules of the two-sorted checker. $\inv = \sortM(0) = \sortA(0)$; a node whose coefficient is zero is normalised to $\inv$. ``Reject'' raises a typing error with the stated reason.}
\label{tab:rulesfull}
\centering
\scriptsize
\begin{tabular}{@{}>{\raggedright\arraybackslash}p{1.7in}>{\raggedright\arraybackslash}p{5.2in}@{}}
\toprule
Operator & Rule \\
\midrule
Price field ($\op{open}$, $\op{high}$, $\op{low}$, $\op{close}$, $\op{vwap}$) & $\sortM(-1)$ \\
$\op{volume}$ & $\sortM(+1)$ \\
Constant, window length & $\inv$ \\
$\op{Abs}$ & $\sortM(s) \to \sortM(s)$; reject $\sortA(a \ne 0)$ \\
$\op{Inv}$ & $\sortM(s) \to \sortM(-s)$; reject $\sortA(a \ne 0)$ \\
$\op{Sign}$ & $\sortM(s) \to \inv$; reject $\sortA(a \ne 0)$ \\
$\op{Log}$ & $\sortM(s) \to \sortA(s)$; reject $\sortA(a \ne 0)$ \\
$\op{SLog1p}$, $\op{Rank}$ & $\inv \to \inv$; reject any non-invariant input \\
$\op{Add}$, $\op{Sub}$, $\op{Greater}$, $\op{Less}$ & Both $\inv$: $\inv$. If exactly one operand is $\inv$, it is read as $\sortM(0)$ or $\sortA(0)$ to match the other operand's sort; its coefficient stays zero. Different sorts with non-zero coefficients: reject. $\sortM(s), \sortM(r)$: require $s = r$ and output $\sortM(s)$ (so $\inv$ with $\sortM(s \ne 0)$ is rejected). $\sortA(a), \sortA(b)$: $\op{Add} \to \sortA(a+b)$, $\op{Sub} \to \sortA(a-b)$; $\op{Greater}$ and $\op{Less}$ require $a = b$ and output $\sortA(a)$ (so $\inv$ compared with $\sortA(a \ne 0)$ is rejected) \\
$\op{Mul}$ & $\sortM(s), \sortM(r) \to \sortM(s+r)$; $\sortA(a)$ with constant $q$ on either side $\to \sortA(aq)$; otherwise reject \\
$\op{Div}$ & $\sortM(s), \sortM(r) \to \sortM(s-r)$; $\sortA(a)$ divided by constant $q \ne 0$ $\to \sortA(a/q)$; otherwise reject \\
$\op{Pow}$ & Additive base with $a \ne 0$: reject. Constant exponent $q$: $\sortM(s) \to \sortM(qs)$. Non-constant exponent: both invariant $\to \inv$, otherwise reject \\
$\op{Ref}$, $\op{TsMean}$, $\op{TsMax}$, $\op{TsMin}$, $\op{TsMed}$, $\op{TsWMA}$, $\op{TsEMA}$ & Type preserved \\
$\op{TsSum}$ (window $w$) & $\sortM(s) \to \sortM(s)$; $\sortA(a) \to \sortA(aw)$ \\
$\op{TsStd}$, $\op{TsMad}$, $\op{TsMaxDiff}$, $\op{TsMinDiff}$, $\op{TsMinMaxDiff}$, $\op{TsDelta}$ & $\sortM(s) \to \sortM(s)$; $\sortA(a) \to \inv$ \\
$\op{TsVar}$ & $\sortM(s) \to \sortM(2s)$; $\sortA(a) \to \inv$ \\
$\op{TsSkew}$, $\op{TsKurt}$, $\op{TsRank}$ & Any typable input $\to \inv$ \\
$\op{TsIr}$, $\op{TsDiv}$, $\op{TsPctChange}$ & $\sortM(s) \to \inv$; reject $\sortA(a \ne 0)$ \\
$\op{TsCov}$ & $\sortM(s'+r')$ with $s' = s$ for an $\sortM(s)$ argument and $0$ for an $\sortA$ argument \\
$\op{TsCorr}$ & Any typable inputs $\to \inv$ \\
Any other operator & Reject \\
Root & Certified if and only if the root type is $\inv$ \\
\bottomrule
\end{tabular}
\end{table*}

\textbf{Proof of Theorems 1 and 2 of the main paper.} \emph{Semantics.} We use the idealised operator semantics of Section~4.2 of the main paper: arithmetic is exact, each operator is the mathematical function stated below, and a value is non-finite exactly where that function is undefined (a non-positive argument of $\op{Log}$, a zero divisor, a negative base with a non-integer exponent, or a window with zero variance where a statistic divides by the standard deviation). The fixed variance floor of the implemented $\op{TsCorr}$ and float32 rounding lie outside this semantics (Section~S8).

\emph{Claim.} Fix $c \in (\R_+)^{|\calU|}$ and write $L_i = \log c_i$. For an expression $E$, write $E' = E(g_c X)$ and $\mathrm{dom}(E)$ for the set of $(t,i)$ at which $E$ is finite. If $\vdash E : \tau$, then (a)~$\mathrm{dom}(E') = \mathrm{dom}(E)$, and (b)~on $\mathrm{dom}(E)$, $E'_{t,i} = c_i^{\,s} E_{t,i}$ if $\tau = \sortM(s)$ and $E'_{t,i} = E_{t,i} + a L_i$ if $\tau = \sortA(a)$. Because $\inv = \sortM(0) = \sortA(0)$, both readings give $E' = E$ for $\inv$, and a node normalised to $\inv$ after its coefficient becomes zero satisfies the same statement. The claim for $\tau = \inv$ is Theorem~1; the general claim is Theorem~2. The proof is by structural induction on the derivation of $\vdash E : \tau$, and each case below assumes the claim for the operands. Two facts are used throughout. First, $c_i^{\,s} > 0$, so multiplication by $c_i^{\,s}$ preserves zeros, signs, order and the conditions $x > 0$ and $x \ne 0$. Second, the action is constant in time, so on a rolling window ending at $t$ every value of an operand of type $\sortM(s)$ or $\sortA(a)$ is transformed by the same map $x \mapsto \alpha x + \beta$, with $(\alpha, \beta) = (c_i^{\,s}, 0)$ or $(1, a L_i)$.

\emph{Leaves.} A price field satisfies $p' = c_i^{-1} p$, so its type is $\sortM(-1)$; volume satisfies $v' = c_i v$, so its type is $\sortM(+1)$. Constants and window lengths do not change.

\emph{Unary operators on $\sortM(s)$.} $|c_i^{\,s} x| = c_i^{\,s} |x|$, $1/(c_i^{\,s} x) = c_i^{-s} (1/x)$, $\mathrm{sign}(c_i^{\,s} x) = \mathrm{sign}(x)$ and $\log(c_i^{\,s} x) = \log x + s L_i$, which give the rules for $\op{Abs}$, $\op{Inv}$, $\op{Sign}$ and $\op{Log}$. The domain conditions $x \ne 0$ and $x > 0$ are preserved. These operators reject $\sortA(a)$ with $a \ne 0$. $\op{SLog1p}$ and $\op{Rank}$ accept only $\inv$, for which $E' = E$ on the whole cross-section of the day, so the outputs agree. $\op{Rank}$ needs this restriction because $c_i$ varies across stocks, so no common factor exists on a day.

\emph{Binary arithmetic.} For $\op{Add}$ and $\op{Sub}$, $c_i^{\,s} x \pm c_i^{\,s} y = c_i^{\,s} (x \pm y)$ and $(x + a L_i) \pm (y + b L_i) = (x \pm y) + (a \pm b) L_i$. An $\inv$ operand is read as $\sortM(0)$ or $\sortA(0)$, which is the same statement with coefficient zero. For $\op{Greater}$ and $\op{Less}$, $\max(c_i^{\,s} x, c_i^{\,s} y) = c_i^{\,s} \max(x, y)$ because $c_i^{\,s} > 0$, and $\max(x + a L_i, y + a L_i) = \max(x, y) + a L_i$; the same holds for $\min$. For $\op{Mul}$ and $\op{Div}$, $c_i^{\,s} x \cdot c_i^{\,r} y = c_i^{\,s+r} xy$ and $c_i^{\,s} x / (c_i^{\,r} y) = c_i^{\,s-r} (x/y)$, and $y' = 0$ if and only if $y = 0$. A constant $q$ does not change, so $(x + a L_i) q = xq + aq L_i$ and $(x + a L_i)/q = x/q + (a/q) L_i$ for $q \ne 0$. For $\op{Pow}$ with a constant exponent $q$ and a base of type $\sortM(s)$, $(c_i^{\,s} x)^q = c_i^{\,sq} x^q$; the base keeps its sign and its zeros, so the domain does not change. With a non-constant exponent both operands are $\inv$ and the output does not change. A base of type $\sortA(a)$ with $a \ne 0$ is rejected.

\emph{Rolling operators.} Let $x_1, \dots, x_w$ be the operand on the window and $x'_k = \alpha x_k + \beta$ as above. $\op{Ref}$ returns a transformed value and keeps the type. $\op{TsMean}$, $\op{TsWMA}$ and $\op{TsEMA}$ are weighted means whose fixed weights sum to one (the implemented $\op{TsWMA}$ and $\op{TsEMA}$ normalise their weights), so $\sum_k \omega_k x'_k = \alpha \sum_k \omega_k x_k + \beta$; $\op{TsMax}$, $\op{TsMin}$ and $\op{TsMed}$ are order statistics, which commute with $x \mapsto \alpha x + \beta$ for $\alpha > 0$. All six keep the type. $\op{TsSum}$ gives $\sum_k x'_k = \alpha \sum_k x_k + w\beta$, hence $\sortM(s)$ or $\sortA(aw)$. $\op{TsDelta}$ ($x_w - x_{w-d}$), $\op{TsMaxDiff}$ ($x_w - \max_k x_k$), $\op{TsMinDiff}$ ($x_w - \min_k x_k$) and $\op{TsMinMaxDiff}$ ($\max_k x_k - \min_k x_k$) are differences of two values transformed by the same map, and $\op{TsStd}$ and $\op{TsMad}$ (mean absolute deviation from the window mean) are dispersions of such differences; each is multiplied by $\alpha$ and does not depend on $\beta$, so $\sortM(s) \to \sortM(s)$ and $\sortA(a) \to \inv$. $\op{TsVar}$ is multiplied by $\alpha^2$, giving $\sortM(2s)$ or $\inv$. $\op{TsSkew}$ and $\op{TsKurt}$ (standardised third and fourth moments) and $\op{TsRank}$ (the rank of $x_w$ within the window) are invariant to every map $x \mapsto \alpha x + \beta$ with $\alpha > 0$, and a window has zero variance before the map if and only if it has zero variance after it, so the output is $\inv$ with the same non-finite set. $\op{TsIr}$ (mean over standard deviation), $\op{TsDiv}$ ($x_w$ over the window mean) and $\op{TsPctChange}$ ($(x_w - x_1)/x_1$) are ratios of two quantities that both scale by $\alpha$, so they are unchanged for $\sortM(s)$ and their zero denominators coincide; a shift does not cancel in them, so $\sortA(a)$ with $a \ne 0$ is rejected. $\op{TsCov}$ is bilinear and shift-free, $\mathrm{cov}(\alpha_1 x + \beta_1, \alpha_2 y + \beta_2) = \alpha_1 \alpha_2 \, \mathrm{cov}(x, y)$, which gives $\sortM(s' + r')$. $\op{TsCorr}$ satisfies $\mathrm{corr}(\alpha_1 x + \beta_1, \alpha_2 y + \beta_2) = \mathrm{corr}(x, y)$ for $\alpha_1, \alpha_2 > 0$, with the same degenerate windows, so its output is $\inv$.

\emph{Root.} The root rule only selects the expressions whose type is $\inv$. Every case preserves the domain, which completes the induction. Operators outside Table~\ref{tab:rulesfull} are rejected, so no case is missing.

\textbf{Proof of Proposition~2 of the main paper (precision of the root rule).} By claim (a) above, $E(g_c X)$ and $E(X)$ are finite at the same $(t,i)$, and there Theorem~2 of the main paper gives $E(g_c X)_{t,i}\!-\!E(X)_{t,i} = a \log c_i$ or $(c_i^{\,s}\!-\!1)\,E(X)_{t,i}$. Any $c$ with $c_i \ne 1$ makes the first non-zero, and the second nonzero unless $E(X)_{t,i} = 0$; under the stated conditions both vanish wherever $E(X)$ is finite. If $E(X)$ is finite nowhere, both conditions hold vacuously.

\textbf{Unit cases.} The two-sorted checker's test suite has 43 cases, all passing (Table~\ref{tab:unitcases}): nineteen cover single-sort behaviour that the two-sorted system must preserve and twenty-four the additive sort; seven further cases, listed below, test the coarse control, 50 in all. The unit cases cover an operator domain larger than the search grammar (for example window 1 and the constant 20), because the checker is defined for every window length and every rational constant. The seven coarse-control cases: it rejects $\op{close} + \op{volume}$ and $\op{Greater}(\log\op{close}, \log\op{volume})$, and it accepts $\op{close} + \op{high}$, $\op{close}$, $\op{close}/\op{volume}$, $\op{Rank}(\op{close})$ and $\op{Log}(\op{close})$.

\begin{table*}[!t]
\caption{Unit cases of the two-sorted checker, grouped by the behaviour they test. All 43 cases pass; the expressions in each group are those the checker must accept and those it must reject.}
\label{tab:unitcases}
\centering
\scriptsize
\begin{tabular}{@{}>{\raggedright\arraybackslash}p{0.9in}>{\raggedright\arraybackslash}p{3.2in}>{\raggedright\arraybackslash}p{2.6in}@{}}
\toprule
Group & Accepted & Rejected \\
\midrule
Single-sort behaviour (19) & amount ($\op{close} \times \op{volume}$); 20-day return; $\op{vwap}/\op{close}$; $\op{Rank}(\op{return})$; $\op{Log}(\op{return})$; $(\op{high} - \op{low})/\op{close}$; $\op{return}^{0.5}$; $\op{TsVar}(\op{close}, 20)/\op{close}^2$; $\op{TsCorr}(\op{close}, \op{volume})$; $\op{TsCov}(\op{close}, \op{volume})$; $\op{Sign}(\op{close})$; $\op{TsRank}(\op{close}, 20)$ & $\op{close}$; $\op{volume}$; $\op{Rank}(\op{close})$; $\op{Greater}(\op{close}, 30)$; $\op{close} + \op{volume}$; $\op{close}^2$; $\op{Abs}(\op{close})$ \\
Additive sort (24) & $\op{TsDelta}(\log\op{close}, 1)$; $\log\op{close} - \log\op{open}$; $\op{TsStd}(\log\op{close}, 20)$; $\log\op{close} - \op{TsMean}(\log\op{close}, 20)$; $\log\op{close} + \log\op{volume}$; $\op{TsCorr}(\log\op{close}, \op{volume})$; $(\log\op{close} - \log\op{open})/2$; $2\log\op{close} - 2\log\op{open}$; $\op{TsSum}(\log\op{close}, 20) - 20\log\op{close}$; $\op{TsDelta}(\op{Greater}(\log\op{close}, \log\op{open}), 1)$; $\op{TsRank}(\log\op{close}, 20)$; $\op{Rank}(\op{TsDelta}(\log\op{close}, 5))$ & $\log\op{close}$; $\op{TsMean}(\log\op{close}, 20)$; $\log\op{close} + \log\op{open}$; $\op{Rank}(\log\op{close})$; $\op{Sign}(\log\op{close})$; $\op{TsIr}(\log\op{close}, 20)$; $2\log\op{close}$; $(\log\op{close})^2$; $\log\op{close} \times \op{return}$; $\op{TsSum}(\log\op{close}, 20)$; $\op{Greater}(\log\op{close}, \log\op{open})$; $\op{SLog1p}(\log\op{close})$ \\
\bottomrule
\end{tabular}
\end{table*}

\begin{figure}[!htb]
\centering
\includegraphics[width=\columnwidth]{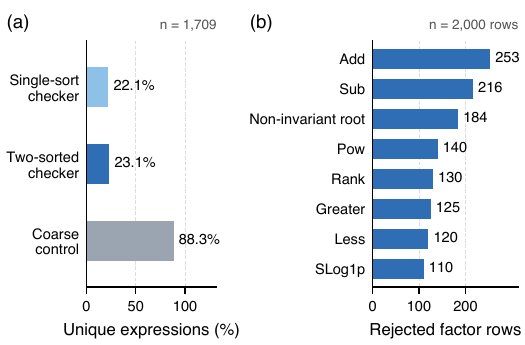}
\caption{Coverage of the checkers on the 1,709 unique expressions in 2,000 factor rows from an earlier library of 40 unrestricted pools (Table~\ref{tab:coverage}). (a)~Share of unique expressions certified by the single-sort checker and by the two-sorted checker, and accepted by the coarse control (accepted, not certified); (b)~the most frequent rejection reasons of the two-sorted checker, counted in factor rows. The two-sorted checker certifies slightly more than the single-sort checker, whereas the coarse control accepts almost everything and so prunes little.}
\label{fig:coverage}
\end{figure}

\textbf{Coverage on mined pools.} The coverage audit uses an earlier library of 40 unrestricted pools mined on the corrected provider by the earlier pipeline (two markets, two reward settings, ten seeds): 2,000 factor rows comprising 1,709 unique expressions (Table~\ref{tab:coverage}; Fig.~\ref{fig:coverage}). On the 40 arm-A pools of the original-membership confirmatory stage (Section~\ref{supp:final-moved}) the shares of unique expressions are 21.6\% (single-sort), 22.5\% (two-sorted) and 89.6\% (coarse control); counted over factor rows, 22.6\% of the arm-A factors are certified. Every expression accepted by the single-sort checker is accepted by the two-sorted one. The two-sorted checker additionally certifies 16 unique expressions, all from the log-shift family. The most frequent rejection reasons are $\op{Add}$ (253 factor rows), $\op{Sub}$ (216), a non-invariant root (184), $\op{Pow}$ (140), $\op{Rank}$ (130), $\op{Greater}$ (125), $\op{Less}$ (120) and $\op{SLog1p}$ (110). The coarse control accepts 1,116 unique expressions that the two-sorted checker rejects.

\begin{table}[!t]
\caption{Coverage of three checkers on 2,000 factor rows (1,709 unique expressions) from an earlier library of 40 unrestricted pools mined by the earlier pipeline.}
\label{tab:coverage}
\centering
\footnotesize
\setlength{\tabcolsep}{4pt}
\begin{tabular}{@{}>{\raggedright\arraybackslash}p{1.5in}cc@{}}
\toprule
Checker & \begin{tabular}[c]{@{}c@{}}Unique expressions\\certified\end{tabular} & \begin{tabular}[c]{@{}c@{}}Factor rows\\certified\end{tabular} \\
\midrule
Single-sort (exponent bookkeeping only) & 22.1\% & 23.5\% \\
Two-sorted (this paper) & 23.1\% & 24.6\% \\
Alpha$^2$-style coarse control (accepted) & 88.3\% & 88.6\% \\
\bottomrule
\end{tabular}
\end{table}

\textbf{Rejected but invariant on the panel.} A value-level audit of the 2,000 arm-A factors of the original-membership confirmatory stage (Table~\ref{tab:confirmatory}, Section~\ref{supp:final-moved}) recomputed every factor under stock-specific re-anchorings. Of the 2,000 factor rows, 81 were rejected by the two-sorted checker but numerically invariant on the panel. Typical examples are the complete expressions $\op{Pow}(\op{Div}(\op{Greater}(0.01,\allowbreak \op{vwap}),\allowbreak \op{close}),\allowbreak -10)$, in which the saturated $\op{Greater}(0.01, \op{vwap})$ equals VWAP and the outer ratio to close cancels the scale; $\op{Sub}(-1,\allowbreak \op{Pow}(\op{TsDiv}(\op{Sub}(\op{Less}(\op{volume},\allowbreak -0.01),\allowbreak \op{volume}),\allowbreak 20),\allowbreak 5))$, in which $\op{Less}(\op{volume}, -0.01)$ is the constant $-0.01$ and the ratio of volumes cancels the scale up to a negligible offset; and $\op{Rank}(\op{Pow}(\op{Sign}(\op{Pow}(\op{open},\allowbreak -5)),\allowbreak \op{high}))$, which is the constant 1 before the rank. Almost all 81 rows are degenerate in this sense, with a sub-expression constant or saturated on the data range, rather than genuine symmetric cancellations. The saturated sub-expressions themselves are not invariant. Value-level invariance on one panel is therefore not a usable certificate on its own.

\section{Size-Matched Analysis of Certified Subsets}\label{supp:sizematched}

Before changing the search, we asked whether, within pools mined without any constraint, the certified factors keep more out-of-sample information than the rest. Comparing the certified subset with the full pool is confounded by subset size. The frozen design therefore compares, for each of the 40 unrestricted pools later used as arm A in the original-membership confirmatory stage (Table~\ref{tab:confirmatory}, Section~\ref{supp:final-moved}), the certified subset with 1,000 random subsets of the same size drawn from the same pool, under both combiners. Here a factor counts as certified if it passes the two-sorted checker and is value-invariant on the panel; the mean subset size is 4.6 of 50 factors. One of the 40 pools has an empty certified subset and cannot be paired, so the analysis has 39 units.

\begin{figure}[!t]
\centering
\includegraphics[width=\columnwidth]{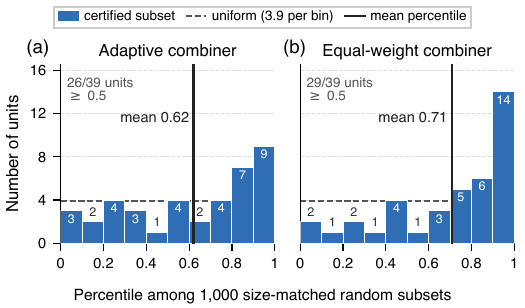}
\caption{Percentile of the certified subset among 1,000 size-matched random subsets of the same pool, over the 39 units, under (a) the adaptive and (b) the equal-weight combiner. The mass sits in the upper bins: 26 of 39 and 29 of 39 units are at or above the median random subset, with mean percentiles of 0.62 and 0.71.}
\label{fig:phase0}
\end{figure}

Table~\ref{tab:sizematched} reports the paired differences. Fig.~\ref{fig:phase0} shows the distribution over the 39 units of the certified subset's percentile among its 1,000 random subsets.

\begin{table}[!t]
\caption{Certified subset minus the median size-matched random subset (39 paired units, folds through 2021).}
\label{tab:sizematched}
\centering
\footnotesize
\begin{tabular}{@{}>{\raggedright\arraybackslash}p{1.8in}cc@{}}
\toprule
 & Adaptive & Equal weight \\
\midrule
$\Delta\mathrm{IC}$ (pooled) & $+0.0118$ & $+0.0183$ \\
Cells with certified $>$ random & 6/8 & 7/8 \\
Units at or above the median random subset & 26/39 & 29/39 \\
Neutralised $\Delta\mathrm{IC}$ & $+0.0068$ & $+0.0114$ \\
Mean percentile of the certified subset among random draws & 0.62 & 0.71 \\
\bottomrule
\end{tabular}
\end{table}

All three frozen criteria were met. Three qualifications apply. First, a five-factor certified subset is not better than the full 50-factor pool in mean IC (0.025 against 0.028, adaptive); the claim is only that same-size random subsets are beaten. Second, the certified subset is steadier across years (adaptive IC between 0.025 and 0.044 in every year 2015--2020, against a range of $-0.006$ to 0.089 for the full pool), but noisier across seeds because it is small. Third, the Alpha$^2$-style coarse subset differs from the full pool by only $+0.0006$ (adaptive) and $-0.0019$ (equal weight). The static two-sorted certificate alone differs from the combined static and value-level certificate by only $+0.0022$ and $-0.0008$, so the static checker suffices as the certificate.

\section{Combiner Ablation}\label{supp:combiner}

\begin{figure}[!t]
\centering
\includegraphics[width=\columnwidth]{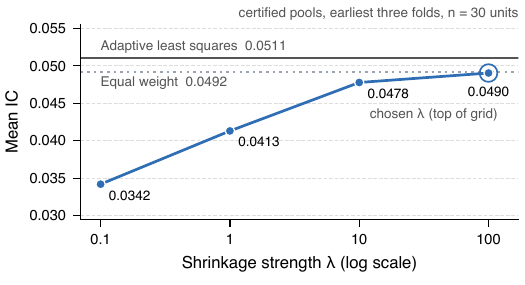}
\caption{Mean IC of the shrinkage combiner on the certified pools of the earliest three folds (30 units) against the shrinkage strength $\lambda$, with the adaptive least-squares (0.051) and equal-weight (0.049) levels on the same units as reference lines. The curve rises monotonically to the top of the grid and stays below both reference lines, so the grid has no interior optimum.}
\label{fig:combiner}
\end{figure}

Equal weighting and adaptive least squares behave differently on the certified pools: least squares is higher in mean, and equal weighting is higher in ICIR and worst 60-day IC (Table~\ref{tab:combiner}). We asked whether a shrinkage combiner toward the equal-weight prior $u$,
\[
w_t = \arg\min_{w \ge 0,\; \mathbf{1}^{\top} w = 1} \; \|y - Xw\|^2 + \lambda\, \frac{\mathrm{tr}(X^{\top} X)}{K}\, \|w - u\|^2 ,
\]
with $\lambda \in \{0.1, 1, 10, 100\}$, has an interior optimum. Under a frozen selection rule on the earliest three folds (30 units), $\lambda$ was chosen at the top of the grid (Fig.~\ref{fig:combiner}): mean IC rises monotonically from 0.034 ($\lambda = 0.1$) through 0.041 and 0.048 to 0.049 ($\lambda = 100$), which is 0.0002 below equal weighting (0.049) and below adaptive least squares (0.051) on the same units. Over all four development folds the corresponding values are 0.026, 0.031, 0.036 and 0.037, against 0.0373 for equal weighting and 0.0387 for least squares. Non-negative ridge without the prior matches least squares in mean (0.039) with a worse worst 60-day IC. The combiner study recomputes both reference combiners from cached factor values with its own implementation. Its four-fold levels (0.0387 and 0.0373) differ from the evaluator's (0.0391 adaptive in Section~\ref{supp:final-moved} and 0.0377 equal weight in the cloud-instance rows of Table~\ref{tab:cloudrows}) by at most 0.0004.

\begin{table}[!t]
\caption{Two reference combiners on the certified development pools (four folds, combiner-study implementation).}
\label{tab:combiner}
\centering
\footnotesize
\begin{tabular}{@{}lcc@{}}
\toprule
 & Adaptive least squares & Equal weight \\
\midrule
Mean IC & 0.0387 & 0.0373 \\
ICIR & 0.30 & 0.36 \\
Cell SD & 0.035 & 0.028 \\
Worst 60-day IC & $-0.077$ & $-0.052$ \\
Turnover & 0.18 & 0.23 \\
\bottomrule
\end{tabular}
\end{table}

The combiner choice does not alter the certified-versus-unrestricted comparison, and we report both endpoints throughout.

\section{Retrospective Comparison with Existing Pools, 2022--2024}\label{supp:retro}

Existing pools of Alpha\-Gen, Alpha\-SAGE, Alpha\-Forge, Alpha\-QCM, genetic programming and random search, five seeds per market, were evaluated with the same evaluator, corrected provider, purged window and combiners as the certified pools mined through 2021-12-03. No pool was re-mined for this table. The external pools were discovered by an earlier benchmark pipeline on the reference snapshot, with training through 2020 and validation on 2021, so their discovery cutoff and data differ from those of the certified pools. The window was exposed during earlier exploratory work. Table~\ref{tab:retro} is descriptive. It was computed on a different machine from Table~\ref{tab:laterfull} (a laptop CPU rather than the cloud instance); for the same unrestricted pools the two evaluations differ by 0.0006 in mean adaptive IC, within the platform sensitivity described in Section~\ref{supp:repro}.

\begin{table*}[!t]
\caption{Unified evaluation on 2022-01-04 -- 2024-12-03 (adaptive combiner; mean over 5 seeds, seed SD in parentheses). Bold marks the best formula pool in each column.}
\label{tab:retro}
\centering
\footnotesize
\begin{tabular}{@{}>{\raggedright\arraybackslash}p{1.9in}ccccc@{}}
\toprule
Method & CSI300 IC & Neutralised & CSI500 IC & Neutralised & \begin{tabular}[c]{@{}c@{}}Certified factors\\(CSI300 / CSI500)\end{tabular} \\
\midrule
Certified grammar (C) & \textbf{0.0519} (0.014) & 0.0152 & 0.0375 (0.013) & 0.0031 & 100\% \\
Unrestricted (A) & 0.0417 (0.014) & 0.0084 & 0.0376 (0.014) & 0.0026 & 30\% / 21\% \\
AlphaSAGE & 0.0508 (0.014) & 0.0144 & \textbf{0.0401} (0.010) & $-0.0021$ & 31\% / 24\% \\
AlphaGen & 0.0506 (0.014) & \textbf{0.0175} & 0.0374 (0.011) & $-0.0004$ & 31\% / 25\% \\
AlphaForge & 0.0410 (0.010) & 0.0105 & 0.0290 (0.008) & $-0.0115$ & 7\% / 4\% \\
AlphaQCM-IQN & 0.0379 (0.016) & 0.0047 & 0.0362 (0.019) & \textbf{0.0044} & 4\% / 4\% \\
Random search & 0.0420 (0.011) & 0.0130 & 0.0295 (0.015) & $-0.0069$ & 18\% / 14\% \\
Genetic programming & 0.0211 (0.019) & 0.0084 & 0.0237 (0.007) & $-0.0006$ & 40\% / 0\% \\
Alpha158 + LightGBM (own head, unpurged 726-day window) & 0.0631 (0.004) & -- & 0.0313 (0.002) & -- & -- \\
\bottomrule
\end{tabular}
\end{table*}

Under the adaptive combiner the certified pools are nominally the best formula pools on CSI300 and within 0.003 of the best on CSI500. No difference from AlphaSAGE or AlphaGen is significant with five seeds per cell (Welch $p \ge 0.7$), so the correct statement is competitiveness. The certified pools retain positive neutralised IC on both markets; on CSI500 most pools are at or below zero, and only AlphaQCM-IQN is also positive on both. Under sign-aligned equal weighting the ranking reverses: AlphaGen (0.062 / 0.061) and random-search pools (0.055 / 0.058) score higher, and the certified pools fall to 0.014 / 0.027. Pool diagnostics explain the reversal. The certified pools have the lowest internal redundancy (mean pairwise absolute correlation 0.064 / 0.072, against 0.16 / 0.18 for AlphaGen). On this window, however, their individual factors are weaker and their signs less persistent: 50\% / 57\% of certified factors keep the sign of their matured IC through 2021, against 62\% / 79\% for AlphaGen. Equal weighting cancels the flipped factors, while the adaptive combiner re-estimates signs daily. The Alpha158 row uses its own prediction head on an unpurged window and is shown for reference only.

\section{Provenance of the Inconsistent Build}\label{supp:provenance}

\begin{figure}[!htb]
\centering
\includegraphics[width=\columnwidth]{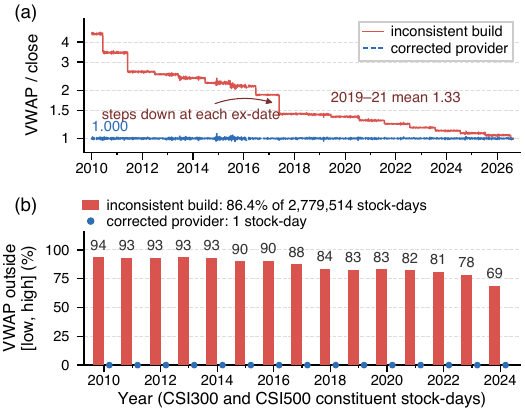}
\caption{The field mismatch in the inconsistent build (Sections~1 and~6.1 of the main paper). (a)~VWAP/close for one large bank from 2010 to 2026 (log axis): it steps down at each ex-date in the inconsistent build and stays at 1.000 in the corrected provider. (b)~Share of CSI300 and CSI500 constituent stock-days per year with VWAP outside the adjusted $[\mathrm{low}, \mathrm{high}]$ range: it falls from 94\% in 2010 to 69\% in 2024 in the inconsistent build, against one stock-day in total in the corrected provider. Index tickers are excluded from both panels; the mislisted SH000905 (Section~7.5 of the main paper) fails the test on all 599 of its in-scope days in both builds.}
\label{fig:leak}
\end{figure}

The incident is described in Section~1 of the main paper and the field-contract checks in Section~6.1. The mismatch had two value-level symptoms, which are distinct (Fig.~\ref{fig:leak}). First, VWAP/close tracked $1/F_{t,i}$: for example, its 2019--2021 level for one large bank was 1.33 and stepped down at each ex-date, against 1.000 in the corrected provider. Second, as a range check on all CSI300 and CSI500 constituent stock-days in 2010--2024, VWAP lay outside the adjusted $[\mathrm{low}, \mathrm{high}]$ interval on 86.4\% of stock-days (2,401,001 of 2,779,514), against one stock-day after correction.

\textbf{Code path.} Code-path inspection showed that the available public and local collection scripts were internally consistent: each divides volume by the adjustment factor before computing VWAP from amount and volume. Nevertheless, the archived provider snapshot exhibited the field-level mismatch described in Section~6.1 of the main paper. The snapshot was generated on the project's server in August 2026 (its calendar ends on 2026-08-21) by a collection pipeline adapted from the AlphaGen script. The generation or post-processing step that introduced the mismatch cannot be determined from the retained files.

\textbf{Public collection scripts.} Section~1 and Section~3.2 of the main paper describe the BaoStock collection script released with AlphaGen and redistributed with AlphaForge and AlphaSAGE. The versions we inspected are listed below; the file is \texttt{data\_collection/fetch\_baostock\_data.py} (under \texttt{src/} for AlphaSAGE), and line numbers refer to that file at the stated commit.
\begin{itemize}
\item AlphaGen (github.com/ICT-FinD-Lab/alphagen), commit 72bea90, 2023-02-28: forward-adjusted query (line 171); prices divided by the forward factor on the reference date (lines 182--184), whose default is 2023-01-15 (line 52) and is not overridden (lines 267--272). The AlphaGen paper states in its footnote~3 that its data are ``forward-dividend-adjusted respected to the adjustment factors on 2023/01/15''.
\item AlphaGen, commit da721c9, 2023-07-28: as above, with the reference date set to the download date (lines 268--274).
\item AlphaGen, commit 1c18754, 2024-12-18 (current master): backward adjustment by default (lines 53, 68, 178), rescaled by the backward factor on an adjustment date (lines 193--196) that the script sets to 2009-01-01 (line 296).
\item AlphaForge (github.com/DulyHao/AlphaForge), commit d0cfc27, 2024-09-01: forward-adjusted query (line 171), prices divided by the forward factor on the reference date (lines 182--184), reference date set to the download date (lines 268--274).
\item AlphaSAGE (github.com/BerkinChen/AlphaSAGE), commit 517467a, 2026-02-26: as AlphaForge (lines 171 and 182--184), reference date set to the download date (lines 282--288).
\end{itemize}
In every version volume is divided by the same factor as the prices before VWAP is computed from amount and volume (lines 185--186; lines 197--198 at 1c18754), so the fields are mutually consistent. With a reference date after the sample, the stored price for a day depends on the corporate actions between that day and the reference date, which is end-anchoring in the sense of Section~3.2 of the main paper. This is a property of the stored values. Ratio factors cancel the anchor, and we make no claim about results reported with these data.

\textbf{Hash linkage and correction scope.} By hash, the archived snapshot is the dataset of the earlier pipeline's formal runs: all 32,877 path and hash pairs in that run's dataset manifest match the archived files. This ties the inconsistency to the numbers that pipeline reported. The correction of Section~6.1 of the main paper changed exactly 3,456 files, all of them VWAP or volume arrays. The corrected provider is frozen with a SHA-256 manifest over all 32,877 files.

\textbf{Consistency with the reference snapshot.} The comparison covers 1,705 shared stocks and 7,555,563 overlapping stock-days. Both snapshots use the same internal adjustment convention: daily close-to-close returns agree to within $2.2\times10^{-5}$ and raw volume to within a relative $2.2\times10^{-7}$, CSI300 and CSI500 membership intervals are identical through 2024-12-31, and absolute price levels differ by a constant per stock because the snapshots use different base dates.

\textbf{Event-level check of the field contract (Section~6.1 of the main paper).} For every stock of the union universe (index tickers excluded) and every trading day from 2010-01-04 to 2026-08-21, the check compares the corrected provider with exchange-reported quantities stored alongside it that do not pass through our adjustment code (\texttt{event\_contract/event\_contract.py}). An ex-date is a day on which a stock's stored adjustment factor changes after a trading day; there are 19,867 (20,530 including those after a suspension). (i) Price: the stored adjusted return $\op{close}_t/\op{close}_{t-1} - 1$ equals the exchange's percentage change, which is computed from the raw close and the ex-rights reference price, within $10^{-3}$ on 99.97\% of ex-dates (median absolute difference $4.3 \times 10^{-7}$, 99th percentile $3.8 \times 10^{-6}$). On the 6.08 million stock-days without a factor change it matches on all but 15. The five exceptions are SH600633 on 2010-03-01, SH601258 on 2020-01-02, SZ000981 on 2022-02-28, SZ001914 on 2019-06-06 and SZ002466 on 2020-01-02; on these days the vendor's factor changes by 1.5--67\% while the exchange's ex-rights reference price implies a different change, so the stored adjusted return differs from the exchange return by 0.015--0.63. (ii) Volume: the implied float share count $\op{volume}_t \cdot F_t / \mathrm{turnover}_t$ is continuous across the 17,161 ex-dates whose factor changes by less than 10\%, a proxy sample chosen because such small changes are typical of cash dividends, which leave the share count unchanged (event types were not matched to an external record): on 99.8\% of them it moves by less than half the factor change (median movement $2.3 \times 10^{-6}$ in log terms, against a median factor change of 0.011), whereas a volume stored without adjustment, the failure of the inconsistent build, would move it by the whole change. (iii) VWAP: the stored VWAP lies inside the stored daily range on every ex-date. The check relies on the exchange's reference prices and turnover rates; it does not verify the cash amount of each dividend.

\textbf{Isolated anomalies.} The full consistency check against the reference snapshot also surfaced anomalies that the correction does not address. The CSI500 membership file of both snapshots lists the index ticker itself (SH000905) as a constituent from 2007-01-15 to 2012-06-25. \textbf{Origin of the line.} The data snapshots distributed through Qlib's download utility (\texttt{qlib.run.get\_data}, which fetches from the releases of the GitHub repository \texttt{SunsetWolf/qlib\_dataset}; accessed 2026-10-31) carry the same line: the CSI500 membership file of releases v2 (\texttt{qlib\_data\_cn\_1d\_latest.zip}, \texttt{qlib\_data\_cn\_1d\_v3.zip}) and v3 (\texttt{qlib\_data\_cn\_1d\_latest.zip}) contains \texttt{SH000905 2007-01-15 2012-06-25} (2,017 rows; the extracted file is identical in the three archives, SHA-256 \texttt{12cf811fe46b4343}\allowbreak\texttt{c1ccaeeda506ccc5}\allowbreak\texttt{440e93c511aea65d}\allowbreak\texttt{1d761b8f592d9129}), while the older releases (v0, v1, v2 \texttt{0.6.1}) and the simple dataset have no CSI500 file. Our file has the same 2,017 (ticker, start) pairs and differs only in the end dates of the 500 open spells, which the AlphaGen collection script (\texttt{fetch\_baostock\_data.py}, \texttt{\_fix\_constituents}) extends to the day of collection; that script takes its membership files from the Qlib base dataset and adds the index series sh.000903, sh.000300, sh.000905 and sh.000852 to the downloaded tickers. In the v2 snapshot SH000905 is active as a CSI500 member on 2011-06-01, one of 500, with a stored close of 4,490.9 and a volume of $2.75 \times 10^9$, against a median of 2.93 and $1.75 \times 10^7$ over 200 other members (maxima 23.8 and $3.15 \times 10^8$); the snapshot has no VWAP field, so the VWAP/close symptom of Section~7.5 of the main paper belongs to our build, not to the public file, whose error is the membership interval. AlphaSAGE's documentation directs users to the Qlib data, AlphaForge's to the AlphaGen script; we did not determine which membership file each published study used, and we found no report of the line in the Qlib issue tracker (one issue, \#1323, concerns the index's own price data). The check reads only the membership and feature files from the release archives by HTTP range requests (\texttt{upstream\_membership/check\_qlib\_official.py} in the repository). \textbf{Factor-level effect by certification.} Removing the ticker from the discovery panel, with the frozen CSI500 pools and the miner's IC (\texttt{sh000905/sh000905\_factor\_level.py}), moves the discovery IC of the 804 non-certified factors of the 20 arm-A pools by 0.00153 on average (11.4\% above 0.005, largest 0.026), of the 196 certified factors of the same pools by 0.00027 (0.5\% above 0.005), and of the 1,000 factors of the 20 arm-C pools by 0.00051 (2.9\% above 0.005, largest 0.031); paired by pool within arm A, the certified factors move less in 19 of 20 pools (Wilcoxon $p = 7 \times 10^{-4}$). Averaged first within each pool, the two figures are 0.00037 and 0.00152, the values quoted in Section~7.5 of the main paper. Among the non-certified factors, those without VWAP move more (0.0018) than those with it (0.0009). These comparisons do not separate the contributions of the index's price and volume level, its returns, the change in the cross-section and, in our build, its VWAP/close mismatch. The comparison is descriptive and concerns fixed pools; the invariance theorem does not cover a change in the set of stocks. We kept the file unchanged in the frozen provider. The ticker has complete price, volume and VWAP fields in this period, but its VWAP field is not on the scale of its close (VWAP/close between 0.0022 and 0.0030), so it breaks the field contract; the range test above excludes it. It sits in the CSI500 discovery cross-section on the 599 trading days from 2010-01-04 to 2012-06-25, which are 50\% of the discovery days of fold f2015\_16 and 23\% of those of fold f2021. On each of these days it is one name among about 500, so it accounts for 0.05--0.10\% of the CSI500 discovery stock-days. It does not appear in CSI300 and enters no validation day, because every validation window starts in 2015 or later, but the evaluator's combiners use factor statistics accumulated from 2010, so it also reaches the validation-window IC through the combiner's history. We measured both channels on the laptop without re-mining; a protocol note was written before the runs. On the discovery windows, removing the ticker changes the discovery IC of individual pool factors by up to 0.03, because its VWAP field is an extreme cross-sectional outlier after day-wise normalisation, while the paired discovery-window difference between arms C and A is unchanged ($-0.0056$ with the ticker and $-0.0058$ without, for the ensemble with the stored weights). In the frozen validation evaluation of the 60 CSI500 pools of arms C, A and B, the per-unit change of adaptive IC has a standard deviation of 0.003--0.005 (at most 0.014), and that of equal-weight IC is at most 0.006. In the laptop environment of Section~\ref{supp:repro}, the paired C $-$ A over the 40 units moves from $+0.0096$ to $+0.0101$ (27 instead of 26 positive units; 6 of 8 cells in both), and C $-$ B and B $-$ A change by less than 0.001. The ticker therefore does not change any reported comparison. Because the fixed-pool checks cannot show whether the ticker changed which expressions the searches found, we re-mined the 40 CSI500 pools of arms A and C on a copy of the provider without the SH000905 line, with the Gate-1 commands unchanged apart from the data path, under a protocol and freeze file written before the runs. Two control runs on the original data (fold f2019\_20, seed 0, one per arm) reproduced the original pools byte for byte on a machine with a different processor and GPU driver, so the re-mined pools differ from the original ones only through the ticker. No re-mined pool is identical to its original: a certified pool keeps on average 26.3 of its 50 expressions (Jaccard overlap 0.37, range 0.17--0.74) and an unrestricted pool 14.6 (0.17, range 0.08--0.32). Evaluated with the frozen evaluator on the same provider copy, the CSI500 difference C $-$ A under the adaptive combiner is $+0.0016$ (95\% CI $[-0.0110, +0.0141]$; 13 of 20 units; 2 of 4 cells; fold means $-0.0125$, $-0.0138$, $+0.0251$ and $+0.0075$), against $+0.0120$ ($[+0.0020, +0.0219]$; 15 of 20; 3 of 4) for the original pools; the neutralised difference is $-0.0049$ against $+0.0067$, and the equal-weight difference $-0.0052$ against $-0.0001$. The unrestricted pools gain ($+0.0065$ per unit, $p = 0.09$) and the certified pools lose ($-0.0039$, $p = 0.32$). The per-unit change in C $-$ A has mean $-0.0104$ and standard deviation 0.026 ($p = 0.09$), the size of the seed-to-seed spread of C $-$ A (standard deviation 0.021 for the original and 0.027 for the re-mined pools). With the re-mined CSI500 pools in place of the original ones and CSI300 unchanged, the 40-unit difference in the laptop environment of Section~\ref{supp:repro} is $+0.0049$ ($[-0.0036, +0.0135]$; 25 of 40 units; 5 of 8 cells), the neutralised difference is $-0.0010$, and both criteria of the main comparison's decision rule ($\ge -0.005$) still hold. This is a post-hoc sensitivity analysis. The first execution of the runs was lost when the cloud host went into scheduled maintenance with 35 of 42 runs complete and not yet downloaded; all 42 runs were restarted from scratch on another machine, and the lost outputs were never used. One stock in the reference snapshot has VWAP outside the range on 2,094 stock-days, and one stock-day is out of range in both.

\section{Protocol Records and Deviations}\label{supp:deviations}

All stage protocols, freeze files and analysis scripts are archived with hashes. The summary table below lists every deviation by stage with its timing relative to the affected outcomes; the grouped entries that follow give the reason for each change and the full record.

\begin{table*}[!t]
\centering
\scriptsize
\renewcommand{\arraystretch}{1.12}
\begin{tabular}{@{}>{\raggedright\arraybackslash}p{1.1in}>{\raggedright\arraybackslash}p{2.25in}>{\raggedright\arraybackslash}p{1.5in}>{\raggedright\arraybackslash}p{1.75in}@{}}
\multicolumn{4}{@{}l}{\small\textbf{Summary of deviations from the frozen stage-level protocols}} \\[2pt]
\toprule
Stage & What changed & When, relative to outcomes & Effect on confirmatory status \\
\midrule
Size-matched analysis & Guard on an empty stack fixes an evaluator crash with zero random draws & Before any outcome was computed & None; no number changed \\
Main comparison & Per-process GPU memory caps; append-mode scheduler log & Execution only & None; search semantics untouched \\
Typed mask & Validity clause amended (every guard rejection must be a length-limit exit); scheduler restart; CPU environment completed; two-unit smoke test & Amendment recorded before the quality analysis; two smoke-test rows seen before the full evaluation & Amendment-conditioned confirmatory analysis; under the original clause all 40 masked runs would be void \\
Later windows & 2025--26 window labelled additional out-of-sample evidence & The window had been opened by an earlier experiment & Not a hold-out \\
Matched baselines & Eleven execution amendments (scheduling, environment, machine allocation); JSON serialisation of cell keys in the analysis script & During execution & None; decision logic, thresholds and inputs unchanged \\
S\&P~500 & Evaluator start-up crash fixed by moving two channel-index lines & Before any data loading or IC computation & None; evaluation re-run \\
Matched-pruning control (R) & Pre-planned extension from 24 to 40 runs per arm & After the 24-run cohort had been inspected, as the protocol allowed & 40-unit result reported; same verdict under both cohorts \\
Single-sort control (S) & Pre-specified as descriptive; rejection rate not matched & In the protocol, before any run & No verdict \\
Numerical re-audits & Float64 diagnosis and relative-floor re-audit run on the laptop, original audit on the cloud instance; same-laptop baseline with the original operator added later & After the original audit & Cloud and laptop pool-level numbers not directly comparable; the operator comparison uses the same-laptop baseline \\
Length-pruning control (P) & Stopping rule triggered; production batch not run & After a logging pilot, before any arm-P run & No arm-P pool produced or evaluated \\
Guard on PPO & Four runs per GPU instead of three; physical copy of the data directory after a failed checksum pre-flight & Before any run & None; run content unchanged \\
SH000905 re-mining (post-hoc) & Concurrency raised (scheduler restart); first execution lost to host maintenance, all runs restarted on another machine & During execution, before any outcome was evaluated & None; lost outputs unused, controls byte-identical \\
Directional audit with real anchors & A launcher fault stopped the first launch after a few minutes; launcher fixed, freeze file re-hashed, one finished unit recomputed & Before any outcome was analysed & None; audit script and analysis unchanged (Section~\ref{supp:realanchor}) \\
Float64 evaluator & Thread-count proxy (E1) left every output bit-identical; the frozen contingency (split reduction, E3) used; memory guard lowered & Before any comparison was computed & None; descriptive (Section~\ref{supp:float64}) \\
Past-anchored baseline and seed extension (post-hoc) & Amendment before the first formal run added CSI300 seeds 5--9, the /4 injection and a union-universe block; the union block (40 runs) withdrawn for budget before any of its runs started; scheduler restarts and a per-GPU job count fix & Amendment before any run; withdrawal during execution, before any outcome was evaluated & None; descriptive stages (Section~\ref{supp:hfq}); two universes only \\
\bottomrule
\end{tabular}
\end{table*}

\begin{enumerate}
\item \textbf{Execution fixes that leave search semantics and decision logic unchanged} (size-matched analysis, main comparison, matched baselines, S\&P~500). These changes were made so that frozen runs could complete or an evaluator could start; none alters what is searched, what is computed or how a verdict is reached. In the size-matched analysis, an evaluator crash with zero random draws was fixed by a guard on an empty stack before any outcome was computed; no number changed. In the main comparison, per-process GPU memory caps and an append-mode scheduler log were added; neither touches search semantics. The matched baselines needed eleven execution amendments, all on scheduling, environment and machine allocation; a few runs that failed to start or ran out of memory were restarted from scratch, and an output-format bug in the analysis script was fixed without touching its logic (each amendment is listed in the stage's protocol record in the repository). The first S\&P~500 evaluation crashed at start-up before loading any data and was re-run after the fix.
\item \textbf{The typed-mask validity clause.} The clause was amended because it contradicted the audited design. The frozen clause required zero guard rejections under the mask, whereas in the audited design termination stays available at the length limit and the guard rejects. All 511 rejections across the 40 masked runs are 19-token length-limit exits with root-type failures. An amended validity check, that every rejection is such an exit, passes all 80 runs. The amendment was recorded before the quality analysis and accepted after an independent code-path recheck, in which both validators were re-run from separately written code. The typed-mask results are therefore an amendment-conditioned confirmatory analysis; under the original clause all 40 masked runs would have been void. Also recorded for this stage: a scheduler restart with a larger memory budget at 59 of 80 runs, with no training process affected; completion of the CPU evaluation environment to the main-comparison package versions; and a two-unit smoke test of the unchanged evaluator, whose two rows were seen before the full evaluation.
\item \textbf{Status of the later windows.} The 2025--26 window had been opened by an earlier experiment, so it is labelled additional out-of-sample evidence rather than a hold-out.
\item \textbf{Control arms R, S and P.} The three controls were set up in protocols written before any of their runs; the entries record how the budget of arm R was reached, why arm S has no verdict, and why arm P was never run. \textit{Arm R.} The protocol and the freeze file were prepared on 2026-10-23 at 01:05 UTC, before any run of this stage. They pre-specified a full budget of 40 runs per arm (seeds 0--4) and a reduced first cohort of 24 runs per arm (seeds 0--2), and they stated that the extension from the reduced to the full budget was pre-planned and permitted only if all 40 units are then used for every arm. Seeds 0--2 were run first, and the 24-pair result was inspected: C $-$ R $= +0.0194$ (95\% CI $[+0.0055, +0.0333]$; 18 of 24 units; 6 of 8 cells) and R $-$ A $= -0.0087$ ($[-0.0183, +0.0010]$; 10 of 24 units). The pre-planned extension to seeds 3--4 was then executed. The decision to extend was taken after the first cohort had been seen, as the protocol allowed. The 40-pair result (C $-$ R $= +0.0181$, R $-$ A $= -0.0085$; Table~4 of the main paper) is the one reported; the frozen verdict is the same under both cohorts, and both are listed here. \textit{Arm S} was pre-specified in the same protocol as descriptive, with no verdict; its rejection rate was not matched and is a reported quantity. \textit{Arm P.} The protocol for arm P (a deterministic operator- and token-count cap, 40 runs, descriptive) and its calibration rule were written before any arm-P run. The calibration rule stated that the production batch would not be launched if the nearest structural rule had an operator cap of zero, a token cap below three, or missed the target rejection rate by more than 0.05. A 2,000-episode logging pilot (excluded from all analyses) gave a nearest rule with a token cap of two and a miss of 0.058, so the stopping condition was triggered and the batch was not run; no arm-P pool was produced or evaluated. An earlier offline calibration on the admitted factors of the arm-A pools had suggested a cap of two operators and four tokens with a rejection rate of 0.89; it was biased, because admitted pool factors are longer than the candidate stream, and on the pilot's candidate stream that rule rejects 73.6\%.
\item \textbf{Platform of the numerical re-audits.} The relative-floor re-audit and the float64 diagnosis both ran on the laptop, whereas the original anchor audit of Section~7.1 of the main paper ran on the cloud instance. This entry is recorded because the platform enters the comparison: their pool-level numbers are not directly comparable with Table~5 of the main paper, owing to the platform sensitivity described in Section~\ref{supp:repro}, and per-factor correlations are compared only within the same environment where available (Section~S8). On 2026-10-24 the frozen audit was re-run on the laptop with the original operator (80 pools; a short protocol note and script hashes were written before the run), so that the relative-floor comparison of Section~S8 is made within one environment.
\item \textbf{Guard on a PPO generator.} The protocol, the guard adapter and the hook were frozen with hashes before any run; it pre-specified a descriptive comparison of 12 runs (C-PPO and A-PPO, CSI300, folds f2019\_20 and f2021, seeds 0--2), no decision rule and no verdict. A rejected expression receives reward 0, the value AlphaGen gives an expression it cannot evaluate; the penalty of $-1$ for truncated episodes is unchanged. Eight unit tests of the hook passed, and a CPU smoke run with the guard off reproduced the matched-stage smoke pool byte for byte. A pre-registered launch check, one guarded run of 40,960 steps, required a non-empty pool; the pool was non-empty, with 98.2\% of unique candidates rejected, and the pilot is excluded from all analyses. Two execution changes were recorded before any run. The first pre-flight failed its checksum test only because the data directory was a symbolic link that the file search did not follow; the directory was replaced by a physical copy of the verified provider (32,877 files) and the pre-flight passed. The scheduler ran four jobs per GPU instead of three, so that every seed pair ran concurrently; no run content changed. All 12 runs completed with full pools at 301,056 steps. The neutralised adaptive difference C-PPO $-$ A-PPO is $+0.0009$; on the same six units the GFlowNet's C $-$ A is $+0.006$, and the difference between the two generators' guard effects is $-0.006$ (95\% CI $[-0.047, +0.035]$). The six A-PPO pools were trained on another machine than the matched-comparison pools they reproduce.
\end{enumerate}

\textbf{Detail for Section~6.5 of the main paper.} The frozen rules of the confirmatory stages other than the main comparison and the matched-pruning control were as follows. \textit{Matched baselines} (Section~7.4 of the main paper): arm C is ahead of a baseline if and only if its four-fold mean adaptive IC is higher in both markets, at least 6 of 8 cells are non-negative, and the neutralised difference is at least $-0.005$ in both markets. \textit{Later windows} (Section~7.4 of the main paper), applied to the 2025--26 window on the native universe under the adaptive combiner (first row of Table~\ref{tab:laterfull}): mean $\Delta\mathrm{IC} \ge 0$, at least 6 of 10 positive pairs, neutralised $\Delta\mathrm{IC} \ge 0$, and both markets $\ge -0.005$. \textit{S\&P~500} (Section~7.4 of the main paper): mean adaptive $\Delta\mathrm{IC} \ge 0$, neutralised $\Delta\mathrm{IC} \ge -0.005$, and at least 6 of 10 seeds positive. \textit{Typed mask} (Section~7.3 of the main paper; Section~\ref{supp:maskpools}): efficiency metrics better in at least 7 of 8 cells, and quality non-inferior overall and per market. \textit{Size-matched analysis} (Section~\ref{supp:sizematched}): certified subsets beat size-matched random subsets in at least 6 of 8 cells under both combiners, with a consistent pooled sign and a non-negative neutralised difference.

\section{Cloud-Instance Evaluation of the C, A and B Pools}\label{supp:cloudrows}

Table~\ref{tab:cloudrows} retains the cloud-instance evaluation of the certified (C), unrestricted (A) and coarse-typed (B) pools, that is, the C, A and B rows of Table~\ref{tab:matchedfull} before the same-environment re-evaluation of Section~\ref{supp:repro}. Per unit, the laptop evaluation differs from the cloud-instance evaluation by a mean of $-0.0004$ (C), $+0.0007$ (B) and $+0.0015$ (A) in adaptive IC, with standard deviations of 0.0051, 0.0095 and 0.0060 and maximum absolute differences of 0.020, 0.041 and 0.022; the adaptive combiner selected a different factor set on at least one day in 38 of the 40 certified units. Under equal weighting the standard deviations are 0.0011--0.0026. With the cloud-instance rows the frozen rule gave ahead against AlphaSAGE ($+0.0118$ and $+0.0100$; 6 of 8 cells; neutralised $+0.0095$ and $+0.0039$) and competitive against AlphaGen ($+0.0019$ and $+0.0011$; 5 of 8 cells; neutralised CSI500 $-0.0076$); in the laptop environment the verdict against AlphaSAGE is mixed (Section~7.4 of the main paper). Arms R and S of Section~7.3 of the main paper were evaluated in the laptop environment, and their A and C rows reuse this re-evaluation (identical evaluator, environment and inputs). The platform sensitivity was also visible during development: across platforms, CPU and GPU evaluations of the same pools differed by about 0.004 IC on average; in one unit, a cross-platform difference changed the adaptive IC from 0.0375 to 0.0449; and on one machine, a re-run of the same evaluation once differed by $3.3\times10^{-6}$ (0.007508205 against 0.007511535) because of multithreaded reductions, while a second re-run reproduced the original value to $10^{-12}$, as did the other 59 units checked.

\begin{table}[!htb]
\caption{Cloud-instance evaluation of the C, A and B pools (their rows of Table~\ref{tab:matchedfull} before the same-environment re-evaluation of Section~\ref{supp:repro}; four-fold mean of cell means).}
\label{tab:cloudrows}
\centering
\scriptsize
\setlength{\tabcolsep}{2pt}
\begin{tabular}{@{}>{\raggedright\arraybackslash}p{1.0in}cccccc@{}}
\toprule
Method & \multicolumn{2}{c}{IC} & \multicolumn{2}{c}{Neutralised IC} & \multicolumn{2}{c}{Equal-weight IC} \\
\cmidrule(lr){2-3}\cmidrule(lr){4-5}\cmidrule(lr){6-7}
 & CSI300 & CSI500 & CSI300 & CSI500 & CSI300 & CSI500 \\
\midrule
C (certified) & 0.0356 & 0.0426 & 0.0214 & 0.0233 & 0.0387 & 0.0366 \\
A (unrestricted) & 0.0235 & 0.0319 & 0.0141 & 0.0185 & 0.0224 & 0.0364 \\
B (coarse-typed) & 0.0222 & 0.0301 & 0.0107 & 0.0169 & 0.0230 & 0.0347 \\
\bottomrule
\end{tabular}
\end{table}

\section{Numerical Diagnosis of the Anchor Audit}\label{supp:numerics}

\textbf{Float64 diagnosis.} A float64 diagnosis of the 244 certified factors of Section~7.1 of the main paper whose stored and re-anchored float32 values correlate below 0.9 under at least one stock-specific anchor, with a seeded control sample of 244 certified factors whose float32 correlation exceeds 0.99, separates the two mechanisms. The diagnosis ran on the laptop. Re-evaluated in float64 under the same re-anchorings, 217 of the 244 (89\%) have their correlation with the stored values restored to at least 0.99, including 169 of the 173 $\op{TsKurt}$ factors, whose float32 values correlate with their own float64 values at a median of only 0.13: their float32 drift is finite precision in the fourth-moment ratio, not anchor dependence. The remaining 22 stay below 0.9 in float64. Twelve of them contain $\op{TsCorr}$ and fail through its variance floor, the threshold semantics of Section~4.2 of the main paper; the others are numerically degenerate expressions, such as $\op{TsStd}(\op{high} \times \op{Inv}(\op{high}), 40)$ or $\op{Rank}(\op{close} \times 0.01/\op{close})$, whose values are rounding noise around a constant and drift as much under the placebo. No control factor falls below 0.9 in either precision. A correlation of at least 0.99 is not elementwise equality, so these factors are described as restored to high correlation, not as invariant.

\textbf{Property-based test of the checker.} We drew 1,500 certified and 1,500 rejected unique random expressions from the search grammar (36 operators, six fields, 14 constants, windows 10--50, at most 20 tokens) and executed each with the core's own operators on a synthetic panel of 500 days and 40 stocks and on its re-anchored copy (stock scales drawn from $\{0.2, 1, 3.7, 25\}$). An expression counts as invariant if the finite masks agree and the scaled maximum error is below $10^{-6}$ (float64) or $10^{-3}$ (float32). With the original operators, 91.2\% of certified expressions are invariant in float64 and 82.9\% in float32; with the relative-floor $\op{TsCorr}$, 98.5\% and 88.9\%. All 106 float64 failures that the relative floor removes contain $\op{TsCorr}$. Of the 22 float64 failures that remain, 21 are ill-conditioned, in that a relative perturbation of $10^{-12}$ of the input already moves the output beyond the tolerance (values that should be exactly zero carry rounding noise that $\op{Sign}$ or a division amplifies), and one is an underflow ($\op{Pow}(\op{low}, -30)$ followed by fourth moments in $\op{TsKurt}$). No failure contradicts a typing rule. Among the rejected expressions, 15--16\% are invariant on this panel, which measures the incompleteness of the checker. The test is a finite check on synthetic inputs, not a proof.

\textbf{Relative-floor re-audit.} As an implementation check, we replaced the absolute variance floor of the implemented $\op{TsCorr}$ ($10^{-6}$) by a floor relative to the window's scale and re-ran the audit of all 80 pools on the laptop with the frozen audit script. Of the 15 $\op{TsCorr}$-containing certified factors that were sensitive, 11 have their correlation with the stored values restored to at least 0.99 under both anchors; of the 4 that remain, three are $\op{TsCorr}(\op{volume}^2, \op{volume})$ factors whose variance product overflows float32, and one is a $\op{TsIr}$ of a $\op{TsCorr}$, where the instability moves into $\op{TsIr}$, which has no floor. The pool-level statistics barely move. To separate the operator change from the platform, we re-ran the same audit on the laptop with the original operator, so that the two laptop runs differ only in $\op{TsCorr}$. On the laptop the original operator reproduces the cloud audit at factor level: for the certified pools the share of factors with correlation below 0.9 is 12.2\% on both platforms and the strict elementwise pass rate is 34.8\% against 34.9\%, and all 15 sensitive factors are sensitive again, so their sensitivity is not platform noise. Its pool-level drift is higher on the laptop (mean stock-specific drift of the certified pools 0.0018 against 0.0011, with a global placebo of 0.0018 against 0.0011), which is the platform sensitivity of Section~\ref{supp:repro}. Against this same-laptop baseline, the relative floor changes the certified pools' mean stock-specific drift from 0.0018 to 0.0020 (global placebo 0.0018 in both), the share of factors with correlation below 0.9 from 12.2\% to 11.7\%, and the strict pass rate from 34.8\% to 35.2\%; the unrestricted pools' mean drift falls from 0.0147 to 0.0141. The stored adaptive IC moves by at most 0.0173 per pool (mean absolute change 0.0031 over the 79 pools that contain $\op{TsCorr}$), and the one pool without a $\op{TsCorr}$ factor is bit-identical. The main experiments keep the original operator, and the dominant sensitivity remains the finite precision of $\op{TsKurt}$, not the threshold.

\textbf{Reference-evaluator audit (Table~5 of the main paper).} The frozen audit was re-run on the same 80 pools with both repairs together, the relative-floor $\op{TsCorr}$ and float64 throughout: the panel is cast to float64 once after loading, and expression evaluation, re-anchoring, day normalisation, ranks and IC statistics then run in float64 (\texttt{ref\_eval/anchor\_audit\_f64.py}, three edits to the frozen script; protocol \texttt{ref\_eval/PROTOCOL\_refeval.md}, laptop). Among the 2,000 certified factors, the share whose values correlate below 0.9 with their stock-specifically re-anchored values falls from 12.2\% to 0.55\% (0.70\% under the global placebo); the share below 0.99 falls from 25.2\% to 1.8\%; and the share matching elementwise under both anchors (equal finite masks, relative error below $10^{-4}$) rises from 34.9\% to 75.1\%. Finite masks still differ somewhere for 13.9\% of certified factors (27.0\% in float32), on a median of about 6,400 cells, but identically under the global placebo, so the difference is numerical: most of these factors contain $\op{TsSkew}$ or $\op{TsKurt}$ on windows of near-constant values. The 11 certified factors that stay below 0.9 are ill-conditioned: algebraic constants evaluated as rounding noise ($\op{TsStd}(\op{high} \times \op{Inv}(\op{high}), 40)$, $\op{Rank}(\op{close} \times 0.01/\op{close})$, $(\op{low}/\op{open})/\op{low} \times \op{open}$ and a product that reduces to the constant 50), powers of a negative base ($\op{Rank}(\op{Pow}(-2, \op{TsDiv}(\op{low}, 40)))$, twice), inverses of near-zero dispersions ($\op{Inv}(\op{TsMad}(\log \op{open}, \cdot))$, three times) and a $\op{TsIr}$ of a $\op{TsCorr}$ and of a doubly ranked $\op{TsKurt}$ (\texttt{ref\_eval/residual\_certified\_below\_0p9.csv}). The unrestricted factors stay anchor-sensitive: 38.9\% correlate below 0.9 (13.6\% under the placebo), and 31\% have excess IC drift above 0.005. At pool level the certified pools' adaptive IC moves by 0.0027 on average under the stock-specific anchors and by 0.0028 under the placebo, against 0.0018 and 0.0018 for the same pools in laptop float32: rounding-level differences still flip threshold selections of the adaptive combiner, whatever the anchor (\texttt{ref\_eval/report\_refeval.json}).

\section{Figures for the Anchor Audit, the Matched Comparison and the Transfer Experiments}\label{supp:figures}

\begin{figure}[!htb]
\centering
\includegraphics[width=\columnwidth]{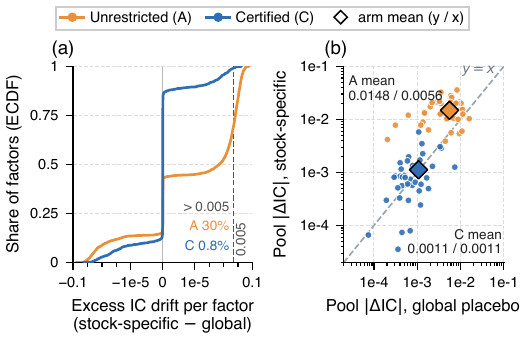}
\caption{Anchor audit of the 40 unrestricted (A) and 40 certified (C) pools of Section~7.1 of the main paper, with each drift taken as the larger of the two re-anchorings. (a)~Empirical cumulative distribution function (ECDF) of factor-level excess IC drift (stock-specific minus global placebo) on a symmetric-log axis, with the 0.005 threshold marked; (b)~pool-level absolute IC drift under stock-specific re-anchoring against the global placebo, one point per pool, with arm means. Certified pools straddle the diagonal, whereas unrestricted pools lie above it, and far more unrestricted than certified factors exceed the 0.005 threshold.}
\label{fig:anchor}
\end{figure}

\begin{figure}[!htb]
\centering
\includegraphics[width=\columnwidth]{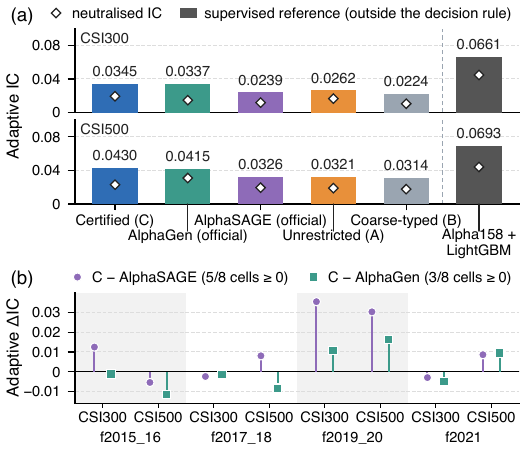}
\caption{Matched comparison on the rolling folds of Section~7.4 of the main paper and Table~\ref{tab:matchedfull}, all rows evaluated in one environment. (a)~Four-fold mean adaptive IC per method and market, with the neutralised IC marked by white diamonds and the supervised reference shown outside the decision rule. (b)~Per-cell differences C $-$ AlphaSAGE and C $-$ AlphaGen. The certified search has the higher mean IC in both markets, with the largest margins in f2019\_20. All pools were mined with the original CSI500 membership file (Section~7.5 of the main paper).}
\label{fig:matched}
\end{figure}

\begin{figure*}[!htb]
\centering
\includegraphics[width=\textwidth]{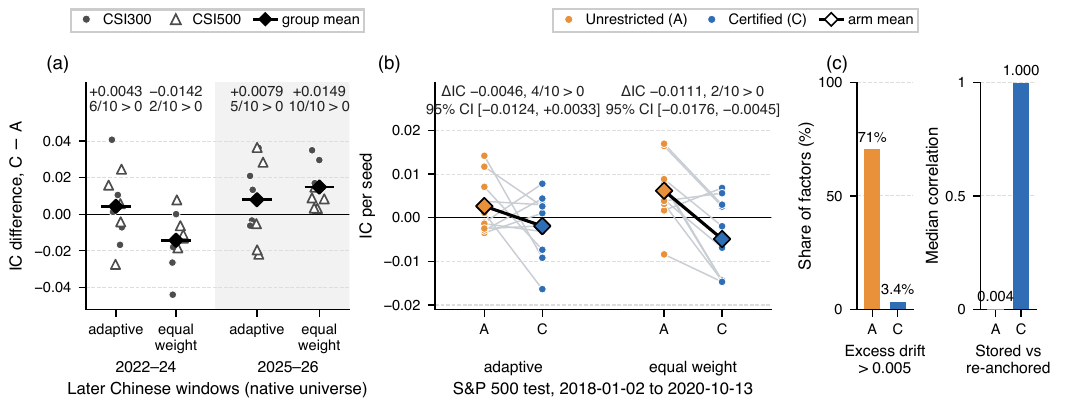}
\caption{Transfer beyond the development folds (Sections~7.1 and~7.4 of the main paper and Tables~\ref{tab:laterfull} and~\ref{tab:us}). (a)~Later Chinese windows, native universe: paired difference C $-$ A for the 10 market $\times$ seed pairs of each window under both combiners. (b)~S\&P~500 test window: per-seed IC of arms A and C under both combiners, with the mean difference and its 95\% CI. (c)~S\&P~500 anchor audit of the newly mined pools: share of factors with excess drift above 0.005 and pool-average median stored-versus-re-anchored correlation. The later-window differences are positive under the adaptive combiner but combiner-sensitive; on the S\&P~500 neither arm carries detectable signal, and the certified pools are far less sensitive to re-anchoring.}
\label{fig:transfer}
\end{figure*}

Fig.~\ref{fig:anchor} accompanies Table~5 and Section~7.1 of the main paper: it shows the distribution of factor-level excess IC drift and the pool-level drift under stock-specific re-anchoring against the global placebo for the 40 unrestricted and 40 certified pools. Fig.~\ref{fig:matched} accompanies Section~7.4 of the main paper and Table~\ref{tab:matchedfull}: it plots the four-fold mean adaptive and neutralised IC of each method and market and the per-cell differences of the certified search against the official AlphaSAGE and AlphaGen re-runs. Fig.~\ref{fig:transfer} accompanies Sections~7.1 and~7.4 of the main paper and Tables~\ref{tab:laterfull} and~\ref{tab:us}: it shows the later-window pairs, the per-seed S\&P~500 IC of both arms and the S\&P~500 anchor audit.


\section{Typed Action Mask: Search Efficiency at a Fixed Budget, Trajectories and Pool Composition}\label{supp:maskpools}

This section holds the full typed-mask results summarised in Section~7.3 of the main paper.

\begin{figure*}[!htb]
\centering
\includegraphics[width=\textwidth]{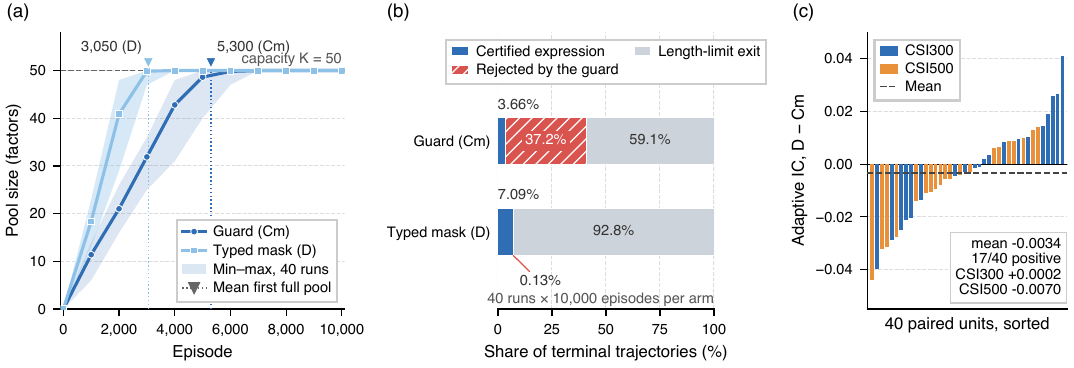}
\caption{Typed action mask (D) against the rejection guard (Cm) of Section~7.3 of the main paper and Table~\ref{tab:maskeff}, 40 runs per arm. (a)~Pool size against episode, mean and min--max range over the 40 runs. (b)~Composition of all terminal trajectories per arm. (c)~Paired difference $\mathrm{D} - \mathrm{Cm}$ in adaptive IC per unit, sorted and coloured by market. The mask fills the pool earlier and almost removes guard rejections, but it converts them into length-limit exits, and the quality difference at the fixed budget is small and negative.}
\label{fig:mask}
\end{figure*}

Arm \textbf{D} (mask plus guard) is compared with arm \textbf{Cm}, a re-run of arm C on the same machine as arm D (byte-identical to arm C; Section~\ref{supp:repro}), with checkpoints every 1,000 episodes. Each arm has 40 pools, interleaved by the same scheduler, and efficiency is measured over all terminal trajectories, including forced exits at the length limit (Table~\ref{tab:maskeff}; Fig.~\ref{fig:mask}).

\begin{table}[!htb]
\caption{Search efficiency of the typed action mask (means over 40 runs; cells are fold $\times$ market). Cm is the re-run of arm C with the rejection guard; D is the typed mask plus guard.}
\label{tab:maskeff}
\centering
\footnotesize
\setlength{\tabcolsep}{4pt}
\begin{tabular}{@{}>{\raggedright\arraybackslash}p{1.8in}rrl@{}}
\toprule
Metric & Cm & D & Cells D better \\
\midrule
legal terminal share & 3.66\% & 7.09\% & 8/8 \\
forced-exit rate at the length limit & 59.1\% & 92.8\% & reported \\
illegal terminal share (guard rejection) & 37.2\% & 0.13\% & -- \\
unique legal expressions per 1,000 episodes & 34.2 & 67.2 & 8/8 \\
unique legal expressions per minute & 2.83 & 4.55 & reported \\
episode at which the pool reaches 50 & 5,300 & 3,050 & D later in 0/8 \\
wall-clock per run (minutes) & 121.0 & 149.6 & -- \\
\bottomrule
\multicolumn{4}{@{}>{\raggedright\arraybackslash}p{\columnwidth}@{}}{\scriptsize Unique legal expressions per minute is the mean of the per-run ratios; the ratio of the mean counts is 4.49 for D and 2.83 for Cm.} \\
\end{tabular}
\end{table}

The mask removes 37.3\% of syntactically applicable operator tokens per state and blocks termination on an ill-typed root about 7,700 times per run. Legal yield and unique legal expressions double, the pool fills 40\% earlier, and unique legal output per minute rises by 60\% despite a 24\% wall-clock premium for per-step type inference; the frozen efficiency criterion is met in every cell. The cost is visible in the composition of terminal trajectories (Fig.~\ref{fig:mask}(b)): with termination blocked and no backtracking, the mask converts guard rejections into length-limit exits, so most trajectories remain unproductive. The residual 0.13\% of guard rejections are length-limit exits on complete 19-token trees whose root is not invariant (511 expressions across 40 runs); the amendment of the frozen validity clause that they required is recorded in Section~\ref{supp:deviations}.

\textbf{Quality at a fixed budget.} The frozen non-inferiority rule required adaptive $\Delta\mathrm{IC} \ge -0.005$ overall and in each market, and neutralised $\Delta\mathrm{IC} \ge -0.005$. It was not met. $\mathrm{D} - \mathrm{Cm}$ is $-0.0034$ over 40 pairs (17 positive, paired $t$-test $p = 0.26$): $+0.0002$ on CSI300 and $-0.0070$ on CSI500, with a neutralised difference of $-0.0017$ (Fig.~\ref{fig:mask}(c)); under equal weighting it is $+0.0019$ (19/40). The mask therefore improves legal-expression yield and pool-filling speed, but does not establish quality non-inferiority at the fixed budget. Relative to the unrestricted baseline, the masked pools keep the certified difference ($\mathrm{D} - \mathrm{A} = +0.0080$, 26/40, paired $t$-test $p = 0.033$). The composition of the masked pools is described below.

\textbf{Trajectories and pool composition.} Fig.~\ref{fig:mask} shows the pool size against the episode, the composition of all terminal trajectories and the paired quality difference $\mathrm{D} - \mathrm{Cm}$ per unit. The masked pools (arm D) contain longer expressions than the guard-only pools (arm Cm), 8.6 against 6.8 tokens, with more ratio constructs, and they have fewer highly collinear factor pairs (0.41\% against 1.26\% of pairs above 0.5). ICIR, worst 60-day IC and turnover differ by less than their noise. Under equal weighting $\mathrm{D} - \mathrm{Cm}$ is $+0.0055$ on CSI500.

\textbf{Proposition S1 (the mask does not change the accepted language).} Let $L_G$ be the set of complete expressions of at most 20 tokens that the generator can emit and the guard accepts, and $L_M$ the set accepted when the typed mask is also on. Then $L_M = L_G$. \emph{Proof.} $L_M \subseteq L_G$ because the guard runs on every completed expression in both settings. Conversely, let $E \in L_G$ with reverse-Polish token sequence $a_1, \dots, a_n$. Every node created along this sequence is a sub-expression of $E$. Typing is compositional: the type of a node is computed from the types of its children by the rule of its operator, and a node that cannot be typed makes every expression containing it untypable. Since $E$ is typable, every sub-expression is typable, so no operator action in the sequence is masked; field, constant and window tokens are not touched by the typed mask, and the syntactic mask is the same in both settings. The final termination action is allowed because the root of $E$ has type $\inv$. Hence the sequence can be generated under the mask, and $E$ passes the guard. $\square$ The proposition concerns the accepted language only: the mask changes which trajectories the policy samples, and with no backtracking a trajectory blocked from an ill-typed exit runs to the length limit, where the guard rejects it.

\section{Compute by Stage}\label{supp:compute}

Table~\ref{tab:compute} lists the compute used by each stage of the main paper. All certified and unrestricted runs used 10,000 episodes, pool capacity 50, a graph neural network encoder, entropy coefficient 0.01, learning rate $10^{-4}$ and hidden size 128. The costs are the rented-instance charges attributable to the completed runs of each stage; the project's total cloud spend, including failed launches, idle time, storage and instances rented for CPU stages, was about USD 200. Per-run wall-clock in the matched comparison was about 2.0--2.3~h for the certified search, 2.4--2.7~h for AlphaSAGE and 8.4--8.6~h for AlphaGen (300,000 PPO steps). Per-run wall-clock in the PPO-guard comparison was 0.7--1.0~h with the guard and 6.7--8.0~h without it. In total, the GPU stages (excluding the arm-P calibration pilot) comprise 372 runs and about 140 wall-clock hours on one to three RTX 4090 GPUs, and the stages with a recorded cloud cost sum to about USD 71.

\begin{table}[!htb]
\caption{Compute by stage of the main paper.}
\label{tab:compute}
\centering
\scriptsize
\setlength{\tabcolsep}{2pt}
\renewcommand{\arraystretch}{0.9}
\begin{tabular}{@{}>{\raggedright\arraybackslash}p{1.08in}>{\raggedright\arraybackslash}p{0.46in}>{\raggedright\arraybackslash}p{0.9in}>{\raggedright\arraybackslash}p{0.32in}>{\raggedright\arraybackslash}p{0.46in}@{}}
\toprule
Stage & Runs & Hardware & Time & Cost \\
\midrule
Size-matched analysis & 40 CPU units & one cloud instance & 20 min & $<$ USD 1 \\
Main comparison (B, C; A reused) & 80 GPU runs & two RTX 4090 instances & 16 h & about USD 12 \\
Later windows & 20 GPU runs & RTX 4090 & 4 h & about USD 6 \\
Combiner ablation & 120 CPU units & laptop & 28 min & -- \\
Typed mask (Cm, D) & 80 GPU runs & one RTX 4090 & 31.5 h & about USD 12 \\
Matched baselines & 80 GPU runs & up to three RTX 4090 instances & 2.1 days & -- \\
S\&P~500 & 20 GPU runs & one RTX 4090 & 7.3 h & -- \\
Same-environment re-evaluation (C, B, A) & 120 CPU units & laptop & 43 min & -- \\
Float64 diagnosis (244 + 244 factors) & 40 CPU units & laptop & 31 min & -- \\
Matched-pruning control (arm R) & 40 GPU runs & two RTX 4090 (one box) & about 11 h & about USD 14 \\
Single-sort control (arm S) & 40 GPU runs & two RTX 4090 (one box) & about 11 h & about USD 14 \\
Mechanism analysis & 40 CPU units & laptop & 30 min & -- \\
Relative-floor re-audit & 80 CPU units & laptop & 1.5 h & -- \\
Same-laptop baseline (original operator) & 80 CPU units & laptop & 70 min & -- \\
SH000905 sensitivity (discovery signal; validation re-evaluation) & 60 + 60 CPU units & laptop & 37 + 29 min & -- \\
SH000905 re-mining (A, C, 2 controls) & 42 GPU runs (+35 lost) & two RTX 4090 (one box) & about 7.7 h (+7.3 h) & about USD 16 \\
Property test, matched attribution, checker timing & -- & laptop & under 1 min each & -- \\
Block bootstrap & -- & laptop & 10 s & -- \\
Length analysis (CPU) & 7 arms, 14,000 pool factors & laptop & 1 min & -- \\
Calibration pilot (arm P, logging only) & 1 GPU run, 2,000 episodes & one RTX 4090 & 11 min & -- \\
Guard on PPO (C-PPO, A-PPO) & 12 GPU runs & two RTX 4090 (one box) & about 9 h & about USD 13 \\
Injected-defect re-mining (2 tickers, 2 controls) & 82 GPU runs & two RTX 4090 (one box) & about 12.5 h & about USD 12 \\
Clean-pool anchor audit & 40 CPU units & laptop & 45 min & -- \\
Guard on GP (C-GP, A-GP) & 20 CPU runs & laptop & about 7 h & -- \\
Guard on LLM (80 runs, 4 blocks) & 80 API runs & laptop + API & about 6 h & USD 19.41 (API) \\
Portfolio backtest & 120 CPU units & laptop & about 1 h & -- \\
Directional audit with real anchors & 80 CPU units & laptop & 3.2 h & -- \\
Float64 evaluation (4 conditions) & 320 CPU units & laptop & about 1.7 h & -- \\
GPU day: past-anchored baseline (AH, CH), seeds 5--9 on both markets, /4 injection, controls & 164 GPU runs & two RTX 4090 (one box) & about 27 h & about USD 20 \\
Evaluation of the GPU-day pools (clean and past-anchored providers) & 206 CPU units & laptop & about 9 h & -- \\
Past-anchored anchor audit; /4 fixed-pool evaluation & 42 + 40 CPU units & laptop & about 1.5 h & -- \\
Float64 re-evaluation of the clean 80-unit pools & 160 CPU units & laptop & 81 min & -- \\
\bottomrule
\end{tabular}
\end{table}

\section{Typing of the 101 Formulaic Alphas}\label{supp:alpha101}

This section gives the details of the Alpha101 check summarised in Section~7.1 of the main paper and Table~\ref{tab:alpha158}.

\textbf{Source and adapter.} The 101 formulas are those of Kakushadze's ``101 Formulaic Alphas'' (Wilmott, 2016, no.~84, pp.~72--81; arXiv:1601.00991). All 101 parse. Fields are mapped as in the main paper: open, high, low, close and VWAP have type $\sortM(-1)$ and volume has type $\sortM(+1)$. Returns are invariant, and $\op{adv}_d$ is read as average daily dollar volume, as the source defines it, and is therefore invariant. The primary mapping assumes that market capitalisation and industry labels are unchanged by re-anchoring; this is a statement about the representation change, not about the historical availability of these fields. The cross-sectional operators $\op{rank}$, $\op{scale}$ and $\op{indneutralize}$ require an invariant input; in particular, $\op{scale}(\op{close})$ does not cancel stock-specific scales. The adapter adds comparisons, boolean conditions, ternary branches, window flooring and literal-zero rules to the rules of Table~2 of the main paper, so it is not the unchanged 36-operator search grammar.

\textbf{Counts and sensitivity readings.} Under a literal reading of every constant, the checker certifies 17 of the 101 Alpha101 formulas and rejects the remaining 84: 77 are untypable and 7 are typable with a root type other than $\inv$. Because the checker is sound but not complete, a rejection is a failure of certification; the numerical check below shows deviations for 82 of the 84 rejected formulas on end-anchored data. These results concern end-anchored data, not the formulas' original usage. The readings below are sensitivity checks, not the headline certificate.
\begin{itemize}
\item Small additive constants read as numerical guards: 18 certified. The extra formula, Alpha\#101 $= (\op{close} - \op{open})/((\op{high} - \op{low}) + 0.001)$, is not exactly invariant (scaled deviation 0.34 in the synthetic check below), so this reading must not be described as an exact certificate.
\item $\op{adv}_d$ read as share volume instead of dollar volume: 19 certified under the literal reading (20 with the guard reading); the additional formulas are Alpha\#7 and Alpha\#21.
\item Market capitalisation typed as a price-level quantity instead of an invariant one: Alpha\#56 is no longer certified (17 under the guard reading).
\end{itemize}

\textbf{The 17 certified formulas.} Alpha\#2, \#6, \#14, \#19, \#26, \#30, \#33, \#35, \#38, \#43, \#48, \#53, \#54, \#56, \#72, \#85 and \#99. They are invariant through ratios of same-stock prices, time-series correlations and ranks, returns, signs, and the dollar-volume normaliser $\op{adv}_d$; Alpha\#56 relies on the market-capitalisation assumption above.

\textbf{Rejection reasons.} Table~\ref{tab:alpha101reasons} lists all rejection categories under the guard reading of constants. A formula can fail at several nodes, so the categories overlap; the second count is the number of formulas for which the category is the only reason.

\begin{table}[!htb]
\caption{Rejection categories of the Alpha101 check on end-anchored data (guard reading of constants; 83 formulas not certified). A formula can have several reasons; ``only reason'' counts the formulas with no other.}
\label{tab:alpha101reasons}
\centering
\scriptsize
\setlength{\tabcolsep}{3pt}
\begin{tabular}{@{}>{\raggedright\arraybackslash}p{2.35in}rr@{}}
\toprule
Category & Formulas & Only reason \\
\midrule
Cross-sectional rank of a price-level quantity & 50 & 34 \\
Rank of a volume-level quantity & 15 & 4 \\
Industry neutralisation of a non-invariant quantity & 14 & 9 \\
Price or volume level at the root & 7 & 7 \\
Comparison of a price- or volume-level quantity with a non-zero constant & 4 & 3 \\
Ternary branches of different type, or a non-invariant condition & 2 & 1 \\
Addition or subtraction across exponents or sorts & 2 & 2 \\
Rank of a price $\times$ volume quantity with non-zero net exponent & 2 & 2 \\
Comparison of quantities with different exponents & 1 & 1 \\
Price- or volume-level quantity plus or minus a non-zero constant & 1 & 1 \\
$\op{scale}$ of a non-invariant quantity & 1 & 1 \\
Power with a non-constant exponent or an additive base & 1 & 1 \\
\bottomrule
\end{tabular}
\end{table}

\begin{table*}[!t]
\caption{Matched comparison on the rolling folds (Section~7.4 of the main paper) with all columns: adaptive and neutralised IC, equal-weight IC, certified share and wall-clock per run (four-fold mean of cell means; adaptive combiner unless stated; all rows evaluated in one environment). EW: equal-weight combiner. Bold marks the best value among the formulaic methods; the supervised reference (italic) sits outside the decision rule.}
\label{tab:matchedfull}
\centering
\footnotesize
\setlength{\tabcolsep}{4pt}
\begin{tabular}{@{}lcccccccc@{}}
\toprule
& \multicolumn{2}{c}{Adaptive IC} & \multicolumn{2}{c}{Neutralised IC} & \multicolumn{2}{c}{EW IC} & Certified & Wall-clock \\
\cmidrule(lr){2-3} \cmidrule(lr){4-5} \cmidrule(lr){6-7}
Method & CSI300 & CSI500 & CSI300 & CSI500 & CSI300 & CSI500 & factors & per run \\
\midrule
Certified grammar (C) & \textbf{0.0345} & \textbf{0.0430} & \textbf{0.0194} & 0.0230 & \textbf{0.0379} & \textbf{0.0368} & 100\% & $\sim$2 h \\
AlphaGen (official) & 0.0337 & 0.0415 & 0.0147 & \textbf{0.0309} & 0.0047 & 0.0312 & 24\% & $\sim$8.5 h \\
AlphaSAGE (official) & 0.0239 & 0.0326 & 0.0118 & 0.0194 & 0.0212 & 0.0337 & 22\% & $\sim$2.6 h \\
Unrestricted (A) & 0.0262 & 0.0321 & 0.0166 & 0.0188 & 0.0229 & 0.0365 & 23\% & $\sim$2 h \\
Coarse-typed (B) & 0.0224 & 0.0314 & 0.0105 & 0.0177 & 0.0226 & 0.0349 & 25\% & $\sim$2 h \\
\midrule
\textit{Alpha158 + LightGBM (supervised reference)} & \textit{0.0661} & \textit{0.0693} & \textit{0.0446} & \textit{0.0441} & -- & -- & -- & minutes (CPU) \\
\bottomrule
\end{tabular}
\end{table*}

\textbf{Synthetic re-anchoring check.} All formulas were recomputed in float64 with a pandas operator adapter on a synthetic panel of 700 days and 16 stocks, discarding the first 300 days, under stock-specific scales $\{0.2, 1, 3.7, 25\}$. The test records finite-mask and non-finite-pattern mismatches and applies \texttt{allclose} with relative tolerance $10^{-8}$ and absolute tolerance $10^{-10}$. All 17 certificates pass, with no finite-mask or non-finite-pattern mismatch and a maximum error, scaled by $\max(1, |f|)$, of $3.2\times10^{-11}$. Alpha\#101 deviates by 0.34. Of the 83 formulas not certified under the guard reading, 81 deviate by more than $10^{-6}$ (the smallest of these deviations is 0.13), one (Alpha\#1) shows no deviation on this panel, and one (Alpha\#96) has no jointly finite values. A milder check with scales $\{0.5, 1, 2\}$ gives the same 17 passes and a deviation of 0.023 for Alpha\#101. These finite tests corroborate the implementation on these inputs; they are not exhaustive soundness proofs, and the adapter is independent of the frozen market evaluator. No Alpha101 predictive score enters this check. The rejection counts measure the coverage of a conservative checker on end-anchored data; they are a statement about dependence on the adjustment anchor, not about predictive value, historical availability or the prevalence of exploitable look-ahead in Alpha101.

\section{Statistical Detail of the Main Comparison and Detail Columns of the Control, Matched and Later-Window Tables}\label{supp:detailtables}

\textbf{Alpha158 + LightGBM segments (Section~7.4 of the main paper).} For each fold, the training segment runs from 2010-01-04 to 21 trading days before the early-stopping segment. The early-stopping segment is the last 12 months of the discovery period, ending at the purged cutoff, and is used only for early stopping. The test segment is the fold's validation window, scored with the same metrics. Features and hyper-parameters are the existing benchmark settings and were not tuned per fold.

\textbf{Stability and laptop details of the confirmatory stage.} These numbers refer to the 40-unit confirmatory stage mined with the original CSI500 membership file, not to the clean pools of Table~6 (for those, see Section~\ref{supp:moved-exp}). The worst 60-day IC is $-0.075$ for arm C against $-0.089$ for arm A, and turnover is 0.183 against 0.175; 2019 and 2021 are the two years in which the yearly IC of the unrestricted pools is negative while that of the certified pools stays positive, and arm C is slightly below arm A in 2016--2017. In the laptop re-evaluation of all 120 pools, C $-$ A $= +0.0096$ (95\% CI $[+0.0019, +0.0172]$; 26 of 40 units; 6 of 8 cells; neutralised $+0.0035$) and B $-$ A $= -0.0023$.

\textbf{Confirmatory stage (original-membership pools; Table~\ref{tab:confirmatory}, Section~\ref{supp:final-moved}).} The decision rule of this stage is an engineering rule ($\Delta\mathrm{IC} \ge -0.005$), and the tests below describe the observed superiority; none of them is the confirmatory criterion. Over the 40 units the paired $t = 3.32$ ($p = 0.002$) and the Wilcoxon $p = 0.001$; over the 8 cells the paired $t$ gives $p = 0.054$ (Wilcoxon 0.055); over the 4 folds it gives $p = 0.19$, all four fold means are positive, and the Wilcoxon $p$ of 0.125 is the smallest value attainable with four pairs. The strength of evidence therefore declines from the unit level to the cell level to the fold level. Because the 20-day labels overlap, the effective number of independent days per unit is about 43 (median; the lag-1 autocorrelation of the daily paired difference is 0.9). The moving-block bootstrap over trading days uses block length 20 (the label horizon) and 5,000 replicates, with blocks drawn within each fold $\times$ market cell and the same day indices for the five seeds of a cell. For C $-$ A it gives the 95\% interval $[+0.0021, +0.0213]$ with a one-sided $p = 0.007$ (the fraction of replicates at or below zero), unchanged in sign at block lengths 40 and 60; a fold-shared variant, with the blocks drawn once per fold and applied to both markets, gives $[-0.0002, +0.0241]$ (one-sided $p = 0.028$). The cluster-robust $t$ interval over the 8 cells, $[-0.0003, +0.0231]$, has $p = 0.027$, the one-sided version of the two-sided $p = 0.054$ above and not independent evidence; over the 4 folds it is $[-0.0101, +0.0329]$ ($p = 0.094$), because f2019\_20 carries most of the effect ($+0.031$) while f2015\_16 and f2017\_18 are near zero. C $-$ B is as robust ($[+0.0036, +0.0216]$ at block length 20); B $-$ A is null under every method; the neutralised C $-$ A ($+0.0060$) is not robust to the block bootstrap ($p = 0.068$).

\textbf{Clean-data main comparison (Table~6 and Fig.~6 of the main paper).} The 80 units are the 40 CSI300 pools of arms A and C (seeds 0--4 of the confirmatory stage, unaffected by the SH000905 line, and seeds 5--9 mined later on the clean provider under the frozen commands) and the 40 CSI500 pools re-mined without SH000905 (seeds 0--4 of Section~\ref{supp:inject}'s clean runs and seeds 5--9), all evaluated in the laptop environment with the frozen evaluator (Section~\ref{supp:hfq}). Adaptive C $-$ A: paired $t = 1.55$ over the 80 units ($p = 0.13$; Wilcoxon $p = 0.059$), paired $t$ over the 8 cell means $p = 0.43$ with interval $[-0.0080, +0.0168]$, over the 4 fold means $p = 0.60$ with interval $[-0.0199, +0.0287]$. Cell means: f2015\_16 $+0.0018$ / $-0.0124$, f2017\_18 $-0.0055$ / $-0.0052$, f2019\_20 $+0.0287$ / $+0.0248$, f2021 $-0.0022$ / $+0.0052$ (CSI300 / CSI500). Neutralised: $t = -0.20$, cell interval $[-0.0116, +0.0105]$. Equal weight: $t = 1.58$, cell interval $[-0.0203, +0.0325]$. Within-cell seed-to-seed SD of C $-$ A (root mean square over the 8 cells): 0.0224 adaptive, 0.0220 neutralised, 0.0181 equal weight; of the C and A levels, 0.0167 and 0.0134 (adaptive). By market, C $-$ A is $+0.0057$ $[-0.0026, +0.0140]$ on CSI300 (23/40) and $+0.0031$ $[-0.0050, +0.0112]$ on CSI500 (25/40); the seeds 0--4 and 5--9 blocks give $+0.0049$ $[-0.0036, +0.0135]$ and $+0.0039$ $[-0.0040, +0.0118]$ (adaptive); the engineering criteria of Section~6.5 of the main paper hold on the 80 units (adaptive $+0.0044$, neutralised $-0.0005$, both $\ge -0.005$). With the observed spreads, the minimum difference detectable at 80\% power with a two-sided test at the 5\% level is 0.0081 over the 80 units (standard error 0.0029), 0.0171 over the 8 cell means (0.0052) and 0.0317 over the 4 fold means (0.0076) (\texttt{clean\_stability/report\_mde.json}).

\begin{figure}[!htb]
\centering
\includegraphics[width=\columnwidth]{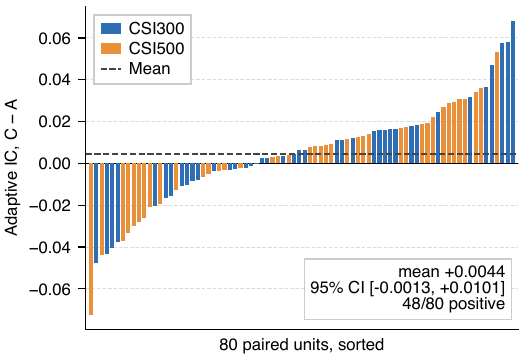}
\caption{Paired difference C $-$ A for the 80 clean-data units of Section~7.2 of the main paper, sorted and coloured by market, with the mean and its 95\% CI (adaptive combiner, laptop evaluation).}
\label{fig:pairs}
\end{figure}

\textbf{Control arms R and S (Table~4 and Section~7.3 of the main paper).} The random guard's probability $p_R = 0.8957$ equals the certified guard's mean rejection rate of unique candidates over the 40 arm-C runs (range 0.884--0.908); rejection is deterministic, by hashing the expression string with the run seed. Provider, folds, 10,000 episodes and pool capacity of 50 are identical to arm C; the realised rejection rate is 0.895 (0.882--0.907), and every pool is full (50/50). Arms R and S were evaluated in the laptop environment of Section~\ref{supp:repro}, reusing its A and C evaluations, so their levels differ from the cloud evaluation of Table~\ref{tab:confirmatory} by the platform sensitivity described there. The adaptive levels are 0.0291 (A), 0.0387 (C) and 0.0206 (R); the neutralised levels are 0.0177, 0.0212 and 0.0060; the worst 60-day IC is $-0.087$, $-0.077$ and $-0.098$; turnover is 0.175, 0.183 and 0.164; the equal-weight levels are 0.0297, 0.0373 and 0.0221; the Wilcoxon $p$ is $<0.001$ for C $-$ R and $0.084$ for R $-$ A. The certified share of the R pools is 24.1\%, against 22.6\% in arm A and 100\% in arm C. The margin over the random guard ($+0.018$) exceeds the margin over unrestricted search ($+0.010$). The pools of arm S are 100\% certified by the single-sort checker and, because the single-sort language is a subset of the two-sorted one, 100\% certified by the two-sorted checker; the share of numerically invariant factors under the strict elementwise test of Section~7.1 of the main paper is 35.4\% in arm S against 34.8\% in arm C. The adaptive level of arm S is 0.0345 (neutralised 0.0199; worst 60-day IC $-0.080$; turnover 0.195), and its equal-weight level is 0.0378. C $-$ S is $+0.0043$ (95\% CI $[-0.0012, +0.0098]$) and S $-$ A is $+0.0053$ ($[-0.0022, +0.0129]$). Under equal weighting, C $-$ R $= +0.0152$ ($[+0.0068, +0.0237]$; 28/40; $p = 0.001$), R $-$ A $= -0.0076$ ($[-0.0147, -0.0005]$; 12/40; $p = 0.038$), C $-$ S $= -0.0004$ ($[-0.0041, +0.0032]$) and S $-$ A $= +0.0081$ ($[-0.0028, +0.0189]$).

Table~\ref{tab:matchedfull} gives the matched comparison of Section~7.4 of the main paper with all columns: adaptive and neutralised IC, equal-weight IC, the certified share of the pool and the wall-clock per run. Table~\ref{tab:laterfull} gives the later-window comparison of Section~7.4 of the main paper under both combiners.

\textbf{Detail for Section~7.4 of the main paper.} Against the official AlphaSAGE, the largest margins are in the f2019\_20 fold ($+0.035$ and $+0.030$; seed-level Welch $t$-test $p = 0.014$ and $0.006$), and the three negative cells are small ($-0.005$, $-0.002$ and $-0.003$). The near-zero equal-weight IC of AlphaGen on CSI300 (0.0047; Table~\ref{tab:matchedfull}) comes from one strongly negative cell (f2019\_20, $-0.039$) and two cells near zero (f2017\_18 and f2021); sign-aligned equal weighting of all 50 factors is sensitive to factors whose sign flips after discovery (Section~\ref{supp:retro}).

\begin{table}[!htb]
\caption{Certified minus unrestricted search on the later Chinese windows of Section~7.4 of the main paper under both combiners (10 market $\times$ seed pairs per window and combiner). Native: index constituents; all stocks: all 1,730 stocks of the union universe. A and C are the mean IC of the unrestricted and the certified pools.}
\label{tab:laterfull}
\centering
\footnotesize
\setlength{\tabcolsep}{4pt}
\begin{tabular}{@{}lrrrrr@{}}
\toprule
Combiner & $\Delta\mathrm{IC}$ & \begin{tabular}[b]{@{}r@{}}pairs\\positive\end{tabular} & \begin{tabular}[b]{@{}r@{}}neutralised\\$\Delta\mathrm{IC}$\end{tabular} & A & C \\
\midrule
\multicolumn{6}{@{}l}{\textit{2025--26, native universe}} \\
adaptive & $+0.0079$ & 5/10 & $+0.0073$ & 0.0055 & 0.0134 \\
equal weight & $+0.0149$ & 10/10 & $+0.0193$ & 0.0047 & 0.0196 \\
\addlinespace
\multicolumn{6}{@{}l}{\textit{2022--24, native universe}} \\
adaptive & $+0.0043$ & 6/10 & $+0.0030$ & 0.0403 & 0.0446 \\
equal weight & $-0.0142$ & 2/10 & $-0.0002$ & 0.0351 & 0.0208 \\
\addlinespace
\multicolumn{6}{@{}l}{\textit{2025--26, all 1,730 stocks}} \\
adaptive & $+0.0015$ & 6/10 & $+0.0030$ & 0.0313 & 0.0328 \\
\addlinespace
\multicolumn{6}{@{}l}{\textit{2022--24, all 1,730 stocks}} \\
adaptive & $+0.0062$ & 6/10 & $+0.0081$ & 0.0456 & 0.0518 \\
\bottomrule
\end{tabular}
\end{table}

\section{Expression Length of Pool Factors}\label{supp:lengths}

This section gives the length analysis behind the learnable pruning control of Section~7.3 of the main paper. Length is the number of tokens of an expression. Table~\ref{tab:lengths}(a) gives the mean length of pool factors by arm and the shares of short and long factors. Certified pools are shorter than unrestricted pools: the pool-mean difference C $-$ A is $-3.3$ tokens over 40 paired units ($p < 10^{-30}$). The masked pools (arm D) average 8.6 tokens and keep a difference over arm A ($\mathrm{D} - \mathrm{A} = +0.0080$; Section~7.3 of the main paper). Table~\ref{tab:lengths}(b) splits the 2,000 factors of arm A into six length bins. The mean validation $|\mathrm{IC}|$ and the certified share fall broadly with length, but the sign-flip share does not rise steadily with length. Over the 2,000 factors, length does not predict a stable IC sign (Spearman 0.015, $p = 0.50$), whereas certification does after controlling for length (logistic coefficient $-0.30$, $p = 0.014$; length-stratified sign-flip share 33.6\% for certified against 40.9\% for non-certified factors).

\begin{table}[!htb]
\caption{Expression length of pool factors. (a) Mean length by arm and the shares of factors with at most 4 and at least 10 tokens. (b) The 2,000 factors of arm A by length bin: number of factors, mean validation $|\mathrm{IC}|$, share with a flipped out-of-sample IC sign, and certified share.}
\label{tab:lengths}
\centering
\footnotesize
\setlength{\tabcolsep}{4pt}
\begin{tabular}{@{}lccc@{}}
\multicolumn{4}{@{}l}{(a) Pool factors by arm} \\
\toprule
Arm & Mean tokens & $\le 4$ tokens & $\ge 10$ tokens \\
\midrule
A (unrestricted) & 10.1 & 9.7\% & 51.6\% \\
R (random guard) & 9.2 & 14.0\% & 43.2\% \\
D (typed mask) & 8.6 & 10.6\% & 35.0\% \\
S (single sort) & 6.9 & 21.8\% & 15.9\% \\
C (certified) & 6.8 & 22.6\% & 16.6\% \\
\bottomrule
\end{tabular}

\vspace{6pt}
\begin{tabular}{@{}lcccc@{}}
\multicolumn{5}{@{}l}{(b) Arm A by length} \\
\toprule
Tokens & $n$ & Mean $|\mathrm{IC}|$ & Sign-flip & Certified \\
 & & (validation) & share & share \\
\midrule
$(0, 4]$ & 194 & 0.025 & 0.335 & 0.515 \\
$(4, 6]$ & 269 & 0.022 & 0.364 & 0.401 \\
$(6, 8]$ & 339 & 0.020 & 0.395 & 0.375 \\
$(8, 10]$ & 346 & 0.019 & 0.428 & 0.205 \\
$(10, 13]$ & 366 & 0.017 & 0.372 & 0.082 \\
$(13, 25]$ & 486 & 0.018 & 0.389 & 0.031 \\
\bottomrule
\end{tabular}
\end{table}

\textbf{Length- and operator-matched attribution.} Within the 2,000 arm-A factors, we matched certified (451) and non-certified (1,549) factors exactly on a length bin (1--3, 4--5, 6--7, 8--9, 10--12, 13--16, 17--20 tokens) and on the set of operator families present (arithmetic, unary, cross-sectional, rolling level, rolling dispersion or difference, rolling normalised, pairwise). The effect on the certified factors is the stratum-weighted difference in sign-flip rate; intervals come from a bootstrap over the 40 pools (2,000 resamples). The matched difference is $-6.9$ percentage points (95\% interval $[-14.4, +2.4]$; 315 of 451 certified factors matched), against $-6.5$ unmatched. Matching on length alone gives $-7.7$ ($[-14.8, +0.2]$; all 451 matched), and on exact token length and operator families $-4.7$ ($[-13.2, +4.6]$; 302 matched). The difference in validation $|\mathrm{IC}|$ is within $\pm 0.003$ in every variant. The point estimate stays negative after matching on length and operator composition, but under pool clustering the interval includes zero, so the matched analysis does not resolve the explanation.

\section{Mechanism Analysis Figure}\label{supp:mechanism}

\begin{figure}[!htb]
\centering
\includegraphics[width=\columnwidth]{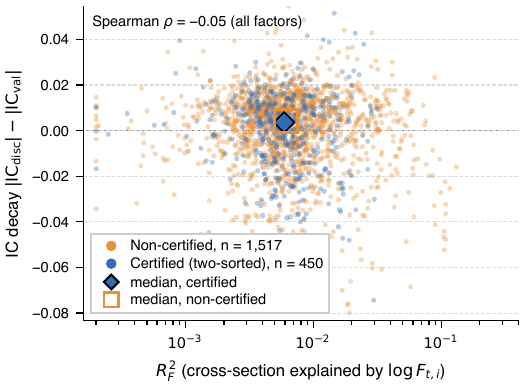}
\caption{Mechanism analysis of Section~7.3 of the main paper: the 1,967 arm-A factors with a defined $R^2$ (450 certified in blue, 1,517 non-certified in orange): cross-sectional $R^2$ on the adjustment factor $\log F_{t,i}$ (log axis) against IC decay, with the group medians marked; 12 points with $R^2$ below the axis range are drawn at its left edge and 7 points lie outside the decay range shown. Both groups sit near $R^2 = 0.01$ and zero decay and the medians coincide, so a simple linear screen on the adjustment factor would not have flagged the non-certified factors.}
\label{fig:mechanismsupp}
\end{figure}

Fig.~\ref{fig:mechanismsupp} shows the factor-level data behind the mechanism analysis of Section~7.3 of the main paper. The $R^2$ statistics are over the 1,967 factors with a defined $R^2$ (450 certified, 1,517 non-certified; 33 degenerate factors have too few finite values), and the IC and sign-flip statistics are over all 2,000 (451 certified, 1,549 non-certified). The mean IC decay is 0.0000 for certified against $+0.0003$ for non-certified factors (Mann--Whitney $p = 0.29$), and the correlation between $R^2$ and decay is slightly negative (Spearman $-0.047$). In the length-pruning pilot of Section~7.3, the nearest non-trivial rule, at most one operator and three tokens, rejects 79.8\% of the 496 unique candidates, and the two-token rule that reaches the target rate rejects 95.4\% and admits 23 candidates. When certification also requires numerical invariance, the factor-level sign-flip rate is 27.0\% against 39.7\% (Fisher $p < 0.001$, exploratory).

\textbf{Detail for Section~7.3 of the main paper.} $R^2$ is about 0.01 for certified factors (mean 0.0072, median 0.0059) and for non-certified ones (mean 0.0105, median 0.0061). Before the cross-sectional regression on $\log F_{t,i}$, each factor's raw value is winsorised at the 1st and 99th percentile, and the discovery window of the IC decay is purged of its last 20 days.

\section{Additional Method and Evaluation Details}\label{supp:moved}

This section holds method and evaluation details that the main paper summarises. Each subsection names the section of the main paper it supports: details of the preliminaries and the typing rules (Section~\ref{supp:moved-grammar}); the overview of the method and details of the generator and the typed action mask (Section~\ref{supp:moved-search}); the rolling folds (Section~\ref{supp:moved-protocol}, Table~\ref{tab:folds}); the full text behind the experimental subsections, with the hand-crafted feature-set table and the S\&P~500 outcome table (Section~\ref{supp:moved-exp}, Tables~\ref{tab:alpha158} and~\ref{tab:us}); and the reproducibility and platform-sensitivity record (Section~\ref{supp:repro}). The matched-comparison and later-window results of Section~7.4 of the main paper are in Tables~\ref{tab:matchedfull} and~\ref{tab:laterfull}, which give them with all columns and rows.

\subsection{Preliminaries and Typing Details (Sections~3 and~4 of the Main Paper)}\label{supp:moved-grammar}

\textbf{Volume adjusted for splits only (Section~3.2 of the main paper).} A provider that adjusts volume for splits but not for cash dividends has two independent per-stock generators: splits (prices $-1$, volume $+1$) and cash dividends (prices $-1$, volume $0$). An expression is certified under that contract if the checker, instantiated once for each scaling, types it to $\inv$ under both assignments; this certifies invariance under the joint action of the two generators, the $k = 2$ case of the next paragraph (script \texttt{check\_volume\_convention.py} in the repository). This is a check of the certificate's coverage; predictive performance was not re-measured. Of the 2,000 factors of the 40 confirmatory certified pools, 1,926 (96.3\%) stay certified, and of the 3,000 of the 60 certified pools mined on clean data (CSI500 seeds 0--9 and CSI300 seeds 5--9; the CSI300 seeds 0--4 of the main comparison are the confirmatory pools of the first count), 2,920 (97.3\%); the 36 and 44 unique expressions lost multiply a price by volume, such as $\op{Mul}(\op{high}, \op{volume})$ or $\op{TsCov}(\op{open}, \op{volume}, 20)$, which are invariant only if volume moves with the dividend factor. Among the unrestricted pools the certified share falls from 22.6\% to 21.7\% (confirmatory) and from 23.9\% to 22.7\% (clean).

\textbf{Several per-entity factors (Section~4.2 of the main paper).} If the fields transform by $k$ independent per-entity factors, each positive and constant in time (for example the adjustment anchor and a per-stock lot size for volume), the sorts become $\mathbb{Q}^k$-graded and every rule and proof step applies componentwise. The case $k = 2$ is the split-only volume check of the previous paragraph, run componentwise with the unchanged checker; a factor that varies in time, such as an exchange rate, falls outside the extension.

\textbf{Operators of the search grammar (Section~3.1).} The 36 operators listed in Table~2 of the main paper are six unary ($\op{Abs}$ to $\op{Rank}$), seven binary ($\op{Add}$ to $\op{Less}$), twenty-one single-input rolling ($\op{Ref}$ to $\op{TsPctChange}$) and two pair rolling ($\op{TsCov}$, $\op{TsCorr}$).

\textbf{Binary operators with one invariant operand (Section~4.1).} When exactly one operand of a binary operator is invariant, it is read as $\sortM(0)$ or $\sortA(0)$, whichever matches the other operand's sort, and the rule for that sort then applies; a zero coefficient is never turned into a non-zero one. A constant added to a price is therefore rejected (exponents $0$ and $s$ differ), a constant added to a log-price is a shifted log-price of type $\sortA(a)$, and a constant compared with a log-price by $\op{Greater}$ or $\op{Less}$ is rejected.

\textbf{Excluded invariant families (Section~4.3).} \textit{Non-affine cancellations.} The expression $\op{TsSum}(\log \op{close}, 10) - 10 \log \op{close}$ is certified, and so is $(\log \op{close} - \log \op{open})^2$. Its expanded form $(\log \op{close})^2 - 2 \log \op{close} \cdot \log \op{open} + (\log \op{open})^2$ is rejected, because squaring an $\sortA$ value and multiplying two $\sortA$ values are untypable, even though the complete polynomial cancels the shift. \textit{Degenerate expressions.} In $\op{Pow}(\op{Div}(\op{Greater}(0.01,\allowbreak \op{vwap}),\allowbreak \op{close}),\allowbreak -10)$, the inner $\op{Greater}(0.01, \op{vwap})$ equals VWAP on every observed stock-day, so the complete expression reduces to $(\op{vwap}/\op{close})^{-10}$ and is invariant on the panel, although the saturated sub-expression on its own is not. The checker rejects it because $\op{Greater}$ requires equal exponents and reads the invariant constant $0.01$ as $\sortM(0)$ beside VWAP's $\sortM(-1)$. Further examples of the 81 rejected but invariant factor rows are in Section~\ref{supp:rules}.

\textbf{The coarse control (Section~4.4).} The Alpha$^2$ rule set reconstructed from the examples in its paper is as follows. Price fields are currency, volume is unit, and constants are neither. $\op{Add}$, $\op{Sub}$, $\op{Greater}$ and $\op{Less}$ require equal categories; $\op{Mul}$ and $\op{Div}$ across categories yield a mixed category; unary and rolling operators pass the category through; $\op{Rank}$ and $\op{Log}$ are unconstrained; there is no exponent bookkeeping and no root constraint.

\subsection{Method Overview, Generator and Typed Action Mask (Section~5 of the Main Paper)}\label{supp:moved-search}

Figs.~2 and~4 of the main paper give the overview of the method and of the search.

\textbf{Checks of the typed action mask (Section~5.3).} Completeness over the 2,451 certified expressions in the arm-C and arm-A pools (1,404 unique) and the absence of dead-end states were also checked before the experiment and monitored during it. The mask never changes which expressions are admissible, so arms C and D differ only in how the policy reaches them.

\textbf{Generator (Section~5.1).} A state is a prefix of at most 20 tokens, parsed into a partial operator tree and encoded by a two-layer relational graph convolutional network. The forward policy gives a distribution over the next token or termination, and grammar masks remove syntactically invalid tokens. The policy is trained for 10,000 episodes with entropy-regularised trajectory balance, one trajectory per episode and an Adam step every 64 episodes.

\subsection{Rolling Folds (Section~6.2 of the Main Paper)}\label{supp:moved-protocol}

Table~\ref{tab:folds} gives the discovery and validation windows of the four rolling development folds of Section~6.2 of the main paper. All method decisions were taken on these folds.

\begin{table}[!htb]
\caption{Rolling folds.}
\label{tab:folds}
\centering
\footnotesize
\begin{tabular}{@{}lll@{}}
\toprule
Fold & Discovery & Validation \\
\midrule
f2015\_16 & 2010-01-04 -- 2014-12-03 & 2015-01-05 -- 2016-12-02 \\
f2017\_18 & 2010-01-04 -- 2016-12-02 & 2017-01-03 -- 2018-11-30 \\
f2019\_20 & 2010-01-04 -- 2018-11-30 & 2019-01-02 -- 2020-12-03 \\
f2021     & 2010-01-04 -- 2020-12-03 & 2021-01-04 -- 2021-12-03 \\
\bottomrule
\end{tabular}
\end{table}

\subsection{Experiments (Section~7 of the Main Paper)}\label{supp:moved-exp}

\textbf{Anchor audit on the Chinese provider (Section~7.1).} Two stock-specific re-anchorings simulate the transformations induced by alternative future corporate-action histories while holding the observations fixed, so a factor's drift between them is the dependence that Proposition~1 of the main paper describes. Every drift statistic takes the larger of the two re-anchorings per factor or pool. A global scaling applies the same $c$ to every stock and is itself in the anchor group, so for a non-invariant expression it changes the values even in exact arithmetic (for example $\op{close} + \op{volume}$, or $\op{Greater}(\op{close}, 30)$). The residual drift of the certified pools is not only rounding. In float64, the implemented $\op{TsCorr}(\op{close}, \op{open}, 10)$ on a window whose sums of squared deviations are of order $10^{-6}$ returns a correlation of $-0.996$, and returns $-6 \times 10^{-8}$ once both prices are divided by 10, because the scaled sums fall below the operator's fixed floor of $10^{-6}$ and the normalisation is replaced by 1. Of the 244 certified factors whose stored and re-anchored values correlate below 0.9 under at least one stock-specific anchor, 96.7\% contain at least one of $\op{TsKurt}$, $\op{TsSkew}$, $\op{TsIr}$, $\op{TsCorr}$, $\op{TsRank}$, $\op{Sign}$, $\op{TsPctChange}$ or $\op{TsDiv}$. These operators divide by a rolling dispersion, apply a fixed floor, or branch on signs and ties. A float64 diagnosis of the 244 sensitive certified factors separates the two mechanisms (Section~\ref{supp:numerics}): for 217 of the 244 (89\%) the correlation with the stored values is restored to at least 0.99 in float64 under the same re-anchorings, and the property-based test executed its 1,500 random certified expressions in float64. A strict elementwise test (identical non-finite patterns, and \texttt{allclose} with relative tolerance $10^{-4}$ and absolute tolerance $10^{-6}$ on the finite values) passes 34.8\% of certified factors under stock-specific anchors and 34.0\% under the placebo; at this tolerance the test measures implementation-level numerical and threshold sensitivity.

\textbf{Re-anchoring stress test on the S\&P~500 bundle (Section~7.1).} Membership coverage is incomplete (price data for 387 index members in 2010 and 502 in 2020; delisted names missing), so the universe is tilted toward survivors. The pre-specified anchor audit reported in Section~7.1 of the main paper is the audit of the newly mined pools.

\textbf{Checking hand-crafted feature sets (Section~7.1).} The full text of the Alpha158 and Alpha101 check follows, with Table~\ref{tab:alpha158}; Section~\ref{supp:alpha101} gives the Alpha101 details. We ran Qlib's 158 Alpha158 features through the checker, with Qlib's operators mapped to the typing rules (Table~\ref{tab:alpha158}). Read literally, with every constant at face value (as for Alpha101 below), 99 features are certified; reading the $10^{-12}$ division guards as numerical guards gives 153. The 5 rejected features are the CORR family, in which the constant 1 breaks the volume scaling (relative deviation under re-anchoring about $10^{-6}$). Of the 153, 108 are invariant only because their designers divided them by close, open or a sum, and 45 are intrinsically invariant ($R^2$, ranks, index positions, counts). Thirty features (SUMP, SUMN, SUMD and their volume versions) need one extra rule, $\max(x, 0)$ for $x$ of type $\sortM(s)$, which is sound because $c_i > 0$.

\begin{table}[!htb]
\caption{Two-sorted check of two hand-crafted feature sets, with operators mapped to the typing rules. Alpha158: the 108 and 45 partition the 153 certified features, and the 30 that need the extra rule are among them. Alpha101 rejection reasons: Table~\ref{tab:alpha101reasons}.}
\label{tab:alpha158}
\centering
\footnotesize
\begin{tabular}{@{}>{\raggedright\arraybackslash}p{2.9in}r@{}}
\toprule
Features & Count \\
\midrule
\multicolumn{2}{@{}l}{\textit{Alpha158 (Qlib)}} \\
All & 158 \\
Certified, division guards read as numerical guards & 153 \\
Certified, every constant literal read literally & 99 \\
Rejected: CORR family, $\op{Corr}(\op{close}, \op{Log}(\op{volume} + 1), d)$ & 5 \\
Invariant only through normalisation by close, open or a sum & 108 \\
Intrinsically invariant & 45 \\
Needing the extra rule $\max(x, 0)$ for $x$ of type $\sortM(s)$ & 30 \\
\addlinespace
\multicolumn{2}{@{}l}{\textit{Alpha101 (101 Formulaic Alphas)}} \\
All (all parse) & 101 \\
Certified, every constant read literally & 17 \\
Certified, small constants read as guards (not exact) & 18 \\
Rejected by the checker, constants literal & 84 \\
\bottomrule
\end{tabular}
\end{table}

All 101 formulas of Kakushadze's ``101 Formulaic Alphas'' (Alpha101) parse in the checker, through an adapter that adds comparisons, ternary branches and the Alpha101 operators ($\op{rank}$, $\op{scale}$, $\op{indneutralize}$, $\op{decay\_linear}$, $\op{ts\_argmax}$, $\op{signedpower}$; $\op{adv}$ as dollar volume, as in the source); it is not the 36-operator search grammar. Under the literal reading of constants, the checker certifies 17 of the 101 formulas and rejects 84; numerical re-anchoring on end-anchored data shows deviations for 82 of the 84. Reading small additive constants as guards gives 18, but the extra formula, Alpha\#101 $= (\op{close} - \op{open})/((\op{high} - \op{low}) + 0.001)$, is not exactly invariant (scaled deviation 0.34 in a synthetic check): 0.001 is not small beside a price difference, unlike Alpha158's $10^{-12}$ guards. The literal count of 17 of 101 is therefore the headline, against 99 of 158 for Alpha158 under the same reading. The most frequent rejection reason is a cross-sectional rank of a price-level quantity (50 formulas, 34 with no other reason; Table~\ref{tab:alpha101reasons}), and all 17 certificates pass a float64 synthetic re-anchoring check (maximum scaled error $3.2\times10^{-11}$). The checker thus tells the two conventions apart: Alpha158 divides almost every feature by a price or a sum, while Alpha101 ranks price levels across stocks. A rejection is a failure of certification, not by itself a proof of non-invariance; the statements concern end-anchored data, not the formulas' original usage or predictive value.

\textbf{Stability and factor-level strength (Section~7.2).} On the clean pools of Table~6 of the main paper (80 units per arm), the certified pools are also steadier across time (Fig.~6(b) of the main paper): the standard deviation of yearly adaptive IC is 0.033 for C against 0.037 for A, and C exceeds A in five of seven years, by most in 2019 ($+0.045$). Under equal weighting the difference is $+0.0061$ $[-0.0016, +0.0137]$ (44/80) and comes from CSI300 ($+0.0170$; CSI500 $-0.0048$). Single certified factors are weaker than unrestricted ones (validation mean $|\mathrm{IC}|$ 0.0138 against 0.0189 over 4{,}000 factors per arm; 34\% against 48\% above 0.015), so the pool-level difference comes from combination, consistent with the lower redundancy of certified pools in the retrospective comparison (Section~\ref{supp:retro}). The confirmatory-stage pools give the same picture (yearly SD 0.034 against 0.038; single-factor $|\mathrm{IC}|$ 0.0144 against 0.0193). Script: \texttt{analysis/clean\_stability/} in the repository.

\textbf{Attribution criterion (Section~7.3).} The pre-specified attribution criterion $(\mathrm{C} - \mathrm{B}) \ge (\mathrm{B} - \mathrm{A}) - 0.002$ holds automatically whenever $\mathrm{B} - \mathrm{A} < 0$ and $\mathrm{C} \ge \mathrm{A}$, so the attribution rests on arm B being within 0.002 of arm A.

\textbf{Length-pruning calibration (Section~7.3).} We pre-specified a learnable pruning control of equal strength (arm P), a deterministic cap on operator and token counts calibrated on the candidate stream of an unguarded pilot run, expecting a length cap to give a fair control. On the 496 unique completed candidates of the pilot, however, the only structural rule that reaches the certified guard's rejection rate of 89.6\% admits expressions of at most two tokens, so under the pre-declared stopping rule no admissible equal-strength rule was found in this family, and arm P was not run (Section~\ref{supp:deviations}). The certified guard rejects at the same rate yet fills every 50-factor pool, with expressions of 6.8 tokens on average against 10.1 in unrestricted pools, so it removes a different set of expressions from any length cap. Length may contribute to the difference; masked pools (arm D, 8.6 tokens; Section~7.3 of the main paper) still keep a positive difference over A. Within the 2,000 unrestricted factors, length does not predict a stable IC sign, whereas certification does after controlling for length (logistic coefficient $-0.30$, $p = 0.014$). Matching on length and operator composition leaves the difference in sign-flip rate unchanged, but its pool-clustered interval includes zero (Section~\ref{supp:lengths}, Table~\ref{tab:lengths}).

\textbf{A second generator (Section~7.4).} We attached the same checker, unchanged, to the official AlphaGen PPO generator of Section~7.4 of the main paper: a rejected expression receives reward 0 and never enters the pool (C-PPO), and the unguarded arm runs the same code with the guard off (A-PPO). The descriptive protocol covers CSI300, folds f2019\_20 and f2021 and seeds 0--2, six pairs (Section~\ref{supp:deviations}). With the guard off, all six pools are byte-identical to the official AlphaGen pools of the matched comparison, so the disabled hook is inert. With the guard on, 95.6--98.5\% of unique candidates are rejected, every pool is full and certified, and a run takes 0.7--1.0~h instead of 6.7--8.0~h. C-PPO $-$ A-PPO is $-0.0004$ under the adaptive combiner ($[-0.014, +0.013]$; 2 of 6 units positive) and $+0.026$ under equal weighting ($[-0.009, +0.060]$; 5 of 6), so the guard carries over to a reinforcement-learning generator with no detectable loss of IC over these six pairs, at about one eighth of the run time.

\textbf{Matched re-discovery of official miners and a supervised reference (Section~7.4).} The full text of the matched comparison follows; its table is Table~\ref{tab:matchedfull}. To separate ``better than our re-implementation'' from ``better than the published method'', we re-ran the official AlphaSAGE and AlphaGen code on the same provider, rolling folds, label maturity, purge, seeds and evaluator: 80 runs, each method in its published configuration and budget, with only the date, provider and output interfaces changed (the diffs are archived). All rows of Table~\ref{tab:matchedfull} were evaluated on the laptop CPU in one software environment, in which the C, A and B pools were re-evaluated (cloud-instance evaluations of the original-membership confirmatory stage: Section~\ref{supp:cloudrows}). As a reference outside the decision rule, we trained Alpha158 + LightGBM on the same folds with the existing benchmark settings (segments: Section~\ref{supp:detailtables}).

Against the official AlphaSAGE the certified search has the higher four-fold mean IC in both markets ($+0.0106$ and $+0.0104$) and the higher neutralised IC ($+0.0076$ and $+0.0036$), but only 5 of 8 cells are non-negative, one short of the frozen criterion, so the frozen verdict is mixed (per-cell margins: Section~\ref{supp:detailtables}). In the cloud environment the count is 6 of 8 (Section~\ref{supp:repro}), so the frozen cell count sits inside the evaluator's platform sensitivity, which is why the mean margins with their sign in both markets are the primary statistic. Our arm A reproduces the official AlphaSAGE closely, with a mean cell difference of $+0.0010$, so the contrast C $-$ A of the original-membership confirmatory stage (Table~\ref{tab:summary} of the main paper) and this contrast are two measurements of the same effect. Against the official AlphaGen the outcome is ``competitive'': the four-fold means are higher by $+0.0008$ and $+0.0015$, 3 of 8 cells are non-negative, and the neutralised IC on CSI500 is lower by 0.0079, outside the margin. Under equal weighting the certified pools are the strongest in both markets (Table~\ref{tab:matchedfull}; AlphaGen's near-zero CSI300 value: Section~\ref{supp:detailtables}). AlphaGen used about four times the wall-clock of the certified search, so we do not claim equal compute (compute of every stage: Table~\ref{tab:compute}).

The supervised reference is the strongest model in the table: 61--92\% higher than the certified pools in raw IC and 1.9--2.3 times as high after exposure neutralisation, at about twice the turnover. It uses 158 handcrafted features and a nonlinear gradient-boosted predictor, provides no compact formula and was not built with an anchor-invariance certificate, although its features pass the checker up to their numerical guards (Section~7.1 of the main paper).

\textbf{Mechanism (Section~7.3).} On the 2,000 factors of the 40 arm-A pools we tested one explanation of the difference: that non-invariant factors act as stock identifiers through the adjustment factor. For each factor we computed the mean over validation days of the cross-sectional $R^2$ of its winsorised raw value on $\log F_{t,i}$, the per-stock adjustment factor, and the IC decay $|\mathrm{IC}_{\mathrm{discovery}}| - |\mathrm{IC}_{\mathrm{validation}}|$ on a purged discovery window (Fig.~\ref{fig:mechanismsupp}). $R^2$ is about 0.01 for certified and non-certified factors alike (Section~\ref{supp:mechanism}); no factor exceeds 0.3, and only 0.6\% of non-certified factors exceed 0.1. This weak linear association with $\log F$ does not show that the dependence is small or rule out non-linear dependence, but a simple linear screen would not have flagged it. The IC decay does not differ between the groups (Mann--Whitney $p = 0.29$), and the data do not support the identifier explanation.

\textbf{Later Chinese windows and the S\&P~500 predictive outcome (Section~7.4).} With the grammar and all hyper-parameters frozen, arm C was re-mined through 2021-12-03 and 2024-12-03 and compared with unrestricted pools mined on the same cutoffs, as retrospective out-of-sample evidence (Section~6.2 of the main paper; Table~\ref{tab:laterfull}; both combiners: Fig.~\ref{fig:transfer}(a)). The frozen confirmation rule (Section~\ref{supp:deviations}) applies to the first row of Table~\ref{tab:laterfull} and was not met: the mean, the neutralised difference and both markets pass, but only 5 of 10 pairs are positive, where 6 are required. The 2022--24 window is mixed, with the equal-weight combiner favouring the unrestricted pools: under equal weighting the native-universe difference is $+0.0149$ (10/10 pairs) on 2025--26 and $-0.0142$ (2/10) on 2022--24 (Table~\ref{tab:laterfull}). Both arms are near zero in absolute terms on 2025--26, and both carried more signal on the all-stock universe than on the index constituents in both windows (0.031--0.052 against 0.005--0.045). The adaptive differences stayed positive but were combiner-sensitive. Section~\ref{supp:retro} compares existing pools of six external miners on 2022--24 descriptively.

In the S\&P~500 experiment of Section~7.1 of the main paper neither arm carries detectable signal in this window: yearly adaptive IC lies between $-0.005$ and $+0.007$ for both arms, and the adaptive combiner selects on average 1.2 eligible factors per day, against 10--20 on the Chinese folds, so the test has little power. The frozen confirmation rule is not met (Table~\ref{tab:us}; Fig.~\ref{fig:transfer}(b)): $\Delta\mathrm{IC} = -0.0046$ with 4 of 10 seeds positive, and under equal weighting the certified pools have a lower IC ($-0.0111$, 2 of 10), with a 95\% CI of the difference that excludes zero.

\begin{table}[!htb]
\caption{S\&P~500, test window 2018-01-02 to 2020-10-13, 10 seeds per arm. A and C are the mean IC of the unrestricted and the certified pools; $\Delta\mathrm{IC} = \mathrm{C} - \mathrm{A}$, with the 95\% CI of the mean difference over seeds.}
\label{tab:us}
\centering
\scriptsize
\setlength{\tabcolsep}{1.4pt}
\begin{tabular}{@{}lrrrcrr@{}}
\toprule
Combiner & A & C & $\Delta\mathrm{IC}$ & 95\% CI & \begin{tabular}[b]{@{}r@{}}seeds\\positive\end{tabular} & \begin{tabular}[b]{@{}r@{}}neutr.\\$\Delta\mathrm{IC}$\end{tabular} \\
\midrule
adaptive & 0.0027 & $-0.0019$ & $-0.0046$ & $[-0.0124, +0.0033]$ & 4/10 & $-0.0026$ \\
equal weight & 0.0062 & $-0.0049$ & $-0.0111$ & $[-0.0176, -0.0045]$ & 2/10 & $-0.0095$ \\
\bottomrule
\end{tabular}
\end{table}

\textbf{Typed action mask (Section~7.3).} The summary of the main paper rests on the following paragraph; the full results are in Section~\ref{supp:maskpools}. Arm \textbf{D} (mask plus guard) is compared with arm \textbf{Cm}, a byte-identical re-run of arm C on arm D's machine (Section~\ref{supp:repro}), 40 pools per arm, under its frozen rule (rule and full results: Sections~\ref{supp:deviations} and~\ref{supp:maskpools}). The mask raises the legal terminal share from 3.66\% to 7.09\% and fills the pool at 3,050 instead of 5,300 episodes, at a 24\% wall-clock premium, and meets the efficiency criterion in every cell. Quality non-inferiority at the fixed budget is not established ($\mathrm{D} - \mathrm{Cm} = -0.0034$, and $-0.0070$ on CSI500, against the margin of $-0.005$), while the masked pools keep the certified pools' positive difference over A ($\mathrm{D} - \mathrm{A} = +0.0080$, 26/40). Forcing a legal completion, or allowing backtracking before the length limit, is a separate experiment we have not run.

\subsection{Reproducibility and Platform Sensitivity}\label{supp:repro}

This subsection gives the reproducibility and platform-sensitivity record summarised in Section~6.4 of the main paper.

\textbf{Byte-identical re-runs.} Arm Cm (Section~7.3 of the main paper) repeated arm C's 40 commands two days later on another machine of the same configuration (fixed seeds and hash seed, deterministic cuBLAS workspace); all 40 pools are byte-identical. The re-mining's two control runs (Section~7.5 of the main paper) reproduced their pools on a machine with another CPU and GPU driver.

\textbf{Platform sensitivity of the evaluator.} Per unit, adaptive IC differs between our two environments with standard deviations of about 0.005--0.010, and equal-weight IC is an order of magnitude more stable, because a floating-point difference of order $10^{-6}$ can flip one threshold selection of the adaptive combiner and change a unit's IC by much more (Section~\ref{supp:cloudrows}). In Table~\ref{tab:confirmatory}, the arm-A rows come from the size-matched stage's evaluation and the B and C rows from the confirmatory stage's, on cloud instances of the same host class and software environment; the byte-identical re-runs show that this pairing is exact.

\textbf{Same-environment re-evaluation.} After the matched comparison had been analysed, we re-evaluated the 120 C, B and A pools, first evaluated on cloud instances, on the laptop with the evaluator and environment of the official baselines of Table~\ref{tab:matchedfull}; the frozen verdict against AlphaSAGE moves from ahead to mixed (Section~7.4 of the main paper), and every other frozen criterion is unchanged. Arms R and S, the float64 diagnosis and the relative-floor re-audit use this laptop environment (per-arm differences and cloud-instance rows: Section~\ref{supp:cloudrows}).

\section{Injected-Defect Control for the Discovery Channel}\label{supp:inject}

\textbf{Design (frozen before the draws).} The defect divides the stored VWAP of one ordinary CSI500 member by 400 on every day from 2007-01-15 to 2012-06-25, the span over which SH000905 is listed. The factor reproduces the VWAP/close ratio of SH000905 on discovery days (1st, 50th and 99th percentiles 0.0023, 0.0026 and 0.0029), not its index-level price and volume, so the control tests an isolated field-level anomaly rather than the full mislisting. Candidates were the members of the clean CSI500 file that (i) belong to the index on all 599 trading days from 2010-01-04 to 2012-06-25, the discovery days on which SH000905 is listed, (ii) leave the index on 2012-06-25, as SH000905 does, so that modified values never sit next to unmodified ones inside the universe, (iii) do not re-enter within 120 trading days, and (iv) have a finite stored VWAP on at least 95\% of the 599 days. Twenty-eight tickers qualify. The first ticker was drawn with a fixed generator seed (SH600966; 1,302 stored values changed, injected VWAP/close 0.00245--0.00257 between the 1st and 99th percentiles); a second, registered before its draw because one draw is one replicate, came from the remaining 27 with another fixed seed (SZ000687; 1,299 values, 0.00244--0.00259). Each defect touches 599 discovery days and 0.05--0.10\% of the discovery stock-days of a fold, the footprint of SH000905.

\textbf{Runs.} Arms A and C were re-mined on each corrupted provider with the commands of the clean re-mining (Section~\ref{supp:provenance}), 4 folds $\times$ 5 seeds per arm and ticker, 80 runs, on one 2$\times$RTX~4090 machine (AMD EPYC 7642); evaluation used the clean data, so the mining data are the only difference from the clean pools. Two control runs repeated clean commands on this machine and reproduced the clean pools, mined on another machine, byte for byte. For the fixed-pool channel the 40 clean pools were re-evaluated on each corrupted provider. Discovery reward and search statistics were added to the reported outcomes after the fixed-pool evaluation had started and before any re-mined result existed.

\begin{table*}[!t]
\caption{Per-unit change in CSI500 C $-$ A, with defect minus without (20 fold $\times$ seed units per defect; pooled: both injected tickers, 40 units). Re-mined: pools mined on the corrupted data, evaluated on clean data. Fixed: clean pools re-evaluated on the corrupted data. Ratio: SD re-mined / SD fixed, with a 95\% bootstrap interval over the 20 or 40 fold $\times$ seed units, which treats them as exchangeable; the fold-level sensitivity in the text (leave-one-fold-out ratios and a bootstrap over the four folds) accounts for the shared folds and is the more conservative reading.}
\label{tab:inject}
\centering
\footnotesize
\setlength{\tabcolsep}{5pt}
\begin{tabular}{@{}lrrrrr@{}}
\toprule
Defect & SD re-mined & SD fixed & Ratio & Mean re-mined [95\% CI] & $>0$ \\
\midrule
\multicolumn{6}{@{}l}{\textit{Adaptive combiner}} \\
SH000905 (real) & 0.0258 & 0.0038 & 6.8 [4.2, 10.5] & $+0.0104$ $[-0.0017, +0.0225]$ & 14/20 \\
SH600966 (injected) & 0.0240 & 0.0016 & 15.1 [9.1, 26.9] & $+0.0031$ $[-0.0081, +0.0144]$ & 10/20 \\
SZ000687 (injected) & 0.0192 & 0.0014 & 13.3 [8.4, 21.4] & $+0.0022$ $[-0.0068, +0.0112]$ & 13/20 \\
Injected, pooled & 0.0215 & 0.0015 & 14.3 [10.2, 20.8] & $+0.0027$ $[-0.0042, +0.0095]$ & 23/40 \\
\addlinespace
\multicolumn{6}{@{}l}{\textit{Adaptive combiner, neutralised IC}} \\
SH000905 (real) & 0.0258 & 0.0042 & 6.2 & $+0.0117$ $[-0.0004, +0.0238]$ & 13/20 \\
Injected, pooled & 0.0200 & 0.0013 & 15.4 & $+0.0064$ $[-0.0001, +0.0128]$ & 27/40 \\
\addlinespace
\multicolumn{6}{@{}l}{\textit{Equal-weight combiner}} \\
SH000905 (real) & 0.0172 & 0.0014 & -- & $+0.0052$ $[-0.0029, +0.0132]$ & 14/20 \\
Injected, pooled & 0.0129 & $\le$0.0001 & -- & $+0.0049$ $[+0.0008, +0.0090]$ & 27/40 \\
\addlinespace
\multicolumn{6}{@{}l}{\textit{Lower-magnitude defect, SH600966 VWAP / 4 (20 units)}} \\
adaptive & 0.0141 & 0.0014 & 10.0 [6.0, 21.1] & $+0.0060$ $[-0.0006, +0.0125]$ & 14/20 \\
adaptive, neutralised & 0.0149 & 0.0012 & 12.6 [8.5, 22.6] & $+0.0104$ $[+0.0034, +0.0174]$ & 14/20 \\
equal weight & 0.0094 & $\le$0.0001 & -- & $+0.0051$ $[+0.0006, +0.0095]$ & 16/20 \\
\addlinespace
\multicolumn{6}{@{}l}{\textit{Small numerical perturbation, SH600966 VWAP $\times (1 + 2^{-20})$ (20 units)}} \\
adaptive & 0.0077 & $2 \times 10^{-5}$ & -- & $+0.0006$ $[-0.0030, +0.0042]$ & 8/20 \\
adaptive, neutralised & 0.0055 & $\le 10^{-5}$ & -- & $+0.0008$ $[-0.0018, +0.0034]$ & 9/20 \\
equal weight & 0.0082 & $3 \times 10^{-5}$ & -- & $-0.0003$ $[-0.0041, +0.0035]$ & 6/20 \\
\bottomrule
\end{tabular}
\end{table*}

\textbf{Results.} Table~\ref{tab:inject} gives both channels. Under the adaptive combiner the re-mined change has a standard deviation of 0.019--0.024 for the injected defects, close to the 0.026 of SH000905, while the fixed-pool change is smaller than for SH000905 (0.0014--0.0016 against 0.0038), so the ratio between the channels is larger. The units are not independent: the five seeds of a fold share its data, and both injections share the folds, so the intervals of Table~\ref{tab:inject} describe these two tickers and four folds and do not estimate a population of random defects. The ratio is nonetheless large in every fold: per fold, the re-mined and fixed standard deviations are 0.014--0.037 and 0.0027--0.0049 for SH000905 and 0.008--0.035 and 0.0007--0.0025 for the injections; leaving out one fold gives ratios of 5.5--7.5 (SH000905), 12.7--16.6 (SH600966) and 12.5--14.8 (SZ000687), and a bootstrap over folds gives 95\% intervals of $[5.3, 9.8]$, $[9.1, 21.6]$ and $[12.4, 16.8]$. The mean adaptive change is not distinguishable from zero for either ticker; the arm levels move by at most 0.003 (C $+0.0029$ and $+0.0011$, A $-0.0002$ and $-0.0011$). Under equal weighting the pooled mean change is $+0.0049$ $[+0.0008, +0.0090]$ (27 of 40), in the direction of SH000905; with three combiner and IC variants and no correction for multiple comparisons we report it as exploratory. The equal-weight combiner has almost no history-dependent selection, so its fixed-pool change is below $10^{-4}$ and the ratio is not informative. The re-mined pools changed in every unit: the certified pools keep 34.7 and 35.2 of their 50 expressions (Jaccard overlap 0.56 and 0.58), the unrestricted pools 18.9 and 19.9 (0.24 and 0.25), and the certified pools keep more in 18 of 20 units for each ticker (Wilcoxon $p = 1.4 \times 10^{-4}$ and $2.9 \times 10^{-4}$). The defect hardly moves the search itself: the mean discovery $|\mathrm{IC}|$ of pool factors changes by less than 0.0005 per arm (C $-$ A pooled $+0.0006$ $[+0.0002, +0.0010]$), certified completions per run are 365 and 366 against 362 on clean data (unrestricted: 508 and 502 against 506), and the share of pool factors that read VWAP stays at 24--25\% (C) and 32\% (A). A small defect therefore changes which expressions the search keeps, and with them the measured difference, without changing how the search behaves on average. Limitations: one kind of defect, two tickers, one market and four folds.

\textbf{Lower-magnitude defect.} Under a protocol amendment written before the runs (block 4 of Section~\ref{supp:hfq}), the first injected ticker's stored VWAP was divided by 4 instead of 400 over the same span (1,302 stock-days), and arms A and C were re-mined on this provider with the same commands (20 CSI500 units per arm), evaluated on clean data; the fixed-pool channel re-evaluates the 40 clean pools on the /4 provider. The re-mining spread of C $-$ A is 0.59 $[0.46, 0.76]$ of the /400 spread under the adaptive combiner (bootstrap over the 20 units) and 10.0 $[6.0, 21.1]$ times the fixed-pool spread (0.0014; 12.6 times on the neutralised IC); the pools changed less than under /400 (mean Jaccard 0.28 for A and 0.57 for C against 0.24 and 0.56; the certified pools kept more expressions in 17 of 20 units, Wilcoxon $p = 3 \times 10^{-4}$), the discovery reward moved by $+0.0005$ (C $-$ A, $[-0.0000, +0.0011]$; /400: $+0.0008$), and the share of VWAP factors was unchanged (0.24 for C, 0.32 for A). The mean change in C $-$ A is positive under all three combiners, with the unrestricted pools losing adaptive IC ($-0.0030$, SD 0.013) and the certified pools gaining ($+0.0029$, SD 0.008); the 20 units share one ticker and four folds, so this is a descriptive observation.

\textbf{Small numerical perturbation.} Under an amendment written before its runs (\texttt{clean\_controls/PROTOCOL\_clean\_controls.md}, Amendment~1), the first injected ticker's stored VWAP was multiplied by $1 + 2^{-20}$ over the same span (1,302 values, each changed by a relative $9.5 \times 10^{-7}$; nothing else changed), and arms A and C were re-mined with the clean commands on the 20 CSI500 units of a second rented machine (Amendment~2: the block was moved there before any of its results existed; one run that had started on the first machine for about 40~s was discarded), with one control run that reproduced the archived clean AX pool byte for byte. Re-mined pools differ from the clean ones in 39 of 40 runs; they keep 48.0 (C) and 40.6 (A) of 50 expressions (minimum 26 and 19), against 26.3/14.6 for the ticker removal and 34.6--35.1/18.9--21.2 for the injections, and the certified pools keep more in 12 of 20 units (4 ties). Table~\ref{tab:inject} gives the change in C $-$ A. The pre-stated reading applies in its first form: the field-scale defects cause changes beyond those of this perturbation, which itself is not zero.

\section{Anchor Audit of the Re-Mined CSI500 Pools}\label{supp:cleanaudit}

The frozen anchor audit of Section~7.1 of the main paper (same script, original core, CPU evaluator, two stock-specific re-anchorings and their matched global placebos) was run, under a protocol written before the runs, on the 40 CSI500 pools re-mined without SH000905 (arms CX and AX of Section~7.5 of the main paper), with the provider they were mined and evaluated on. Table~\ref{tab:cleanaudit} compares them with the 40 original CSI500 pools audited on the same laptop with the same core.

\begin{table*}[!t]
\caption{Anchor audit of the CSI500 pools: re-mined without SH000905 (CX, AX) and original (C, A), 20 pools per arm; last two rows: the pools mined on the past-anchored provider (Section~\ref{supp:hfq}), both markets, audited on that provider. Pool $|\Delta\mathrm{IC}|$: adaptive combiner, larger of the two re-anchorings; stock: stock-specific; placebo: matched global scaling. Excess: factor $|\Delta\mathrm{IC}|$ under the stock-specific anchors minus that under the placebo.}
\label{tab:cleanaudit}
\centering
\footnotesize
\setlength{\tabcolsep}{5pt}
\begin{tabular}{@{}lrrrrr@{}}
\toprule
Arm & Pool $|\Delta\mathrm{IC}|$ stock & placebo & ratio & Corr $<$ 0.9 stock / placebo & Excess $>$ 0.005 \\
\midrule
CX, certified, re-mined & 0.00094 & 0.00098 & 0.96 & 12.9\% / 13.4\% & 0.8\% \\
C, certified, original & 0.00092 & 0.00081 & 1.13 & 12.6\% / 13.3\% & 0.7\% \\
AX, unrestricted, re-mined & 0.0139 & 0.0029 & 4.8 & 54.8\% / 19.0\% & 34.2\% \\
A, unrestricted, original & 0.0148 & 0.0047 & 3.2 & 43.2\% / 18.3\% & 33.0\% \\
\addlinespace
AH, unrestricted, mined on past-anchored data (40) & 0.0151 & 0.0048 & 3.2 & 44.0\% / 19.7\% & 24.3\% \\
CH, certified, mined on past-anchored data (2) & 0.0004 & 0.0009 & 0.5 & 15.0\% / 15.0\% & 0.0\% \\
\bottomrule
\end{tabular}
\end{table*}

The last two rows are the AH and CH pools of Section~\ref{supp:hfq}, audited with the same script on the past-anchored provider they were mined on (equal weight: 0.0089 / 0.0047 for AH, 0.0013 / 0.0018 for CH). Under equal weighting the pool drift is 0.0019 (stock) and 0.0013 (placebo) for CX and 0.0130 and 0.0032 for AX. The pattern of the main audit holds for the re-mined pools: the certified pools drift as much under stock-specific re-anchoring as under the placebo, and the unrestricted pools drift several times more, with about a third of their factors above an excess of 0.005.

\section{Certified Guard on an LLM Generator}\label{supp:llm}

\textbf{Design.} The checker, adapter and guard of the PPO experiment (Section~7.4 of the main paper) were attached unchanged to the LLM alpha generator released with AlphaGen (\texttt{scripts/llm\_only.py}): the model proposes expressions in AlphaGen's syntax, the pool is re-fitted, and the next prompt reports the pool and its scores. A parsed expression that is not certified is dropped before it reaches the pool and receives no IC, through the same path that the official loop uses for unparsable lines, which the model is never told about. The prompts are identical in both arms and never mention invariance. Settings follow the official loop (20 iterations, mean-squared-error pool), with a pool of 50 instead of 20, the Anthropic Messages API instead of the OpenAI client, CPU evaluation with exact memoisation, a cap of five retries when a response yields nothing usable, and two crashes of the official code handled (an empty pool, and nested windows beyond the 100-day backtrack, now skipped as the official pool loader already does). The models are Claude Haiku~4.5 and Claude Opus~5.5 at API defaults, with no fine-tuning, no system prompt about the checker and no access to it; the same models assisted the authors with text and scripts (Acknowledgments of the main paper), which has no bearing on the generator runs, whose transcripts are the record. The API is not deterministic; every request and response is archived and is the reproducibility record. Units are fold $\times$ replicate pairs on the folds f2019\_20 and f2021: CSI300 replicates 0--4 for each model (Haiku: six pairs in a first protocol and four in a separately frozen extension) and CSI500 replicates 0--4 for Haiku and, in a third frozen protocol, for Opus, on the membership without SH000905. A spend ledger enforced hard caps in code; the whole experiment cost USD 19.41 (Haiku 2.91, Opus 16.50 including a pilot).

\begin{table*}[!t]
\caption{Guarded minus unguarded LLM generator (C-LLM $-$ A-LLM), mean IC difference with 95\% paired-$t$ CI and pairs positive; GFlowNet and PPO rows on the same units for context. Neut.: exposure-neutralised IC, adaptive combiner.}
\label{tab:llm}
\centering
\footnotesize
\setlength{\tabcolsep}{5pt}
\begin{tabular}{@{}lrrrr@{}}
\toprule
Block (pairs) & Adaptive & $>0$ & Neut. & Equal weight \\
\midrule
Haiku, CSI300, first protocol (6) & $+0.0105$ $[-0.0049, +0.0260]$ & 4/6 & $+0.0058$ & $+0.0040$ $[-0.0038, +0.0119]$ \\
Haiku, CSI300, pooled (10) & $+0.0079$ $[-0.0013, +0.0172]$ & 6/10 & $+0.0007$ & $-0.0008$ $[-0.0092, +0.0075]$ \\
Haiku, CSI500 (10) & $+0.0220$ $[+0.0147, +0.0293]$ & 9/10 & $+0.0138$ & $+0.0028$ $[-0.0020, +0.0075]$ \\
Opus, CSI300 (10) & $+0.0040$ $[-0.0065, +0.0146]$ & 5/10 & $+0.0028$ & $+0.0019$ $[-0.0030, +0.0067]$ \\
Opus, CSI500 (10) & $+0.0073$ $[-0.0077, +0.0224]$ & 6/10 & $+0.0046$ & $+0.0047$ $[-0.0011, +0.0105]$ \\
\addlinespace
GFlowNet C $-$ A, CSI300 (10) & $+0.0097$ $[-0.0157, +0.0351]$ & 6/10 & $+0.0024$ & $+0.0231$ $[-0.0067, +0.0529]$ \\
GFlowNet CX $-$ AX, CSI500 (10) & $+0.0163$ $[+0.0052, +0.0274]$ & 9/10 & $+0.0082$ & $+0.0019$ $[-0.0187, +0.0225]$ \\
PPO, CSI300 (6) & $-0.0004$ $[-0.0138, +0.0131]$ & 2/6 & $+0.0009$ & $+0.0257$ $[-0.0085, +0.0598]$ \\
\bottomrule
\end{tabular}
\end{table*}

\textbf{Results.} Table~\ref{tab:llm} gives the comparisons. On CSI300 neither model shows a difference whose interval excludes zero. On CSI500 the guarded Haiku pools are ahead under the adaptive combiner ($+0.0220$, 9 of 10 pairs; neutralised $+0.0138$ $[+0.0054, +0.0222]$), in the direction and of the size of the GFlowNet difference on the same units, but not under equal weighting. With four blocks and two combiners and no correction for multiple comparisons, this is one result among eight. On the same CSI500 units Opus gives $+0.0073$ $[-0.0077, +0.0224]$ (6 of 10): its guarded arm scores the same as Haiku's (0.0273 against 0.0275) while its unguarded arm is far stronger (0.0200 against 0.0055; difference $+0.0145$ $[+0.0075, +0.0215]$, 10 of 10), so the guard effect for Opus minus that for Haiku is $-0.0147$ $[-0.0287, -0.0006]$ on CSI500 and $-0.0039$ $[-0.017, +0.009]$ on CSI300. Arm levels (adaptive IC): Haiku CSI300 C 0.0351, A 0.0271; Opus CSI300 C 0.0353, A 0.0313; Haiku CSI500 C 0.0275, A 0.0055; Opus CSI500 C 0.0273, A 0.0200. All 2,000 guarded factors are certified by the frozen evaluator. Without the guard the share of certified pool factors is 52--64\% for Haiku and 82--85\% for Opus, against 22--29\% for the unguarded GFlowNet, so the stronger model rarely proposes anchor-dependent expressions and the guard changes its pools less. Per run, 97--100\% (Haiku) and 88--99\% (Opus) of proposed lines are valid expressions, and among unique proposals 58--82\% pass certification in the guarded Haiku runs and 87--91\% in the guarded Opus runs; a run makes 21--25 API calls. Limitations: two folds, 6--10 pairs per block, two models, and API non-determinism; the two arms of a CSI500 pair ran one after the other. The models were trained on public text that may include formulaic alphas and market history; both arms use the same model and prompts, so such exposure affects them alike and the paired difference remains interpretable, but the absolute levels of the LLM pools should not be read as out-of-sample discovery.

\section{Portfolio Backtest}\label{supp:backtest}

\textbf{Design (frozen before any backtest result).} Each pool's daily combined score is reproduced by importing the frozen evaluator unchanged; for all 120 backtested pools and both combiners the IC recomputed from the extracted scores matches the evaluator to within $6 \times 10^{-12}$. The strategy follows the TopK-Dropout convention of this literature: hold the 50 highest-scored stocks, replace at most five a day, equal cash per new position, score at day $t$'s close, trade at $t{+}1$'s close and earn the return over $t{+}2$. A stock cannot be bought when suspended or up at least 9.5\% on the trade day, nor sold when suspended or down at least 9.5\%. Costs are 0.05\% on buys and 0.15\% on sells (x1), with zero (x0) and double (x2) as a pre-specified sensitivity. The benchmark is the equal-weight average of the index members at $t$ that trade on $t{+}1$, a descriptive benchmark adjusted for realised tradability rather than a portfolio fixed at $t$. The order of events is: scores are computed at the close of $t$ from data through $t$; at the close of $t{+}1$ orders execute for stocks that are neither suspended nor at the price limit on $t{+}1$, which assumes that the limit status is known when the order is placed, a simplification of real execution; the position then earns the return over $t{+}2$. No information from after $t$ enters a score or the choice of stocks; returns are the stored daily percentage changes, which agree with the adjusted closes on 99.999\% of stock-days. Annualisation is 252 times the daily mean. An auxiliary cost-free long-short portfolio (top minus bottom decile, re-set daily) isolates signal quality. All scores use only information matured by the score date; no cost, threshold or rule was tuned.

\begin{table*}[!t]
\caption{Portfolio backtest on the validation windows (adaptive combiner; auxiliary analysis). TopK-Dropout: top 50, 5 swaps per day, orders at the next day's close, excess return over the equal-weight investable universe, annualised as a fraction; costs x1 = 0.05\% on buys and 0.15\% on sells, x0 = none, x2 = double. LS: top minus bottom decile, re-set daily, cost-free. CSI500 rows use the membership without SH000905; CX $-$ AX: CSI500 pools re-mined without it. $\Delta$IC: frozen evaluator. Units and cells $>0$ count positive paired differences.}
\label{tab:backtest}
\centering
\footnotesize
\setlength{\tabcolsep}{3pt}
\begin{tabular}{@{}lcccccccccc@{}}
\toprule
Pair & $\Delta$IC & $\Delta$Excess & 95\% CI & Units & Cells & $\Delta$Excess & $\Delta$Excess & $\Delta$IR & $\Delta$LS & $\Delta$LS \\
 & & x1 & & $>0$ & $>0$ & x0 & x2 & x1 & return & Sharpe \\
\midrule
C $-$ A, CSI300 & $+0.0083$ & $+0.0203$ & $[-0.0202, +0.0607]$ & 11/20 & 3/4 & $+0.0201$ & $+0.0204$ & $+0.242$ & $+0.0405$ & $+0.113$ \\
C $-$ A, CSI500 & $+0.0120$ & $+0.0070$ & $[-0.0504, +0.0645]$ & 12/20 & 2/4 & $+0.0060$ & $+0.0080$ & $+0.305$ & $+0.0635$ & $+0.294$ \\
C $-$ A, both & $+0.0101$ & $+0.0136$ & $[-0.0199, +0.0472]$ & 23/40 & 5/8 & $+0.0130$ & $+0.0142$ & $+0.274$ & $+0.0520$ & $+0.203$ \\
\addlinespace
CX $-$ AX, CSI500 & $+0.0016$ & $-0.0040$ & $[-0.0521, +0.0441]$ & 11/20 & 1/4 & $-0.0041$ & $-0.0040$ & $+0.057$ & $+0.0445$ & $+0.221$ \\
CSI300 + CX $-$ AX & $+0.0049$ & $+0.0081$ & $[-0.0221, +0.0384]$ & 22/40 & 4/8 & $+0.0080$ & $+0.0082$ & $+0.150$ & $+0.0425$ & $+0.167$ \\
\midrule
Arm level & IC & Excess & Turnover & & Excess & Excess & IR & LS & LS & LS \\
 & & x1 & & & x0 & x2 & x1 & return & Sharpe & turnover \\
\midrule
A, CSI300 & 0.0262 & $-0.0317$ & 0.089 & & $+0.0133$ & $-0.0767$ & $-0.372$ & $+0.1267$ & $+0.736$ & 0.231 \\
C, CSI300 & 0.0345 & $-0.0115$ & 0.089 & & $+0.0334$ & $-0.0563$ & $-0.130$ & $+0.1672$ & $+0.849$ & 0.232 \\
A, CSI500 & 0.0329 & $-0.0572$ & 0.091 & & $-0.0113$ & $-0.1030$ & $-0.666$ & $+0.2554$ & $+1.366$ & 0.234 \\
C, CSI500 & 0.0448 & $-0.0502$ & 0.089 & & $-0.0053$ & $-0.0950$ & $-0.361$ & $+0.3189$ & $+1.659$ & 0.264 \\
AX, CSI500 & 0.0394 & $-0.0437$ & 0.090 & & $+0.0019$ & $-0.0892$ & $-0.460$ & $+0.2729$ & $+1.462$ & 0.225 \\
CX, CSI500 & 0.0410 & $-0.0477$ & 0.090 & & $-0.0021$ & $-0.0933$ & $-0.403$ & $+0.3174$ & $+1.684$ & 0.293 \\
\bottomrule
\end{tabular}
\end{table*}

\textbf{Results.} Table~\ref{tab:backtest} gives the paired differences and the arm levels. Every interval of the TopK net excess difference includes zero; the differences barely change with the cost level because both arms turn over about 9\% of the book a day. Net of costs every arm trails the equal-weight benchmark, as costs of about 0.045 a year exceed the gross excess. The cost-free long-short difference is $+0.052$ $[+0.0007, +0.103]$ for C $-$ A (25 of 40) and $+0.045$ $[-0.033, +0.122]$ for CX $-$ AX, with Sharpe differences of $+0.20$ $[-0.09, +0.50]$ and $+0.22$. Across units the IC difference correlates with the TopK excess difference at $+0.48$ (CSI300 $+0.68$, CSI500 $+0.35$, re-mined CSI500 $+0.09$) and with the long-short difference at $+0.75$. A post-hoc diagnostic, which changed no rule, shows that in CSI500 7.5--13.7\% of the top-50 candidates are suspended on the trade day (CSI300 about 3\%), so buys fail and 5--11\% of the book sits in cash; the certified scores point at suspended stocks more often (12.2\% against 7.5\% for the re-mined pools), and the difference in cash share correlates with the excess difference at $-0.74$ (C $-$ A) and $-0.57$ (CX $-$ AX). The CSI500 TopK comparisons therefore partly measure how often each score selects untradable stocks. Under the equal-weight combiner the TopK differences are positive (C $-$ A $+0.048$ $[+0.021, +0.076]$, CX $-$ AX $+0.035$ $[+0.005, +0.066]$) although its IC differences are small, partly through the same cash effect. Limitations: validation windows of one to two years that were already used for IC, eight fold-by-market cells, simplified execution, and no correction for multiple comparisons.

\section{Certified Guard on a Genetic-Programming Generator}\label{supp:gp}

\textbf{Design (frozen before the formal runs).} The checker, adapter and guard of the PPO experiment were attached to the genetic-programming baseline released with AlphaGen (\texttt{gp.py}, gplearn's symbolic regressor with the IC on the discovery window as fitness). All official settings are kept: population 1,000, 40 generations, tournament 600, the official crossover and mutation probabilities, function set, constants and target, the token cap of 20 parentheses (fitness $-1$), and the pool rule (cached programs by discovery IC, admitted if their absolute mutual IC with admitted factors is at most 0.7, mean-squared-error weights). In the guarded arm a program that is not certified receives the official floor fitness $-1$, is cached, and gets no IC. Deviations: a pool of 50 instead of 20; the official internal test split is dropped; the pool is built once after generation 40 instead of every fourth generation for logging; in both arms only cached programs with fitness above $-1$ may enter the pool. Seven tests pass, among them behaviour identity (unguarded, the script reproduces the official program's cache, 274 entries in order and value, and its pool) and byte-identical repeats. Units are CSI300 $\times$ \{f2019\_20, f2021\} $\times$ seeds 0--4, 10 pairs, on the discovery and validation windows of the PPO experiment. All 20 runs ran on one laptop CPU (two threads each, at most two at a time); the protocol records four changes of computing venue, all made before any formal run started.

\begin{table*}[!t]
\caption{Guarded minus unguarded GP (C-GP $-$ A-GP) on 10 CSI300 pairs, with the GFlowNet C $-$ A on the same units; mean IC difference, 95\% paired-$t$ CI, pairs positive.}
\label{tab:gp}
\centering
\footnotesize
\setlength{\tabcolsep}{4pt}
\begin{tabular}{@{}lrrr@{}}
\toprule
Comparison & Adaptive & Neutralised & Equal weight \\
\midrule
C-GP $-$ A-GP & $+0.0117$ $[-0.0082, +0.0316]$, 7/10 & $+0.0087$ & $+0.0290$ $[-0.0039, +0.0619]$, 7/10 \\
GFlowNet C $-$ A & $+0.0097$ $[-0.0157, +0.0351]$, 6/10 & $+0.0024$ & $+0.0231$ $[-0.0067, +0.0529]$, 5/10 \\
\bottomrule
\end{tabular}
\end{table*}

\textbf{Results.} Table~\ref{tab:gp} gives the comparison; both intervals include zero, and most of the difference sits in f2019\_20 ($+0.0244$ adaptive) rather than f2021 ($-0.0010$). Arm levels (adaptive IC): C-GP 0.0221, A-GP 0.0104, against 0.0227 and 0.0130 for the GFlowNet arms. The guard rejected 31.4\% of unique programs (28--35\% per run), so a guarded run computed 5,263 IC values on average against 7,079 and took 12.9 minutes against 20.7. Every C-GP factor is certified by the frozen evaluator; 41.6\% of A-GP factors are, against 28.8\% for the unguarded GFlowNet. Per run, 33--82\% of unique candidates are valid programs (parsable, within the token cap and inside the data range), every pool holds 50 factors, and none of the 1,000 pool factors has a discovery IC of exactly zero, the mark of the constant programs that the official pool rule would admit. With a tournament of 600 out of 1,000, both arms follow similar search paths when the fittest programs are certified, and they then differ mainly in what reaches the pool. Limitations: CSI300 only, two folds, 10 pairs, and budgets that differ from those of the other generators.

\section{Directional Audit with Real Anchors}\label{supp:realanchor}

\textbf{Design (frozen before the run).} Section~7.1 audits anchor sensitivity with simulated stock-specific re-anchorings and reports absolute drift. Here the second anchor is real: for each fold the corrected provider is re-anchored at the fold's discovery cutoff $T_c$ (2014-12-03, 2016-12-02, 2018-11-30, 2020-12-03), so that $p'_t = p_t / c_i$ and $v'_t = v_t c_i$ with $c_i = A_{T_c,i} / A_{T_{\mathrm{end}},i}$, read from the provider's end-anchored factor file (whose last value is 1.0 for every stock; stocks listed after $T_c$ keep $c_i = 1$). This is the vintage a snapshot taken at $T_c$ would hold for days up to $T_c$ and the same anchor continued afterwards; it is not the sequence of values available day by day. A global placebo scales every stock by the median $c_i$. The frozen audit script of Section~7.1 was extended by these two scenarios only (the a50/a90 scenarios reproduce the laptop baseline: 78 of 80 pool values and 79 of 80 factor summaries within $10^{-6}$, the rest within $3 \times 10^{-4}$), and the signed change $\Delta\mathrm{IC} = \mathrm{IC}(\text{stored}) - \mathrm{IC}(\text{re-anchored})$ is reported for the 40 arm-A and 40 arm-C pools.

\begin{table*}[!t]
\caption{Signed change in validation IC, stored end-anchored snapshot minus cutoff-anchored vintage, 40 pools per arm (mean, 95\% paired-$t$ CI, pools positive, Wilcoxon $p$).}
\label{tab:realanchor}
\centering
\footnotesize
\setlength{\tabcolsep}{4pt}
\begin{tabular}{@{}llrrrr@{}}
\toprule
Arm & Scenario & Adaptive & $>0$ & Equal weight & $>0$ \\
\midrule
A & cutoff, stock-specific & $+0.0062$ $[+0.0031, +0.0092]$ & 26/40 & $+0.0077$ $[+0.0048, +0.0106]$ & 31/40 \\
A & cutoff, global placebo & $+0.0006$ $[-0.0010, +0.0021]$ & 24/40 & $+0.0008$ $[-0.0001, +0.0017]$ & 23/40 \\
C & cutoff, stock-specific & $+0.0007$ $[-0.0002, +0.0015]$ & 24/40 & $+0.0010$ $[+0.0001, +0.0018]$ & 25/40 \\
C & cutoff, global placebo & $-0.0001$ $[-0.0010, +0.0008]$ & 21/40 & $+0.0001$ $[-0.0007, +0.0008]$ & 24/40 \\
\bottomrule
\end{tabular}
\end{table*}

\textbf{Results.} Table~\ref{tab:realanchor} gives the pool-level changes (Wilcoxon $p = 7 \times 10^{-4}$ and $3 \times 10^{-6}$ for arm A under the adaptive and equal-weight combiners; $0.15$ and $0.05$ for arm C). The change is concentrated in the two earliest folds (arm A, adaptive: $+0.0137$ and $+0.0125$ on f2015\_16 and f2017\_18, $-0.0039$ on f2019\_20, $+0.0023$ on f2021), where the cutoff lies furthest from the snapshot's end and $c_i$ departs most from 1. At factor level the signed change of the 2,000 arm-A factors has mean $-0.00003$ and standard deviation $0.0085$ (31\% beyond $\pm 0.005$; 4\% among the 451 certified factors against 39\% among the rest), so the pool-level gain is a property of the pools the search assembled, not of a one-sided shift of every factor. Limitations: one provider, four cutoffs, and a comparison of two vintages, which bounds but does not reproduce the day-by-day history.

\textbf{Factor families.} For the 1,549 non-certified factors of the 40 arm-A pools we split the signed change by the feature that makes them non-certified: a cross-sectional rank of a price-level expression (393 factors), a comparison of a price with a constant (350), a power of a price (281), a price added to or subtracted from a constant or another quantity (228), a logarithm of a price (102), a bare field (10) and the rest (185). The mean signed change is within $\pm 0.001$ of zero in every family except the bare fields ($-0.009$), while the share of factors that lose more than 0.005 of absolute IC under the cutoff anchor is 16\% for the ranks, 18\% for the comparisons, 21\% for the powers, 13\% for the sums, 21\% for the logarithms, 24\% for the rest and 6 of the 10 bare fields, against 1\% of the 451 certified factors. No family loses IC on average; the pool-level overstatement of Table~\ref{tab:realanchor} arises when the combiner selects and weights factors whose values moved, which is why it is larger than any factor-level mean. The split is by regular expression on the stored formula, and a factor is counted once, in the first family that matches.

\textbf{Published methods (Section~7.1 of the main paper).} The frozen real-anchor script (\texttt{real\_anchor/anchor\_audit\_cutoff.py}, unchanged) was run on the existing pools of Section~\ref{supp:retro}: five seeds per market of each external miner and of our certified and unrestricted arms, discovery through 2021-12-03 and evaluation on 2022-01-04 -- 2024-12-03, cutoff anchor $T_c = $ 2021-12-03, original operators in float32 (protocol \texttt{external\_anchor/PROTOCOL\_external\_anchor.md}, written before the run; 80 pools, no failure). Table~\ref{tab:externalanchor} gives the re-anchoring sensitivity, IC on the stored representation minus IC on the cutoff-anchored one. Over the 50 pools of the five published methods it is $+0.0050$ $[+0.0030, +0.0070]$ under the adaptive combiner (42 of 50 positive; 13.6\% of their mean IC of 0.0368) and $+0.0076$ $[+0.0061, +0.0092]$ under equal weighting (49 of 50; 24.3\% of 0.0315), against $+0.0010$ and $+0.0011$ under the global placebo. The external pools were discovered by an earlier benchmark pipeline on the reference snapshot (Section~\ref{supp:retro}), so the comparison measures how their stored performance depends on the representation, not how the original papers' reported numbers would change.

\begin{table*}[!t]
\caption{Re-anchoring sensitivity of existing pools on 2022--2024: IC on the stored end-anchored representation minus IC on the cutoff-anchored one (mean over 10 pools, 95\% paired-$t$ CI, pools positive). Stored: mean adaptive IC on the stored representation.}
\label{tab:externalanchor}
\centering
\scriptsize
\setlength{\tabcolsep}{2pt}
\begin{tabular}{@{}lrrrrr@{}}
\toprule
Pools & Stored & Adaptive & $>0$ & Equal weight & $>0$ \\
\midrule
AlphaGen & 0.0440 & +0.0065 $[+0.0018, +0.0112]$ & 8/10 & +0.0101 $[+0.0061, +0.0142]$ & 10/10 \\
AlphaSAGE & 0.0454 & +0.0025 $[+0.0002, +0.0047]$ & 8/10 & +0.0086 $[+0.0034, +0.0137]$ & 9/10 \\
AlphaForge & 0.0350 & +0.0101 $[+0.0054, +0.0148]$ & 10/10 & +0.0057 $[+0.0025, +0.0089]$ & 10/10 \\
AlphaQCM (IQN) & 0.0371 & +0.0023 $[-0.0053, +0.0099]$ & 8/10 & +0.0084 $[+0.0048, +0.0120]$ & 10/10 \\
Genetic programming & 0.0224 & +0.0037 $[+0.0013, +0.0062]$ & 8/10 & +0.0055 $[+0.0027, +0.0083]$ & 10/10 \\
Random search & 0.0358 & +0.0039 $[+0.0012, +0.0065]$ & 9/10 & +0.0075 $[+0.0045, +0.0104]$ & 10/10 \\
Unrestricted (ours, A) & 0.0396 & +0.0029 $[-0.0001, +0.0060]$ & 7/10 & +0.0025 $[+0.0006, +0.0043]$ & 9/10 \\
Certified (ours, C) & 0.0447 & -0.0010 $[-0.0019, -0.0001]$ & 3/10 & +0.0002 $[-0.0005, +0.0009]$ & 6/10 \\
\bottomrule
\end{tabular}
\end{table*}

\section{Past-Anchored Provider, Seed Extension and the Unrestricted Baseline}\label{supp:hfq}

\textbf{Design (frozen before the runs).} The protocol, written before any run, asked three questions: what the unrestricted grammar finds on data re-anchored in the past, whether a certified search returns the same pool on both representations, and how the main comparison behaves with ten seeds per cell. An amendment before the first formal run added two blocks: CSI300 seeds 5--9 on the clean provider, so that both markets have ten seeds, and the lower-magnitude injection of Section~\ref{supp:inject}; a third block, both arms on the union universe of all 1,730 stocks (40 runs), was withdrawn before any of its runs started, to stay within the budget, and is the reason the paper tests two universes only. The runs use the commands of the confirmatory stage with only the provider path, guard, seed and output interfaces changed, as in Sections~\ref{supp:provenance} and~\ref{supp:inject}.

\textbf{Past-anchored provider.} \texttt{MainProviderHfqClean} is the clean provider (corrected provider without SH000905 in the CSI500 membership file) transformed by the anchor group action with $c_i = A_{\mathrm{first},i} / A_{T_{\mathrm{end}},i}$, the stored end-anchored factor on stock $i$'s first valid day (its last value is 1 for every stock): open, high, low, close and VWAP divided by $c_i$, volume multiplied by $c_i$, and the factor file rewritten to $A_{t,i}/A_{\mathrm{first},i}$, which equals 1 on the first day and no longer changes when a later corporate action is appended, as in a BaoStock backward-adjusted (\emph{hfq}) download. Validation on all 1,730 stocks: the ratio of each rewritten price field to the original is constant per stock to $6 \times 10^{-8}$; the first close equals the raw close; amount $=$ VWAP $\times$ volume holds wherever it held before; the VWAP range test of Section~6.1 of the main paper is unchanged except for 658 boundary cells that now pass; every other file is byte-identical. The median of $A_{T_{\mathrm{end}}}/A_{\mathrm{first}}$ over stocks is 4.7 (quartiles 2.4--9.5), so the two representations differ by a factor that is large for most stocks.

\textbf{Runs and evaluation.} 164 runs on one 2$\times$RTX~4090 machine (AMD EPYC~9654): AH, the unrestricted grammar on the past-anchored provider, 4 folds $\times$ 2 markets $\times$ 5 seeds; CH, the certified grammar on the past-anchored provider, f2019\_20, seed 0, both markets; KC and KA, the original certified and unrestricted commands on the corrected provider (f2019\_20, CSI500, seed 0), which reproduced the confirmatory-stage pools byte for byte, so the machine and software stack reproduce the earlier runs; CX5/AX5 and C5/A5, the certified and unrestricted grammars on the clean provider with seeds 5--9 (CSI500 and CSI300, 40 runs each); CJ4/AJ4, the /4 injection (40 runs). Every pool was evaluated in the laptop environment with the frozen evaluator on the clean end-anchored provider; the AH and CH pools and the 80 pools of the confirmatory stage also on the past-anchored provider. A paired comparison never mixes evaluation providers.

\begin{table*}[!t]
\caption{Past-anchored baseline (adaptive combiner unless stated; 40 fold $\times$ market $\times$ seed units; mean, 95\% paired-$t$ CI, SD, units positive). Drift: IC on the past-anchored provider minus IC on the end-anchored clean provider for the same pool. AH: unrestricted grammar mined on the past-anchored provider; A and C: the pools of the confirmatory stage (CSI300) and the clean re-mining (CSI500).}
\label{tab:hfq}
\centering
\footnotesize
\setlength{\tabcolsep}{4pt}
\begin{tabular}{@{}llrrr@{}}
\toprule
Quantity & Evaluated on & Mean [95\% CI] & SD & $>0$ \\
\midrule
\multicolumn{5}{@{}l}{\textit{Representation drift of fixed pools}} \\
C pools, drift & both & $+0.0006$ $[-0.0007, +0.0020]$ & 0.0042 & 21/40 \\
A pools, drift & both & $+0.0001$ $[-0.0075, +0.0077]$ & 0.0237 & 20/40 \\
AH pools, drift & both & $+0.0011$ $[-0.0051, +0.0073]$ & 0.0193 & 23/40 \\
C pools, drift, equal weight & both & $-0.0004$ $[-0.0019, +0.0010]$ & 0.0045 & 21/40 \\
A pools, drift, equal weight & both & $-0.0006$ $[-0.0120, +0.0108]$ & 0.0357 & 15/40 \\
AH pools, drift, equal weight & both & $+0.0002$ $[-0.0040, +0.0044]$ & 0.0131 & 22/40 \\
\addlinespace
\multicolumn{5}{@{}l}{\textit{Paired differences}} \\
AH $-$ A & past-anchored & $-0.0015$ $[-0.0079, +0.0049]$ & 0.0199 & 21/40 \\
AH $-$ A & end-anchored & $-0.0025$ $[-0.0086, +0.0036]$ & 0.0192 & 19/40 \\
AH $-$ A, neutralised & past-anchored & $-0.0036$ $[-0.0099, +0.0027]$ & 0.0197 & 21/40 \\
AH $-$ A, equal weight & past-anchored & $+0.0061$ $[+0.0004, +0.0117]$ & 0.0176 & 29/40 \\
C $-$ AH & past-anchored & $+0.0070$ $[+0.0010, +0.0129]$ & 0.0187 & 29/40 \\
C $-$ AH & end-anchored & $+0.0074$ $[-0.0002, +0.0151]$ & 0.0240 & 26/40 \\
C $-$ AH, neutralised & past-anchored & $+0.0050$ $[-0.0010, +0.0109]$ & 0.0185 & 23/40 \\
C $-$ AH, equal weight & past-anchored & $-0.0010$ $[-0.0075, +0.0055]$ & 0.0202 & 23/40 \\
C $-$ A & past-anchored & $+0.0055$ $[-0.0021, +0.0131]$ & 0.0238 & 23/40 \\
C $-$ A & end-anchored & $+0.0049$ $[-0.0036, +0.0135]$ & 0.0267 & 25/40 \\
\bottomrule
\end{tabular}
\end{table*}

\textbf{Results.} Table~\ref{tab:hfq} gives the drift and the paired differences; levels on the past-anchored provider are 0.0329 (A), 0.0314 (AH) and 0.0384 (C). Per fold, C $-$ AH on the past-anchored provider is $+0.0025$, $-0.0021$, $+0.0205$ and $+0.0073$ (f2015\_16 to f2021), the same pattern as C $-$ A. The AH pools contain 14.7 certified factors of 50 on average (arm A: 11.5): the certified share rises modestly, but most AH factors remain uncertified. The frozen anchor audit of Section~\ref{supp:cleanaudit} on the AH pools, run on the past-anchored provider, gives a stock-specific adaptive pool $|\Delta\mathrm{IC}|$ of 0.0151 against 0.0048 under the matched global placebo (ratio 3.2), 44.0\% of factors with correlation below 0.9 under stock-specific anchors against 19.7\% under the placebo, and 24.3\% with excess drift above 0.005, the pattern of arm A; the two CH pools give 0.0004 against 0.0009 (ratio 0.5), 15.0\% / 15.0\% and 0\% (Table~\ref{tab:cleanaudit}).

\textbf{Certified-search identity test.} In the idealised semantics every certified expression has the same value on the two providers (Theorem~1 of the main paper), so with deterministic updates and representation-invariant rewards, exact arithmetic would yield identical search trajectories; Theorem~1 alone guarantees only expression-level invariance, and the two CH runs measure the float32 trajectory: neither pool is byte-identical to its end-anchored counterpart (the confirmatory-stage C pool for CSI300, the clean CX pool for CSI500); they share 25 and 27 of 50 expressions (Jaccard 0.33 and 0.375; the CSI500 pool stores one expression twice, so its set has 49), every expression in them is certified, and their adaptive IC on the past-anchored provider is 0.0546 against 0.0618 (CSI300) and 0.0380 against 0.0526 (CSI500) for the end-anchored pools evaluated on the same provider, with the AH pools of the same cells at 0.0452 and 0.0286. The two KC/KA controls show that the machine reproduces the end-anchored commands byte for byte, so the divergence is consistent with float32 differences between the two representations propagating through the reward and the sampler, of the order the evaluator shows between float32 and float64 (Section~\ref{supp:float64}).

\textbf{Seed extension.} The 40 CX5/AX5 and C5/A5 pools enter the main comparison of Section~7.2 of the main paper (Table~6; Section~\ref{supp:detailtables}). Over the 40 new units alone C $-$ A is $+0.0039$ $[-0.0040, +0.0118]$ (23/40) under the adaptive combiner, $-0.0000$ neutralised and $+0.0073$ under equal weighting.

\section{Float64 Evaluator and the Platform Sensitivity of the Combiner}\label{supp:float64}

\textbf{Design (frozen before the runs).} Section~6.4 of the main paper reports that the adaptive IC of a unit differs between our cloud and laptop environments by about 0.005--0.010. The pre-specified test looked for a laptop-side proxy of the cross-platform difference: the thread count of the linear-algebra libraries (E1) left every output bit-identical, so the frozen contingency, a split-order reduction of the cross-sectional statistics (E3), was used against the native reduction (E2). The 80 pools of the confirmatory stage (40 C, 40 A) were evaluated under four conditions, float32 and float64 $\times$ E2 and E3, with the frozen evaluator (the float64 build casts the panel and every operator to double; the combiner's thresholds are unchanged). The float32 E2 condition reproduces the archived laptop rows exactly (unit rows exact in 80/80, daily IC bitwise).

\begin{table}[!htb]
\caption{Per-unit differences of adaptive IC under the four evaluator conditions (80 units of the confirmatory stage; SD and largest absolute difference; units in which the combiner's mean number of selected factors differs). Environment: E3 minus E2 at fixed precision. Precision: float64 minus float32 at fixed environment. Cloud: archived cloud float32 rows minus laptop float32 E2 rows, for reference.}
\label{tab:float64}
\centering
\scriptsize
\setlength{\tabcolsep}{2.5pt}
\begin{tabular}{@{}llrrr@{}}
\toprule
Contrast & Combiner & SD & Max $|\Delta|$ & Sel. differs \\
\midrule
Environment, float32 & adaptive & 0.00016 & 0.0009 & 12/80 \\
Environment, float64 & adaptive & 0.00001 & 0.0001 & 4/80 \\
Environment, float32 & equal weight & 0.00003 & 0.0002 & -- \\
Precision, E2 & adaptive & 0.0052 & 0.022 & 45/80 \\
Precision, E2 & neutralised & 0.0049 & 0.015 & 45/80 \\
Precision, E2 & equal weight & 0.0048 & 0.019 & -- \\
Cloud $-$ laptop, float32 & adaptive & 0.005--0.006 & 0.022 & 45/80 \\
\bottomrule
\end{tabular}
\end{table}

\textbf{Results.} Table~\ref{tab:float64} gives the contrasts. The environment proxy moves the adaptive IC by at most 0.0009 in float32 (0.0001 in float64), two orders of magnitude below the cloud--laptop difference, so the question of what differs between the two platforms is not answered by this experiment. The precision contrast is as large as the cloud--laptop difference: float64 and float32 differ by 0.005 per unit (one SD), the daily selected set of the adaptive combiner differs on at least one day in 76 of 80 units (on 311 validation days per unit on average), and the equal-weight sign of at least one factor flips in 54 units. The share of factor-days with a matured $|\mathrm{RankIC}|$ or $|\mathrm{RankICIR}|$ within $10^{-5}$ of a threshold is $6 \times 10^{-4}$, enough for the selection to differ somewhere in every unit. The paired comparison is stable: under float64 the 40-unit C $-$ A is $+0.0090$ $[+0.0009, +0.0171]$ (26/40, 6/8 cells; neutralised $+0.0031$) against $+0.0096$ $[+0.0019, +0.0172]$ in float32, and $+0.0104$ against $+0.0076$ under equal weighting. The value-level invariance classification used in Section~\ref{supp:numerics} is itself precision-dependent (mean 11.0 certified factors per pool classified invariant in float32 against 22.3 in float64, differing in 79 of 80 units), which is why the paper's certification is symbolic. The run time is 33 s per unit in float32 and 44 s in float64. We recommend a float64 reference evaluator for threshold-selecting combiners, with the float32 result reported beside it.

\textbf{The clean 80-unit pools.} After the confirmatory-stage test, the 160 pools behind Table~6 of the main paper (arms A and C, both markets, seeds 0--9, clean provider) were re-evaluated under float64 at the archived environment (E2; 81 min on the laptop, 160 CPU units). Per unit, float64 minus float32 adaptive IC has mean $-0.0003$ and SD 0.0067 (86 of 160 units differ by more than 0.002, and the combiner's mean number of selected factors differs in 92), neutralised IC SD 0.0049 and equal-weight IC SD 0.0044. The largest change is a single unrestricted pool on the short f2021 fold (A5, CSI300, seed 6: IC $+0.0063$ in float32, $-0.0545$ in float64, over 223 validation days), where the combiner's daily selection differs under the two precisions; the next two are $-0.022$ and $-0.017$, so the per-unit stability claimed here is for the paired comparison, not for every pool. The paired comparison does not move: C $-$ A is $+0.0048$ $[-0.0010, +0.0106]$ (48/80 units, 5/8 cells, 2/4 folds, cell-level interval $[-0.0078, +0.0174]$) against $+0.0044$ in float32, the neutralised difference $-0.0001$ against $-0.0005$, and the equal-weight difference $+0.0075$ $[-0.0005, +0.0155]$ against $+0.0061$; the engineering criteria hold under both precisions, and the per-fold pattern is the same (f2019\_20 $+0.0266$, the other three folds within $\pm 0.007$ of zero).

\section{Additional Paired Comparisons}\label{supp:final-moved}

This section holds the comparisons and paragraphs that the main paper summarises in one sentence or one table row.

\textbf{Clean-data control arms (Section~7.3 of the main paper).} Under a protocol written before the runs (\texttt{clean\_controls/PROTOCOL\_clean\_controls.md}), the frozen arm-B and arm-S commands were re-run on CSI500 with the membership file without SH000905 (4 folds $\times$ 5 seeds per arm, one rented 2$\times$RTX~4090 machine), with only the provider path, identifiers and output root changed; two control runs of the original commands on the original data reproduced the archived B and S pools byte for byte. The CSI300 pools of B and S are unaffected by the ticker and are reused. Table~\ref{tab:cleancontrols} pairs each arm only with the A and C pools of the same fold, market and seed (laptop float32 evaluation, the environment of Table~6 of the main paper); the original-file column uses the same 40 units evaluated in the same environment. Levels on clean data (adaptive): A 0.0328, B 0.0288, S 0.0349, C 0.0377. The attribution criterion $(\mathrm{C} - \mathrm{B}) \ge (\mathrm{B} - \mathrm{A}) - 0.002$ holds on clean data under all three combiners; its application to clean data is post hoc.

\begin{table*}[!t]
\caption{Control arms on clean and original data (40 units each; mean, 95\% paired-$t$ CI, units positive).}
\label{tab:cleancontrols}
\centering
\scriptsize
\setlength{\tabcolsep}{2pt}
\begin{tabular}{@{}lll@{}}
\toprule
Pair & Clean data & Original file \\
\midrule
Adaptive, C $-$ B & $+0.0089$ $[+0.0003, +0.0176]$, 26/40 & $+0.0119$ $[+0.0043, +0.0194]$, 25/40 \\
Adaptive, B $-$ A & $-0.0040$ $[-0.0094, +0.0014]$, 13/40 & $-0.0023$ $[-0.0079, +0.0034]$, 18/40 \\
Adaptive, C $-$ S & $+0.0029$ $[-0.0034, +0.0091]$, 23/40 & $+0.0043$ $[-0.0012, +0.0098]$, 24/40 \\
Adaptive, S $-$ A & $+0.0021$ $[-0.0055, +0.0096]$, 25/40 & $+0.0053$ $[-0.0022, +0.0129]$, 25/40 \\
\addlinespace
Neutralised, C $-$ B & $+0.0035$ $[-0.0043, +0.0114]$, 26/40 & $+0.0072$ $[+0.0009, +0.0135]$, 25/40 \\
Neutralised, B $-$ A & $-0.0046$ $[-0.0092, -0.0000]$, 12/40 & $-0.0036$ $[-0.0085, +0.0012]$, 15/40 \\
Neutralised, C $-$ S & $+0.0001$ $[-0.0044, +0.0047]$, 22/40 & $+0.0013$ $[-0.0025, +0.0050]$, 21/40 \\
Neutralised, S $-$ A & $-0.0012$ $[-0.0082, +0.0059]$, 17/40 & $+0.0022$ $[-0.0039, +0.0084]$, 22/40 \\
\addlinespace
Equal weight, C $-$ B & $+0.0065$ $[-0.0038, +0.0167]$, 20/40 & $+0.0086$ $[-0.0012, +0.0183]$, 22/40 \\
Equal weight, B $-$ A & $-0.0016$ $[-0.0061, +0.0029]$, 20/40 & $-0.0009$ $[-0.0051, +0.0032]$, 16/40 \\
Equal weight, C $-$ S & $-0.0016$ $[-0.0059, +0.0026]$, 19/40 & $-0.0004$ $[-0.0041, +0.0032]$, 19/40 \\
Equal weight, S $-$ A & $+0.0065$ $[-0.0052, +0.0181]$, 21/40 & $+0.0081$ $[-0.0028, +0.0189]$, 21/40 \\
\bottomrule
\end{tabular}
\end{table*}

\textbf{Confirmatory stage of the main comparison (original-membership pools; Table~\ref{tab:summary} of the main paper).} Table~\ref{tab:confirmatory} gives the paired differences of the confirmatory stage, mined with the original CSI500 membership file and evaluated in the cloud environment, including the coarse-typed arm B. Levels were 0.0277 (A), 0.0262 (B) and 0.0391 (C); the three criteria of the decision rule were met, and the cell-level interval $[-0.0003, +0.0231]$ already included zero. In the laptop environment the same pools give C $-$ A $= +0.0096$. Under equal weighting the confirmatory difference was $+0.0083$ (23/40), concentrated in CSI300.

\begin{table}[!htb]
\caption{Paired differences of pool IC on the validation windows in the confirmatory stage (adaptive combiner, 40 units, original CSI500 membership file, cloud evaluation) between the certified (C), coarse-typed (B) and unrestricted (A) arms.}
\label{tab:confirmatory}
\centering
\scriptsize
\setlength{\tabcolsep}{2pt}
\begin{tabular}{@{}lccccc@{}}
\toprule
Pair & $\Delta$IC & 95\% CI & Units $>0$ & Cells $>0$ & Neut. $\Delta$IC \\
\midrule
C $-$ A & $+0.0114$ & $[+0.0045, +0.0184]$ & 29/40 & 7/8 & $+0.0060$ \\
B $-$ A & $-0.0015$ & & 19/40 & 2/8 & $-0.0025$ \\
C $-$ B & $+0.0129$ & & 27/40 & 6/8 & $+0.0085$ \\
\bottomrule
\end{tabular}
\end{table}

\textbf{The mislisted constituent (Section~7.5 of the main paper).} Table~\ref{tab:sh000905} gives the re-evaluation and the re-mining side by side.

\begin{table}[!htb]
\caption{Effect of the mislisted index ticker SH000905 on the CSI500 searches (laptop environment; Section~7.5 of the main paper). Both columns are evaluated on the provider without the ticker and differ only in the data used for mining; the 40-unit rows add the unchanged CSI300 units. Brackets: 95\% CI; fractions: units positive. The 40-unit value with the ticker ($+0.0101$) is a laptop evaluation on the provider without it; the same pools give $+0.0096$ (laptop) and $+0.0114$ (cloud) on the corrected provider.}
\label{tab:sh000905}
\centering
\scriptsize
\setlength{\tabcolsep}{3pt}
\begin{tabular}{@{}>{\raggedright\arraybackslash}p{0.33\columnwidth}>{\raggedright\arraybackslash}p{0.30\columnwidth}>{\raggedright\arraybackslash}p{0.30\columnwidth}@{}}
\toprule
 & Mined with the ticker (original pools) & Mined without the ticker \\
\midrule
CSI300, C $-$ A, adaptive (not re-mined) & $+0.0083$ $[-0.0043, +0.0209]$, 12/20 & same pools \\
CSI500, C $-$ A, adaptive & $+0.0120$ $[+0.0020, +0.0219]$, 15/20 & $+0.0016$ $[-0.0110, +0.0141]$, 13/20 \\
CSI500, C $-$ A, neutralised & $+0.0067$ & $-0.0049$ \\
CSI500, C $-$ A, equal weight & $-0.0001$ & $-0.0052$ \\
40 units, C $-$ A, adaptive & $+0.0101$ $[+0.0025, +0.0178]$, 27/40 & $+0.0049$ $[-0.0036, +0.0135]$, 25/40 \\
40 units, C $-$ A, neutralised & $+0.0048$ & $-0.0010$ \\
Decision-rule criteria ($\ge -0.005$) & met & met \\
CSI500 mean adaptive IC, A / C & 0.0329 / 0.0448 & 0.0394 / 0.0410 \\
Expressions kept from the original pool, C / A & 50 / 50 & 26.3 / 14.6 \\
\bottomrule
\end{tabular}
\end{table}

\textbf{S\&P~500 stress test, pool level (Section~7.1 of the main paper).} Adaptive IC drifts by 0.0045 under stock-specific re-anchoring against 0.0023 under the global placebo for the certified pools and by 0.0264 against 0.0177 for the unrestricted pools: the certified drift is far below the unrestricted drift but above the placebo.

\textbf{Mechanism (Section~7.3 of the main paper, full text).} On the 2,000 factors of the 40 arm-A pools, the data do not support the explanation that non-invariant factors act as stock identifiers through the adjustment factor: their cross-sectional $R^2$ on $\log F_{t,i}$ and their IC decay do not differ from those of certified factors (Sections~\ref{supp:mechanism} and~\ref{supp:moved-exp}). What does differ is sign stability: the out-of-sample IC changes sign less often for certified factors ($p = 0.031$, pool-level paired Wilcoxon test over the 40 pools, our primary statement). Pooled over factors, the flip rate is 33.5\% for certified against 40.0\% for non-certified factors (Fisher $p = 0.013$; exploratory, as factors within a pool are dependent). The association does not identify the mechanism of the difference.

\textbf{Certified-search identity runs (Section~7.5 of the main paper, full text).} Two certified-search runs on the past-anchored provider (arm CH, f2019\_20, one seed per market) tested the search itself: in exact arithmetic every certified expression takes the same value on both providers, but the float32 trajectories diverged, the pools sharing 25 and 27 of 50 expressions with their end-anchored counterparts, a turnover comparable to that under the ticker removal of Section~7.5 of the main paper (adaptive IC within 0.007 and 0.015 of theirs), while two control runs of the original commands reproduced their pools byte for byte on the same machine. The certificate is a statement about expressions, not about a training trajectory; the divergence is consistent with finite-precision differences propagating through the reward and the sampler, of the kind Section~6.4 of the main paper measures for the evaluator, and a float64 search, the direct test, was not run (Section~\ref{supp:hfq}).

\section{Seed-to-Seed Reference for the Discovery Channel}\label{supp:seedref}

Section~7.5 of the main paper compares the change that a data defect causes with the change between two seeds. The like-for-like reference is not the SD of C $-$ A across seeds but the SD of the difference in C $-$ A between two pools of one cell mined with different seeds, $D_{s'} - D_s$ with $D_s = \mathrm{IC}_{C,s} - \mathrm{IC}_{A,s}$, since a perturbation compares $D$ after and before the change for the same seed. Over the 10 seeds of the clean main comparison, within each fold $\times$ market cell, it is 0.0309 over the 180 CSI500 seed pairs and 0.0317 over all 360 pairs under the adaptive combiner (neutralised 0.0345, equal weight 0.0243 on CSI500), close to $\sqrt{2}$ times the within-cell SD of $D$ (0.0220). The pairs share data and seeds and are not independent; the figure is a scale, not a test. Two pools of one arm and cell mined with different seeds share on average 0.5 of 50 expressions for the certified search and 0.2 for the unrestricted search (CSI500; 0.6 and 0.2 over all cells), so the higher retention of the certified pools under a data change in Section~7.5 does not come from a higher overlap between seeds. Script and output: \texttt{seed\_reference.py}, \texttt{REPORT\_seed\_reference.md} in the repository.

\end{document}